\documentclass[journal,draftclsnofoot,onecolumn,12pt,oneside]{IEEEtran} % for single column no footer
\usepackage{cite}
\ifCLASSINFOpdf
  \usepackage[pdftex]{graphicx}
  \usepackage{epstopdf} % remove this
\else
\fi
\usepackage{amsmath}
\usepackage{algorithm}
\usepackage{algorithmic}

  \usepackage[caption=false,font=footnotesize]{subfig}
\usepackage{url}
\usepackage{amsthm}

\theoremstyle{definition}

\newtheorem*{remark}{Takeaways}

\usepackage{diagbox}

\usepackage{makecell}
\newcommand{\comment}[1]{{}}

\usepackage{amssymb}
\usepackage{xspace}
\usepackage{bm}
\usepackage{bbm}

\newcommand{\set}[1]{\ensuremath{\mathcal{#1}}\xspace} % caligraphic set notation
\newcommand{\mat}[1]{\ensuremath{\mathbf{#1}}\xspace} % matrices
\renewcommand{\vec}[1]{\ensuremath{\mathbf{#1}}\xspace} % vectors
\newcommand{\parens}[1]{\ensuremath{\left(#1\right)}\xspace}
\newcommand{\brackets}[1]{\ensuremath{\left[#1\right]}\xspace}
\newcommand{\braces}[1]{\ensuremath{\left\{#1\right\}}\xspace}
\newcommand{\bars}[1]{\ensuremath{\left\vert#1\right\vert}\xspace}

\newcommand{\complex}{\ensuremath{\mathbb{C}}\xspace}

\newcommand{\card}[1]{\bars{#1}}

\newcommand{\setcomplex}{\ensuremath{\complex}}

\newcommand{\setmatrix}[3]{\ensuremath{#1^{#2 \times #3}}\xspace}

\newcommand{\setmatrixcomplex}[2]{\setmatrix{\setcomplex}{#1}{#2}}

\newcommand{\inv}{\ensuremath{^{-1}}\xspace}

\newcommand{\trans}{\ensuremath{^{\mathrm{T}}}\xspace}

\newcommand{\distgauss}[2]{\ensuremath{\mathcal{N}\parens{#1,#2}}\xspace} % Gaussian distribution
\newcommand{\distgamma}[2]{\ensuremath{\mathrm{Gamma}\parens{#1,#2}}\xspace}
\newcommand{\ev}[1]{\ensuremath{\mathbb{E}\brackets{#1}}\xspace}

\newcommand{\erf}[1]{\ensuremath{\mathrm{erf}\parens{#1}}\xspace}

\newcommand{\prob}[1]{\ensuremath{\mathbb{P}\parens{#1}}\xspace}

\newcommand{\todB}[1]{\ensuremath{\brackets{#1}_{\mathrm{dB}}}}
\newcommand{\todBm}[1]{\ensuremath{\brackets{#1}_{\mathrm{dBm}}}}

\newcommand{\powertx}{\ensuremath{P_{\mathrm{tx}}}\xspace}

\newcommand{\powernoise}{\ensuremath{P_{\mathrm{noise}}}\xspace}
\newcommand{\powerdes}{\ensuremath{P_{\mathrm{des}}}\xspace}

\newcommand{\powersi}{\ensuremath{P_{\mathrm{SI}}}\xspace}

\newcommand{\snr}{\ensuremath{\mathsf{SNR}}\xspace}

\newcommand{\sinr}{\ensuremath{\mathsf{SINR}}\xspace}

\newcommand{\inr}{\ensuremath{\mathsf{INR}}\xspace}

\newcommand{\idx}[1]{\ensuremath{^{\parens{#1}}}\xspace}

\newcommand{\II}{\ensuremath{\mathcal{I}}\xspace}
\newcommand{\IIij}{\ensuremath{\mathcal{I}_{ij}}\xspace}

\newcommand{\inrij}{\ensuremath{\inr_{ij}}\xspace}
\newcommand{\inrijmax}{\ensuremath{\inrij^{\mathrm{max}}}\xspace}
\newcommand{\inrijmin}{\ensuremath{\inrij^{\mathrm{min}}}\xspace}
\newcommand{\inrijrng}{\ensuremath{\inrij^{\mathrm{rng}}}\xspace}

\newcommand{\inrmax}{\ensuremath{\inr^{\mathrm{max}}}\xspace}
\newcommand{\inrmin}{\ensuremath{\inr^{\mathrm{min}}}\xspace}

\newcommand{\DDmin}{\ensuremath{\mathcal{D}_{\mathrm{min}}}\xspace}
\newcommand{\DDmax}{\ensuremath{\mathcal{D}_{\mathrm{max}}}\xspace}

\newcommand{\nbr}{\ensuremath{\parens{\Delta\theta,\Delta\phi}}\xspace}
\newcommand{\nbrinr}{\ensuremath{\parens{\Delta\theta,\Delta\phi,\inr}}\xspace}
\newcommand{\nbrinrij}{\ensuremath{\parens{\Delta\theta,\Delta\phi,\inrij}}\xspace}
\newcommand{\nbrzerozero}{\ensuremath{\parens{0^\circ,0^\circ}}\xspace}
\newcommand{\nbrzeroone}{\ensuremath{\parens{0^\circ,1^\circ}}\xspace}
\newcommand{\nbronezero}{\ensuremath{\parens{1^\circ,0^\circ}}\xspace}
\newcommand{\nbroneone}{\ensuremath{\parens{1^\circ,1^\circ}}\xspace}

\newcommand{\nbrtwotwo}{\ensuremath{\parens{2^\circ,2^\circ}}\xspace}
\newcommand{\nbrthreethree}{\ensuremath{\parens{3^\circ,3^\circ}}\xspace}

\newcommand{\LLij}{\ensuremath{\mathcal{L}_{ij}}\xspace}

\newcommand{\alpharng}{\ensuremath{\alpha_{\mathrm{rng}}}\xspace}
\newcommand{\betarng}{\ensuremath{\beta_{\mathrm{rng}}}\xspace}
\newcommand{\alphamin}{\ensuremath{\alpha_{\mathrm{min}}}\xspace}
\newcommand{\betamin}{\ensuremath{\beta_{\mathrm{min}}}\xspace}
\newcommand{\alphamax}{\ensuremath{\alpha_{\mathrm{max}}}\xspace}
\newcommand{\betamax}{\ensuremath{\beta_{\mathrm{max}}}\xspace}

\newcommand{\mumin}{\ensuremath{\mu_{\mathrm{min}}}\xspace}
\newcommand{\varmin}{\ensuremath{\sigma^2_{\mathrm{min}}}\xspace}
\newcommand{\stdmin}{\ensuremath{\sigma_{\mathrm{min}}}\xspace}
\newcommand{\mumax}{\ensuremath{\mu_{\mathrm{max}}}\xspace}
\newcommand{\varmax}{\ensuremath{\sigma^2_{\mathrm{max}}}\xspace}

\newcommand{\txdirsetmeas}{\ensuremath{\set{T}_{\mathrm{tx}}}\xspace}
\newcommand{\rxdirsetmeas}{\ensuremath{\set{T}_{\mathrm{rx}}}\xspace}

\newcommand{\txdirsetcb}{\ensuremath{\set{A}_{\mathrm{tx}}}\xspace}
\newcommand{\rxdirsetcb}{\ensuremath{\set{A}_{\mathrm{rx}}}\xspace}

\newcommand{\thetatx}{\ensuremath{\theta_{\mathrm{tx}}}\xspace}
\newcommand{\phitx}{\ensuremath{\phi_{\mathrm{tx}}}\xspace}

\newcommand{\thetarx}{\ensuremath{\theta_{\mathrm{rx}}}\xspace}
\newcommand{\phirx}{\ensuremath{\phi_{\mathrm{rx}}}\xspace}

\newcommand{\anglediff}[1]{\ensuremath{\measuredangle\parens{#1}}\xspace}

\newcommand{\Ntx}{\ensuremath{N_{\mathrm{tx}}}\xspace}
\newcommand{\Nrx}{\ensuremath{N_{\mathrm{rx}}}\xspace}

\def\vf{{\vec{f}}}

\def\vw{{\vec{w}}}

\def\mH{{\mat{H}}}

\usepackage{xspace}
\usepackage[acronym,nogroupskip,nonumberlist,nopostdot]{glossaries}
\makeglossaries

\usepackage{xcolor}

\newacronym{snr}{SNR}{signal-to-noise ratio}
\newacronym{sinr}{SINR}{signal-to-interference-plus-noise ratio}
\newacronym{inr}{INR}{interference-to-noise ratio}
\newacronym{sir}{SIR}{signal-to-interference ratio}
\newacronym{sqr}{SQR}{signal-to-quantization-noise ratio}
\newacronym{sqnr}{SQNR}{signal-to-quantization-plus-noise ratio}
\newacronym{ian}{IAN}{interference as noise}
\newacronym{ber}{BER}{bit error rate}
\newacronym{pn}{PN}{pseudorandom noise}
\newacronym{bfsk}{BFSK}{binary frequency shift keying}
\newacronym{fh}{FH}{frequency-hopped}
\newacronym{fh-bfsk}{FH-BFSK}{frequency-hopped binary frequency shift keying}
\newacronym{crc}{CRC}{cyclic redundancy check}
\newacronym{isi}{ISI}{intersymbol interference}
\newacronym{dsss}{DSSS}{direct-sequence spread spectrum}
\newacronym{ofdm}{OFDM}{orthogonal frequency-division multiplexing}
\newacronym{ofdma}{OFDMA}{orthogonal frequency-division multiple access}
\newacronym{sdr}{SDR}{software-defined radio}
\newacronym{tx}{TX}{transmitter}
\newacronym{rx}{RX}{receiver}
\newacronym{fdd}{FDD}{frequency-division duplexing}
\newacronym{tdd}{TDD}{time-division duplexing}
\newacronym{fdma}{FDMA}{frequency-division multiple access}
\newacronym{tdma}{TDMA}{time-division multiple access}
\newacronym{sdma}{SDMA}{space-division multiple access}
\newacronym[plural=MPCs]{mpc}{MPC}{multipath component}
\newacronym{mui}{MUI}{multi-user interference}

\newacronym{qam}{QAM}{quadrature amplitude modulation}
\newacronym{mqam}{MQAM}{M-ary quadrature amplitude modulation}

\newacronym{ls}{LS}{least-squares}
\newacronym{lms}{LMS}{least mean squares}
\newacronym{rls}{RLS}{recursive least-squares}
\newacronym{rzf}{RZF}{regularized zero-forcing}
\newacronym{mmse}{MMSE}{minimum mean square error}
\newacronym{lmmse}{LMMSE}{linear minimum mean square error}
\newacronym{mse}{MSE}{mean square error}
\newacronym{fft}{FFT}{fast Fourier transform}
\newacronym{dft}{DFT}{discrete Fourier transform}
\newacronym{dtft}{DTFT}{discrete-time Fourier transform}
\newacronym{ctft}{CTFT}{continuous-time Fourier transform}
\newacronym{ml}{ML}{machine learning}
\newacronym[plural=NNs]{nn}{NN}{neural network}
\newacronym[plural=RNNs]{rnn}{RNN}{recurrent neural network}
\newacronym[plural=ADCs]{adc}{ADC}{analog-to-digital converter}
\newacronym[plural=DACs]{dac}{DAC}{digital-to-analog converter}
\newacronym[plural=FPGAs]{fpga}{FPGA}{field-programmable gate array}
\newacronym{evm}{EVM}{error vector magnitude}
\newacronym{enob}{ENOB}{effective number of bits}
\newacronym{zf}{ZF}{zero-forcing}
\newacronym{rv}{r.v.}{random variable}
\newacronym{omp}{OMP}{orthogonal matching pursuit}
\newacronym{svd}{SVD}{singular value decomposition}

\newacronym{sdp}{SDP}{semidefinite programming}
\newacronym{psd}{PSD}{positive semidefinite}
\newacronym{nsd}{NSD}{negative semidefinite}

\newacronym{ks}{K-S}{Kolmogorov-Smirnov}

\newacronym{mad}{MAD}{median absolute deviation around the median}

\newacronym{agc}{AGC}{automatic gain control}
\newacronym{rf}{RF}{radio frequency}
\newacronym{if}{IF}{intermediate frequency}
\newacronym{los}{LOS}{line-of-sight}
\newacronym{nlos}{NLOS}{non-line-of-sight}
\newacronym{ple}{PLE}{path loss exponent}
\newacronym[plural=dB,firstplural=decibels (dB)]{db}{dB}{decibel}
\newacronym[plural=dBm,firstplural=decibel milliwatts (dBm)]{dbm}{dBm}{decibel milliwatts}
\newacronym{pa}{PA}{power amplifier}
\newacronym{lna}{LNA}{low noise amplifier}
\newacronym{cw}{CW}{continuous wave}
\newacronym{papr}{PAPR}{peak-to-average power ratio}
\newacronym{usrp}{USRP}{Universal Software Radio Peripheral}
\newacronym{irr}{IRR}{image rejection ratio}
\newacronym{lo}{LO}{local oscillator}
\newacronym{vm}{VM}{vector modulator}
\newacronym{mmwave}{mmWave}{millimeter wave}
\newacronym{eirp}{EIRP}{effective isotropic radiated power}
\newacronym{rsrp}{RSRP}{reference signal received power}

\newacronym{csma}{CSMA}{carrier-sense multiple access}
\newacronym{csmaca}{CSMA/CA}{carrier-sense multiple access with collision avoidance}
\newacronym{csmacd}{CSMA/CD}{carrier-sense multiple access with collision detection}
\newacronym{mac}{MAC}{medium access control}
\newacronym{phy}{PHY}{physical layer}
\newacronym{4g}{4G}{fourth generation}
\newacronym{lte}{LTE}{Long-Term Evolution}
\newacronym{4glte}{4G LTE}{\gls{4g} \gls{lte}}
\newacronym{5g}{5G}{fifth generation}
\newacronym{nr}{NR}{New Radio}
\newacronym{5gnr}{5G NR}{5G New Radio}
\newacronym{ieee}{IEEE}{Institute of Electrical and Electronics Engineers}
\newacronym{wifi}{Wi-Fi}{IEEE 802.11}
\newacronym{lan}{LAN}{local area network}
\newacronym{wlan}{WLAN}{wireless local area network}
\newacronym[plural=BSs]{bs}{BS}{base station}
\newacronym[plural=SBSs]{sbs}{SBS}{small-cell base station}
\newacronym[plural=FD-SBSs]{fdsbs}{FD-SBS}{\gls{fd}-enabled \gls{sbs}}
\newacronym[plural=MBSs]{mbs}{MBS}{macrocell base station}
\newacronym[plural=UEs]{ue}{UE}{user equipment}
\newacronym{ul}{UL}{uplink}
\newacronym{dl}{DL}{downlink}
\newacronym{qos}{QoS}{Quality of Service}
\newacronym{fcc}{FCC}{Federal Communications Commission}
\newacronym{iab}{IAB}{integrated access and backhaul}
\newacronym{fab}{FAB}{fixed access and backhaul}
\newacronym{hetnet}{HetNet}{heterogeneous network}

\newacronym{siso}{SISO}{single-input single-output}
\newacronym{mimo}{MIMO}{multiple-input multiple-output}
\newacronym{sumimo}{SU-MIMO}{single-user \gls{mimo}}
\newacronym{mumimo}{MU-MIMO}{multi-user \gls{mimo}}
\newacronym{bf}{BF}{beamforming}
\newacronym{ca}{CA}{constant amplitude}
\newacronym{ula}{ULA}{uniform linear array}
\newacronym{upa}{UPA}{uniform planar array}
\newacronym[\glslongpluralkey={angles of arrival}]{aoa}{AoA}{angle of arrival}
\newacronym[\glslongpluralkey={angles of departure}]{aod}{AoD}{angle of departure}
\newacronym{dof}{DoF}{degrees of freedom}
\newacronym{csi}{CSI}{channel state information}
\newacronym{csit}{CSIT}{\gls{csi} at the transmitter}
\newacronym{csir}{CSIR}{\gls{csi} at the receiver}
\newacronym{cs}{CS}{compressed sensing}

\newacronym{fd}{FD}{in-band full-duplex}
\newacronym{hd}{HD}{half-duplex}
\newacronym{si}{SI}{self-interference}
\newacronym{sic}{SIC}{self-interference cancellation}
\newacronym{soi}{SoI}{signal of interest}
\newacronym{asic}{A-SIC}{analog \acrlong{sic}}
\newacronym{dsic}{D-SIC}{digital \gls{sic}}
\newacronym{star}{STAR}{simultaneous transmit and receive}
\newacronym{warp}{WARP}{Wireless Open-Access Research Platform}
\newacronym{bfc}{BFC}{beamforming cancellation}
\newacronym{ipi}{IPI}{inter-panel-interference}
\newacronym{ipic}{IPIC}{inter-panel-interference cancellation}

\newacronym{qcqp}{QCQP}{quadratically-constrained quadratic programming}
\newacronym{pdf}{PDF}{probability density function}
\newacronym{cdf}{CDF}{cumulative density function}
\newacronym{iid}{i.i.d.}{independently and identically distributed}

\newacronym{elf}{ELF}{extremely low frequency}
\newacronym{slf}{SLF}{super low frequency}
\newacronym{ulf}{ULF}{ultra low frequency}
\newacronym{vlf}{VLF}{very low frequency}
\newacronym{lf}{LF}{low frequency}
\newacronym{mf}{MF}{medium frequency}
\newacronym{hf}{HF}{high frequency}
\newacronym{vhf}{VHF}{very high frequency}
\newacronym{uhf}{UHF}{ultra high frequency}
\newacronym{shf}{SHF}{super high frequency}
\newacronym{ehf}{EHF}{extremely high frequency}
\newacronym{thf}{THF}{tremendously high frequency}

\newacronym{wncg}{WNCG}{Wireless Networking and Communications Group}
\newacronym{linc}{LINC}{Laboratory of Informatics, Networks, and Communications}
\newacronym{ut}{UT Austin}{The University of Texas at Austin}
\newacronym{uiuc}{UIUC}{University of Illinois at Urbana-Champaign}
\newacronym{usc}{USC}{University of Southern California}
\newacronym{mit}{MIT}{Massachusetts Institute of Technology}
\newacronym{berkeley}{UC Berkeley}{University of California, Berkeley}
\newacronym{osu}{OSU}{Ohio State University}

\newcommand{\mmwave}{\gls{mmwave}\xspace}
\newcommand{\mimo}{\gls{mimo}\xspace}

\newcommand{\iab}{\gls{iab}\xspace}

\newcommand{\ginr}{\gls{inr}\xspace}
\newcommand{\gcdf}{\gls{cdf}\xspace}
\newcommand{\gpcdf}{\glspl{cdf}\xspace}

\newcommand{\upa}{\gls{upa}\xspace}
\newcommand{\upas}{\glspl{upa}\xspace}

\newcommand{\iid}{\gls{iid}\xspace}

\newcommand{\secref}[1]{Section~\ref{#1}}
\newcommand{\subsecref}[1]{Subsection~\ref{#1}}
\newcommand{\tabref}[1]{Table~\ref{#1}}
\newcommand{\figref}[1]{\figurename~\ref{#1}}

\usepackage{multirow}

\newcommand{\twocolfigheightfrac}{0.27}
\newcommand{\threecolfigheightfrac}{0.38}

\begin{document}
	
% \title{28 GHz Self-Interference Measurements for Millimeter Wave Full-Duplex\\Using Phased Arrays}
% \title{28 GHz Self-Interference Measurements for Millimeter Wave Full-Duplex}
% \title{28 GHz Self-Interference Measurements Using Phased Arrays}
% \title{Self-Interference Channel Measurements at 28~GHz: Spatial Characteristics, Statistical Modeling, and Impacts on Full-Duplex Performance}
% \title{Self-Interference Channel Measurements at 28~GHz: Spatial Characteristics, Statistical Modeling, and Full-Duplex Potential}
% \title{Self-Interference Channel Measurements at 28~GHz: Spatial Characteristics and\\Statistical Modeling}
% \title{Self-Interference Channel Measurements at 28~GHz: Spatial Insights and Angular Spread}
\title{Beamformed Self-Interference Measurements at 28~GHz: Spatial Insights and Angular Spread}

\author{%
    Ian~P.~Roberts,~%
    Aditya Chopra,~%
    Thomas Novlan,~\\%
    Sriram Vishwanath,~%
    and Jeffrey~G.~Andrews%
    \thanks{Ian~P.~Roberts (ipr@utexas.edu), Sriram~Vishwanath, and Jeffrey~G.~Andrews are with the 6G@UT Research Center and the Wireless Networking and Communications Group at the University of Texas at Austin. Thomas Novlan is with the Advanced Wireless Technologies Group at AT\&T Labs, Austin, TX 78759 USA. Aditya Chopra was with the Advanced Wireless Technologies Group at AT\&T Labs during this work. He is currently with Project Kuiper at Amazon. The work of Ian P.~Roberts was supported by the National Science Foundation Graduate Research Fellowship Program under Grant DGE-1610403.}%
    \thanks{Related code is available at: {https://ianproberts.com/bfsi}.}%
}

\markboth{~}%
 {Roberts \MakeLowercase{\textit{et al.}}: Self-Interference Channel Measurements at 28 GHz}

% \IEEEpubid{\copyright~2020~IEEE}


\maketitle

% \pagebreak

% \setcounter{tocdepth}{1} % 0,1,2,...
% \tableofcontents

% \pagebreak

\begin{abstract}
% We present measurements of the 28 GHz self-interference channel for full-duplex.
% We measure the isolation between the input of a transmitting phased array and the output of a colocated receiving phased array.
% Abstract here.
% We present measurements and analysis of the 28 GHz self-interference channel for multi-panel \mmwave full-duplex communication systems.
We present measurements and analysis of self-interference in multi-panel \mmwave full-duplex communication systems at 28 GHz. % , which is essential in characterizing and evaluating full-duplex operation.
% The degree of self-interference incurred by a full-duplex system largely dictates its ability to successfully receive a desired signal while transmitting in-band.
% The degree of self-interference incurred by an in-band full-duplex communication system largely dictates its ability to success
In an anechoic chamber, we measure the self-interference power between the input of a transmitting phased array and the output of a colocated receiving phased array, each of which is electronically steered across a number of directions in azimuth and elevation.
These self-interference power measurements shed light on the potential for a full-duplex communication system to successfully receive a desired signal while transmitting in-band.
% In total, nearly 6.5 million measurements were taken in an anechoic chamber to densely inspect the directional nature of the direct coupling between 256-element phased arrays.
% Our nearly 6.5 million measurements illustrate that slight shifts in steering direction of the transmit and receive beams (on the order of one degree) can lead to significant and seemingly random variations in self-interference power.
Our nearly 6.5 million measurements illustrate that more self-interference tends to be coupled when the transmitting and receiving phased arrays steer their beams toward one another but that slight shifts in steering direction (on the order of one degree) can lead to significant fluctuations in self-interference power.
% Our measurements illustrate that the general steering directions of the transmit and receive arrays describe how much the incurred self-interference tends to be but simultaneously indicate that slight shifts in steering direction can lead to significant and seemingly random changes in self-interference.
We analyze these measurements to characterize the spatial variability of self-interference to better quantify and statistically model this sensitivity. % , which can motivate future work involving beam steering/selection for \mmwave full-duplex applications.
Our analyses and statistical results can be useful references when developing and evaluating \mmwave full-duplex systems and motivate a variety of future topics including beam selection, beamforming codebook design, and self-interference channel modeling.
\end{abstract}

% \pagebreak 

% \input{sec-peer-review-title.tex}

% \input{sec-keywords.tex}

\glsresetall

% \pagebreak

\section{Introduction} \label{sec:introduction}

Research on full-duplex \mmwave systems has been motivated by the potential for dense antenna arrays to strategically steer transmit and receive beams in a way that reduces self-interference \cite{roberts_wcm,xia_2017}.
Proposed solutions for \mmwave full-duplex (e.g., \cite{xia_2017,liu_beamforming_2016,prelcic_2019_hybrid,satyanarayana_hybrid_2019,cai_robust_2019,zhu_uav_joint_2020,roberts_bflrdr}) typically make use of hybrid digital/analog beamforming to shape transmit and receive beams that do not couple across the \mimo self-interference channel present between separate transmit and receive arrays of a full-duplex transceiver.
% Proposed solutions for \mmwave full-duplex (e.g., \cite{xia_2017,liu_beamforming_2016,lopez_prelcic_2019_analog,prelcic_2019_hybrid,satyanarayana_hybrid_2019,roberts_2019_bfc,cai_robust_2019,zhu_uav_joint_2020,roberts_bflrdr}) typically make use of hybrid digital/analog beamforming to shape transmit and receive beams that do not couple across the \mimo self-interference channel present between separate transmit and receive arrays of a full-duplex transceiver.
With enough isolation between its transmit and receive beams, a full-duplex \mmwave transceiver could simultaneously transmit and receive in-band, offering improvements at the physical layer, reduced latency, new approaches for medium access, and cost-effective solutions for network deployment \cite{roberts_wcm}.
Full-duplex \iab, for instance, could serve a downlink user while simultaneously receiving backhaul in-band, making better use of \mmwave spectrum while reducing latency between the network core and its edge \cite{3GPP_IAB_2,iab_ericsson,gupta_fdiab_arxiv}.
% For example, a pole-mounted \iab node with full-duplex capability could serve a user on the ground while simultaneously maintaining wireless backhaul to a fiber-connected node, all using the same \mmwave spectrum.

% The need for beam alignment is present in full-duplex which means that conventional beams

%In an effort to evaluate proposed solutions, existing work has typically relied on the spherical-wave \mimo channel \cite{spherical_2005} as an attempt to model the direct self-interference coupled between the transmit and receive arrays [cite those that rely on SW channel] of a full-duplex transceiver.
%Some of these [cite those that do] have also incorporated a ray-based model to capture reflections that may stem from the environment \cite{rajagopal_2014,li_2014}.
%These approaches to model and simulate the self-interference channel are certainly a good start but have not been verified with measurement and are problematic for a variety of reasons.
%The highly idealized nature of these approaches make them sensitive to errors such as those in array geometry and neglect relevant practical factors such as device enclosures and non-isotropic antenna elements. 

% In an effort to evaluate proposed \mmwave full-duplex solutions, existing work has typically relied on idealized \mimo channel models to capture self-interference coupled between the transmit and receive arrays of a full-duplex transceiver.
% These idealized self-interference channel models, however, have not been verified with measurement and may lead to misleading results.
Given the highly directional nature of \mmwave communication, understanding the spatial characteristics of self-interference is critical to evaluating proposed \mmwave full-duplex solutions.
% This is especially true considering practical \mmwave systems rely so heavily on beam alignment to establish links by steering energy in a highly directional manner.
This is especially true when considering analog-only beamforming systems where digital beamforming cannot be relied on to further mitigate self-interference.
Currently, however, there is not a strong understanding of self-interference in \mmwave systems.
% Works [cite] consider beam alignment and rely on digital beamforming for the bulk of self-interference mitigation.
% Furthermore, since self-interference can saturate transmit and receive chains, it is 
% Furthermore, \mmwave systems equipped with analog-only beamforming, which are more practical, are particularly susceptible to the directionality of self-interference since they do not have digital precoding/combining to fine-tune their beams and cancel self-interference.
% While many proposed solutions rely on digital beamforming, 
% \mmwave systems equipped with analog-only beamforming, which are more practical, are particularly susceptible to the directionality of self-interference since they do not have digital precoding/combining to fine-tune their beams and cancel self-interference.
% Research and development of full-duplex \mmwave systems would benefit from measurement-backed self-interference models and from insights on expected self-interference power levels, particularly when beamformed phased arrays are employed.
Research and development of full-duplex \mmwave systems would benefit from measurement-backed insights on self-interference power levels and spatial characteristics, particularly when beamformed phased arrays are employed.

\subsection{Prior Work Measuring and Modeling mmWave Self-Interference}

% There has been limited work characterizing the self-interference channel at \mmwave.
One of the earliest known attempts to characterize \mmwave self-interference was in \cite{rajagopal_2014}, where a beam-sweeping approach was taken to measure the received self-interference power for a combination of transmit and receive beams.
A relatively low number of beams were swept using 28 GHz 8 $\times$ 8 \upas in indoor and outdoor environments, which offered limited characterization of the spatial characteristics and the distribution of self-interference. 
Nonetheless, this work provided a valuable first look at the expected self-interference power levels seen by a multi-panel \mmwave full-duplex system.
In \cite{kohda_2015}, using phased array transceivers, a proof-of-concept 60 GHz, short-range, full-duplex communication link was established.
The authors observed performance improvements when varying the angular difference between the two transceivers, suggesting that there exist noteworthy spatial characteristics in 60~GHz self-interference.
% While this work does not strictly investigate the self-interference channel, they found that there was a clear performance improvement when the angular difference between the two transceivers was varied, suggesting that the self-interference channel at 60 GHz does in fact have noteworthy spatial characteristics. 
%
The work of \cite{lee_2015} presents self-interference channel measurements at 28 GHz using a pair of directional horn antennas for transmission and reception as well as omnidirectional dipole antennas.
% The horn antennas had a gain of $15.4$ dBi with a $30^\circ$ half-power beamwidth, and the dipole offered an omnidirectional pattern with $-2$ dBi of gain. 
% The power delay profile of the self-interference channel in their results indicate that reflections from the environment are significant for over $100$ nanoseconds in indoor environments and well beyond that in outdoor settings, especially when vehicles are present.
Little was shown on the variability with changes in transmit and receive steering direction, though it was noted that some steering combinations offered starkly less self-interference than others.
% \edit{It is worth noting that the measurements conducted in \cite{lee_2015} were with $20$ cm of separation between the transmitter and receiver, meaning they were beyond the far-field distance.}

In \cite{yang_2016} and \cite{he_2017}, the authors conducted measurements of the 60 GHz self-interference channel with a rotating channel sounder comprised of horn antennas used for transmission and reception, which were fixed to the sounder. % , meaning their relative steering was also fixed.
Measurements in \cite{yang_2016} and \cite{he_2017} showed that large variations in self-interference power can be seen as the sounder rotated in azimuth due to indoor features such as large furniture.
% The measurement campaign in \cite{he_2017}, which was conducted indoors using the same system as \cite{yang_2016}, explored polarization and elevation in addition to azimuth rotations.
Clear gains in isolation were had with cross polarization and the self-interference power delay profile saw variability across azimuth and elevation of the sounder.
% Same folks: \cite{yang_2016}, \cite{he_2017}
In \cite{haneda_2018}, self-interference channel measurements at 70 GHz were conducted using lens antennas, which primarily provided insights on the degree of self-interference in various settings. % that offer $28.3$ dBi of gain.
% Measurements were conducted in an anechoic chamber, indoors, and outdoors in multiple configurations.
% The self-interference channel's directional characteristics were not inspected, though this campaign did provide useful insights on the level of self-interference in various settings.
% \edit{It is again worth noting that the measurements conducted in \cite{haneda_2018} were with co-located antennas separated beyond the far-field distance.}

% \subsection{Approaches to Model mmWave Self-Interference}
Measurements of \mmwave self-interference in \cite{rajagopal_2014,kohda_2015,lee_2015,yang_2016,he_2017,haneda_2018} are certainly useful but do not offer a means to evaluate proposed \mmwave full-duplex solutions since they provide neither a \mimo self-interference channel model nor adequate beam-based measurements.
%  a \mimo self-interference channel model and do not 
% of proposing or verifying \mimo self-interference channel models that \mmwave full-duplex research relies so heavily on.
To evaluate beamforming-based \mmwave full-duplex solutions thus far, researchers have primarily used highly idealized channel models.
For instance, the spherical-wave \mimo channel model \cite{spherical_2005} has been used most widely to capture coupling between the arrays of a full-duplex \mmwave system.
This extremely idealized geometric model is sensitive to small errors in the arrays' relative geometry due to the small wavelength at \mmwave, and it does not capture significant artifacts of practical systems such as enclosures, mounting infrastructure, and non-isotropic antenna elements. % \footnote{In fact, we show herein that this model does not align with measurements taken in this campaign.}
In fact, we show herein that this model does not align with the measurements taken in this campaign.
To account for environmental reflections---which were observed in \cite{rajagopal_2014}---it has been common for researchers to combine a ray-based model with the spherical-wave model in a Rician fashion \cite{li_2014,satyanarayana_hybrid_2019}.
While this inches the model closer to a seemingly more practical one, it has yet to be verified with measurement. % , leading to uncertainty in which dominates: the spherical-wave model or the reflective one.
Finally, we remark that the severity of inaccurate self-interference channel models can lead to highly misleading results and conclusions for beamforming-based \mmwave full-duplex solutions.
This is because, in theory, only very few spatial degrees of freedom are needed to execute \mmwave full-duplex and highly idealized channel models may readily offer these degrees of freedom whereas practical ones may not.

% ---

% This is especially true for those [cite] that conduct beam alignment and then execute digital beamforming to achieve full-duplex.
% In fact, beam alignment has proven to be prevalent in \mmwave systems and will understanding the spatial characteristics is particularly important 

% extremely narrow beams 

% ---

% What levels of SI to expect, particularly with beamforming or what range of isolation these beams can provide

% What the spatial/directional characteristics are of SI, \mmwave communication is highly directional meaning the directional characteristics of SI are extremely important

% Existing models are highly idealized

% Consequences of SI on full-duplex performance, then are extremely uncertain

% No existing datasets or model for the SI

% \textit{What limitations does the SI channel impose on full-duplex performance?}
% ---

\subsection{Contributions}

\comment{
Presented herein, we have conducted a measurement campaign in an attempt to better understand the characteristics of realistic \mmwave self-interference.
Using our collected measurements, we aim to answer many questions in this paper concerning \mmwave self-interference.
\textit{What levels of self-interference are expected in practical full-duplex \mmwave systems?}
\textit{Do highly directional \mmwave beams establish enough isolation on their own for full-duplex operation?}
\textit{What spatial insights can be drawn about self-interference?}
\textit{How does self-interference vary with slight shifts in transmit and receive direction?}
% \textit{What are the spatial characteristics of self-interference and can they be statistically modeled?}
% \textit{Based on our measurements, does beamforming alone have the potential to mitigate self-interference to levels sufficient for full-duplex?}
% In this work, we aim to answer all of these questions using our collected \mmwave self-interference channel measurements that we first presented in \cite{roberts_2021_att_measure}.
In an effort to answer these, the contributions of this work are as follows.
}

\textbf{Measuring and characterizing \mmwave self-interference.} 
Extending our work in \cite{roberts_2021_att_measure}, we present the first set of spatially dense measurements of self-interference at 28 GHz using finely steered phased arrays. 
Our nearly 6.5 million measurements shed light on the levels of self-interference that a practical \mmwave full-duplex system could expect using conventional beam steering.
This can drive the design requirements for full-duplex systems, including those relying solely on beam steering and those with additional self-interference cancellation measures.
Importantly, we compare our measurements of self-interference against what one would expect based on the idealized spherical-wave model in \cite{spherical_2005} to show that such a channel does not align with reality---motivating the need for new, measurement-backed models.
% A spatial inspection of our measurements uncovers large-scale (global) trends in self-interference as well as noteworthy small-scale (local) variability.
A spatial inspection of our measurements uncovers large-scale trends in self-interference based on general transmit and receive steering directions, as well as noteworthy small-scale variability when these steering directions undergo small shifts (on the order of one degree).
% and (ii) the spatial characteristics of self-interference. 

\textbf{Quantifying and statistically modeling the angular spread of self-interference.}
To explore this small-scale variability further, we investigate the angular spread of self-interference.
We examine how self-interference varies over small spatial neighborhoods to quantify the range in \ginr, minimum \ginr, and maximum \ginr when small shifts are made to the transmit and receive steering directions.
We fit distributions to these quantities and tabulate the fitted parameters for various spatial neighborhood sizes to supply engineers with tools for conducting statistical analyses and evaluation of full-duplex \mmwave systems.
% Importantly, 
These findings on the angular spread of self-interference shed light on the efficacy of \mmwave full-duplex and excitedly motivate a variety of future work including beam codebook design and beam selection.

This paper is organized as follows.
In \secref{sec:setup}, we describe our 28 GHz self-interference measurement platform and methodology.
In \secref{sec:summary}, we provide a summary of our measurements along with high-level spatial insights.
In \secref{sec:angular-spread}, we illustrate how small shifts in transmit and receive steering directions can lead to significant variations in self-interference.
We conclude this paper in \secref{sec:conclusion}.

\section{Measurement Setup \& Methodology} \label{sec:setup}

%\begin{figure}
%    \centering
%    \includegraphics[width=\linewidth,height=0.3\textheight,keepaspectratio]{visio/chamber.pdf}
%    \caption{A side- and top-view of the phased array measurement platform in the anechoic chamber.}
%    \label{fig:chamber}
%\end{figure}

In this section, we summarize the setup and methodology we used to collect measurements of the 28 GHz self-interference channel.
% ; please refer to \cite{roberts_2021_att_measure} for further details and presentation of the collected measurements beyond what is included herein.
This campaign sought to measure self-interference between the input of a transmitting phased array and the output of a colocated receiving phased array.
Since the degree of self-interference coupled between the phased arrays depends on the steering directions of their beams, it was measured across a variety of steering combinations.

%\begin{figure*}
%    \centering
%    \includegraphics[width=\linewidth,height=0.25\textheight,keepaspectratio]{fig/experiments_att_system_type_06.pdf}
%    \caption{A simplified block diagram of our measurement setup using phased arrays to inspect the 28 GHz self-interference channel. The degree of self-interference depends on the particular choice of transmit beam and receive beam.}
%    \label{fig:setup}
%\end{figure*}

\begin{figure*}
    \centering
    \subfloat[Simplified block diagram of our measurement setup.]{\includegraphics[width=\linewidth,height=0.22\textheight,keepaspectratio]{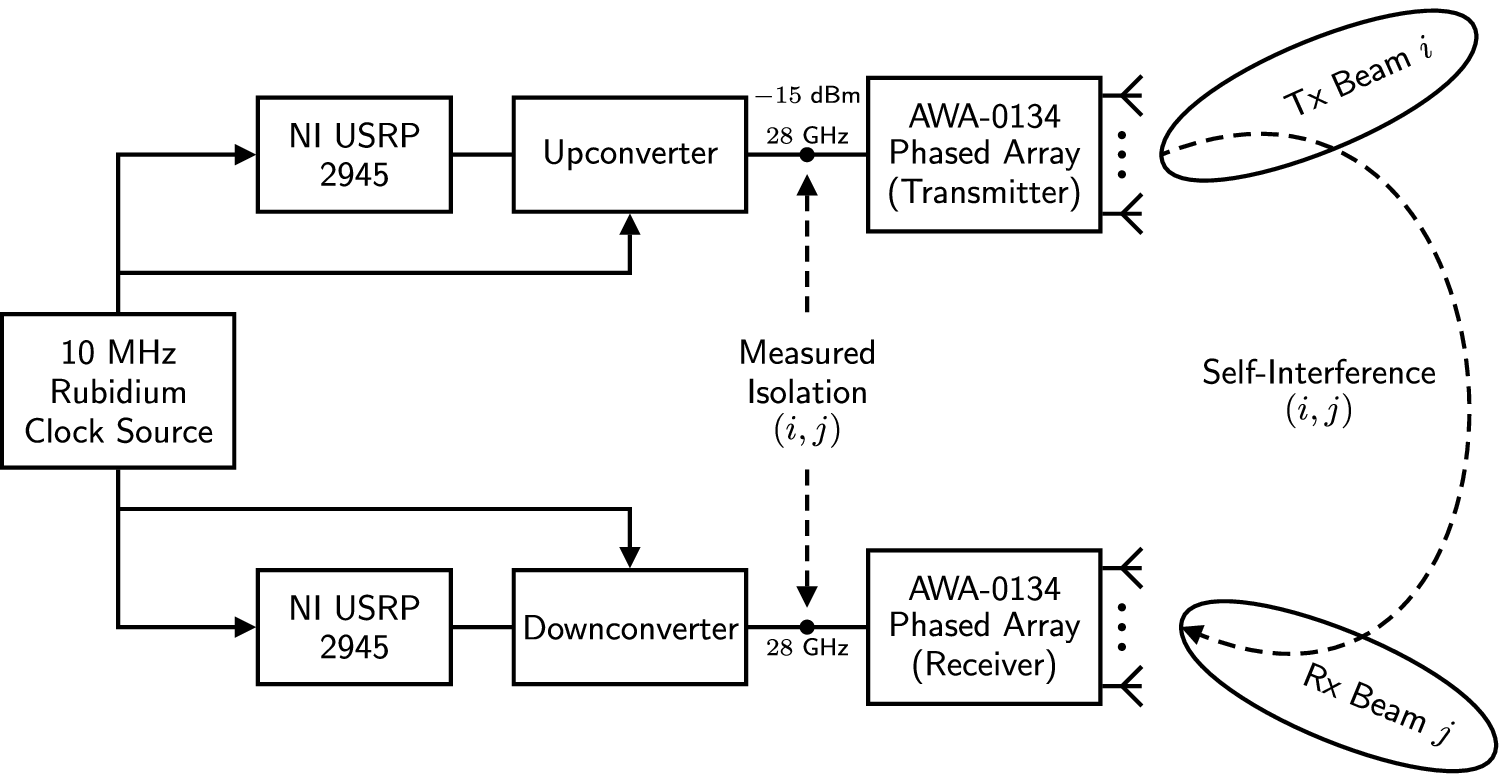}
        \label{fig:setup}}
    \quad
    \subfloat[Phased array measurement platform.]{\includegraphics[width=\linewidth,height=0.22\textheight,keepaspectratio]{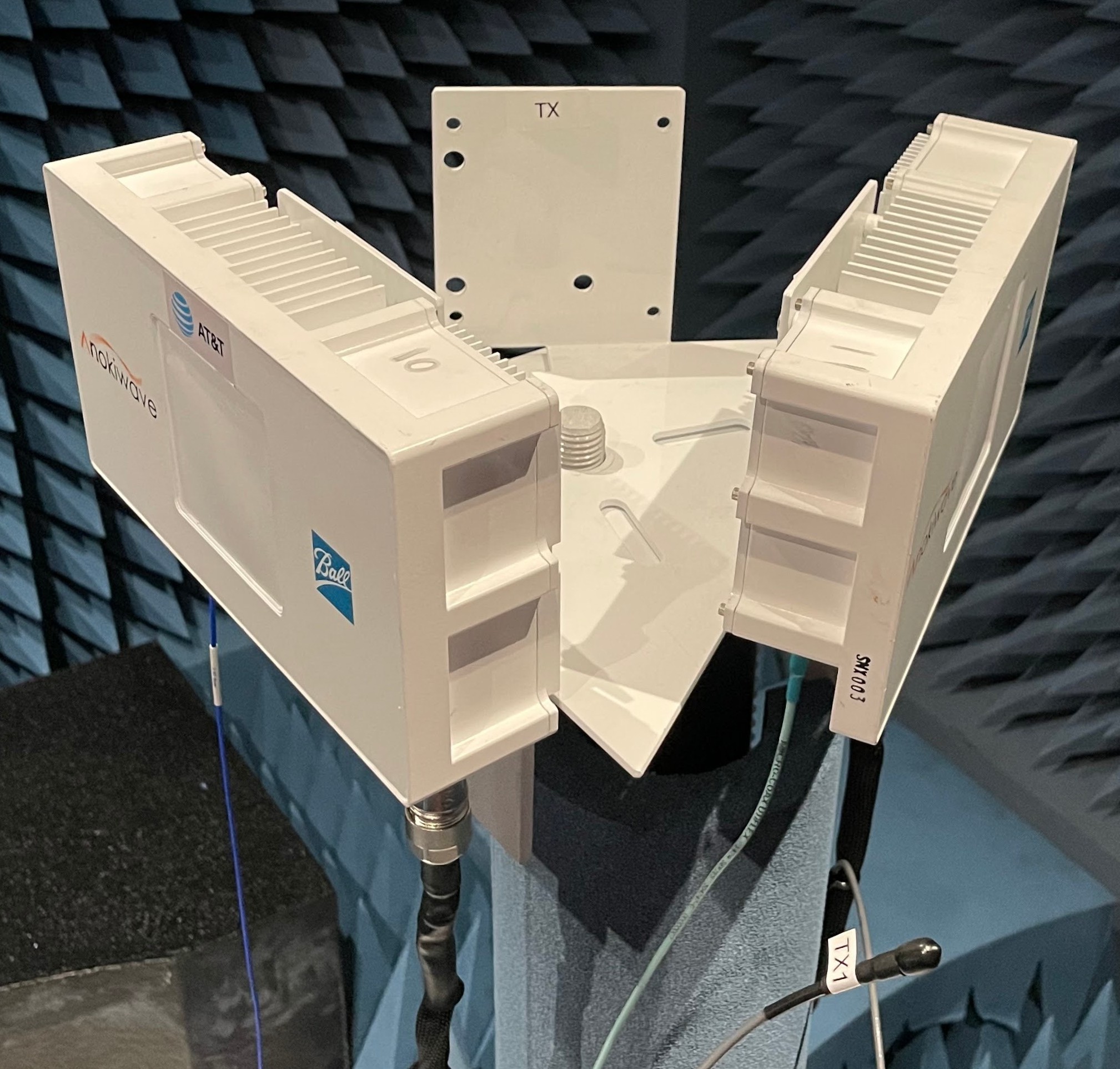}
        \label{fig:chamber}}
    \caption{(a) A simplified block diagram of our measurement setup using phased arrays to inspect self-interference at 28 GHz. The degree of self-interference depends on the particular choice of transmit beam and receive beam. (b) The phased array measurement platform in an anechoic chamber; receive array on left and transmit array on right.}
    \label{fig:subfigs}
\end{figure*}

% \subsection{Setup}
Our \mmwave self-interference measurement system, illustrated as a block diagram in \figref{fig:setup}, is comprised of two identical Anokiwave AWA-0134 28 GHz phased array modules \cite{anokiwave}: one for transmission and one for reception.
Each phased array module consists of a $16 \times 16$ half-wavelength \upa, offering high spatial resolution in azimuth and elevation.
The transmit and receive arrays were mounted to separate sides of a $60^\circ$ equilateral triangular platform, as shown in \figref{fig:chamber}, where the centers of the arrays were separated by $30$ cm.
This configuration aligns with practical, multi-panel (sectorized) full-duplex \mmwave deployments, such as for full-duplex \iab as proposed in 3GPP \cite{3GPP_IAB}.
The measurement platform was placed in an anechoic chamber free from significant reflectors; valuable future work would investigate the impact of reflections.

Each array can be electronically steered by a network of digitally-controlled analog beamforming weights, allowing us to form narrow transmit and receive beams to inspect the directional characteristics of the direct coupling between the transmit and receive arrays.
% By steering our transmit and receive arrays across a number of directions, we could inspect the spatial characteristics of the direct coupling that exists between transmit and receive arrays of a \mmwave full-duplex system.
The transmit array is driven by an upconverted 28 GHz Zadoff-Chu sequence with 100 MHz of bandwidth and a power level of $-15$ dBm.
The amplified and beamformed transmit signal is radiated by the transmit array at an \gls{eirp} of $60$ dBm before coupling over the air with the receive array.
The beamformed signal captured by the receive array is internally amplified and downconverted before being digitized.
Separate \acrlong{sdr} platforms \cite{usrp} were used to generate signals in the transmit chain and capture signals in the receive chain.
In baseband, the transmit and receive signals were processed to estimate the isolation from the input of the transmit array to the output of the receive array.
In an effort to more reliably measure isolation, we employed a single Rubidium oscillator and a custom lossless dual upconversion/downconversion system built by Mi-Wave \cite{miwave}. 
Correlation-based processing of the Zadoff-Chu signals was used to estimate isolation levels that extend well below the noise floor at the output of our receive array.
We calibrated and verified our measurement capability using high-fidelity test equipment \cite{pwrMeter} and stepped attenuators \cite{stepAtten} to ensure our isolation measurements had low error (typically less than $1$ dB) across a broad range of received power levels (roughly from $-20$ dBm to $-110$ dBm).
Moreover, we confirmed the repeatability of our measurements in both the short-term (milliseconds) and long-term (minutes).

\begin{figure*}
    \centering
    \subfloat[3-D geometry.]{\includegraphics[width=0.46\linewidth,height=\textheight,keepaspectratio]{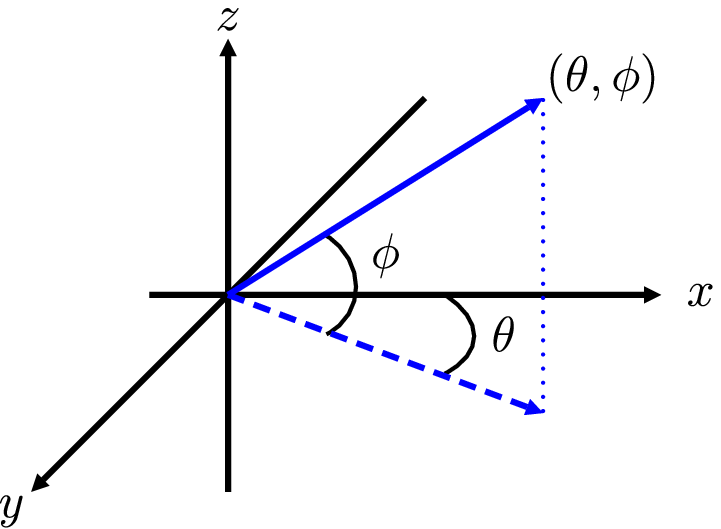}
        \label{fig:geometry}}
    \qquad
    \subfloat[Beam pattern.]{\includegraphics[width=0.46\linewidth,height=\textheight,keepaspectratio]{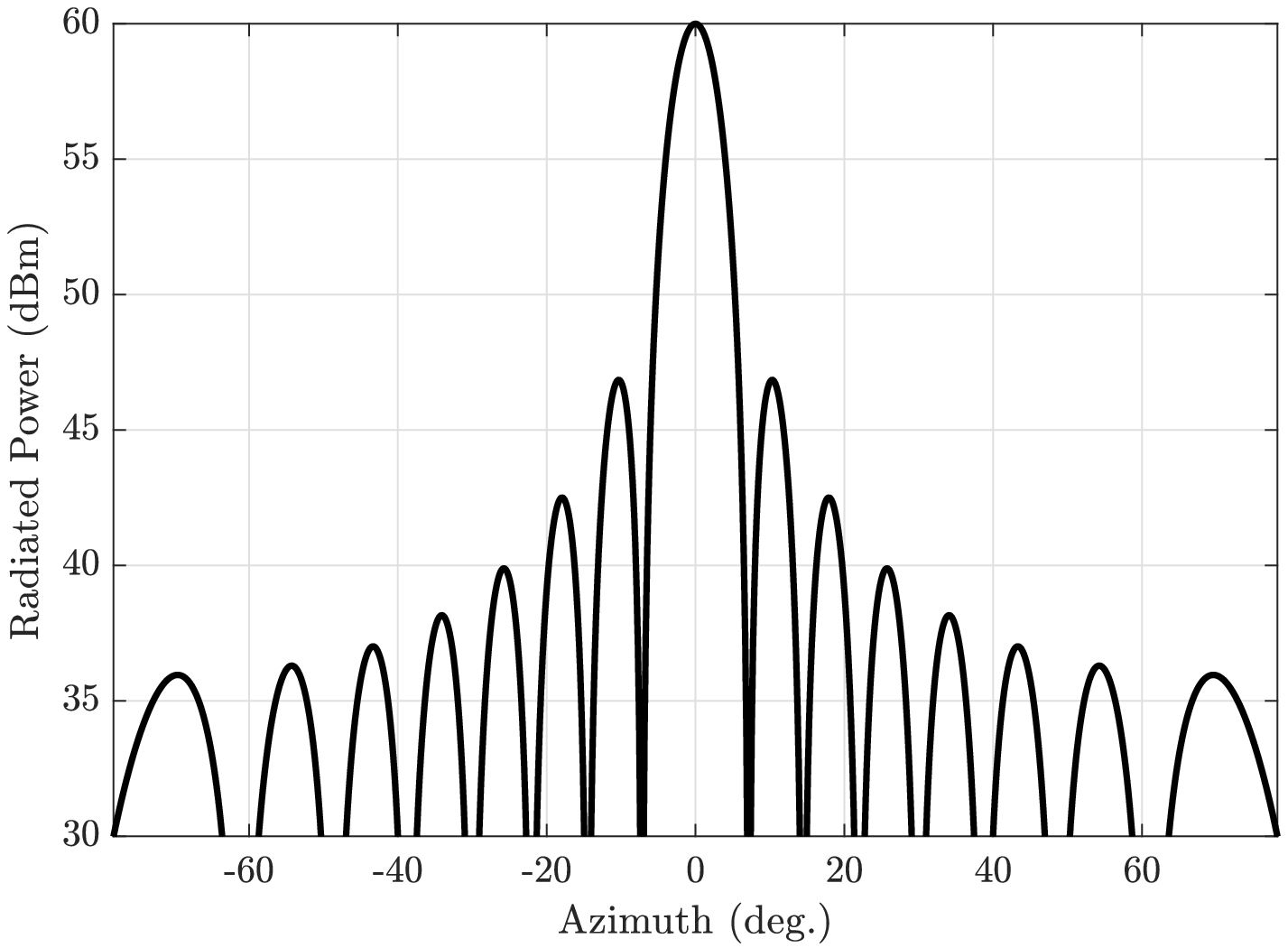}
        \label{fig:beam-pattern}}
    \caption{(a) The 3-D geometry observed by each phased array. (b) The idealized azimuth radiation pattern of our $16 \times 16$ transmit \upa steered broadside. The elevation pattern is identical. The \gls{eirp} is $60$ dBm, and the $3$ dB beamwidth is roughly $7^\circ$. The pattern of the receive \upa steered broadside is identical.}
    \label{fig:subfigs-array}
\end{figure*}

To describe the arrays' steering directions in 3-D, we use an azimuth-elevation convention as illustrated in \figref{fig:geometry}.
We assume independent coordinate systems for each array, where each array is centered at the origin facing the positive $x$ axis.
From each array's perspective, broadside corresponds to $0^\circ$ in azimuth and elevation (along the positive $x$ axis), upward is an increase in elevation $\phi$ (positive $z$ direction), and rightward (positive $y$ direction) is an increase in azimuth $\theta$.
% Let $\dirset$ be the set of all azimuth-elevation pairs $(\theta,\phi)$ where $\theta \in [0,2\pi)$ and $\phi \in [0,\pi)$ relative to each array in our measurement setup.
% A vector in the direction $(\theta,\phi)$ from the center of an array is described by azimuth $\theta$ being the angle between the positive $x$ axis and the vector's orthogonal projection onto the $x$-$y$ plane and elevation $\phi$ being the angle from the $x$-$y$ plane to the vector itself.
Each array can be independently steered toward a relative azimuth-elevation $(\theta,\phi)$ via beamforming weights.
In this work, we employ conjugate beamforming (i.e., matched filter beamforming), where a beam is formed in a particular direction by setting beam weights equal to the conjugate of the array response in that direction.
% The power level of the upconverted signal injected into the transmit array was $\Ptx = -15$ dBm, which delivers an \gls{eirp} of $60$ dBm when steered broadside.
For context, the idealized pattern of a transmit beam steered broadside is shown in \figref{fig:beam-pattern}, which has a $3$ dB beamwidth of around $7^\circ$ in both azimuth and elevation; the shape of a broadside receive beam is identical.
Naturally, practical beam patterns will not be as well-defined nor exhibit the deep nulls as that shown in \figref{fig:beam-pattern}.
% Likewise, the receive array can be independently steered using weights $\vw(\theta,\phi) \in \setvectorcomplex{256}$.

%\begin{figure}
%    \centering
%    \includegraphics[width=\linewidth,height=0.23\textheight,keepaspectratio]{plots/radiated_power_pattern}
%    \caption{The idealized azimuth radiation pattern of our $16 \times 16$ transmit \upa steered broadside. The elevation pattern is identical. The \gls{eirp} is $60$ dBm, and the $3$ dB beamwidth is roughly $7^\circ$. The pattern of the receive \upa steered broadside is identical.}
%    \label{fig:beam-pattern}
%\end{figure}

%\begin{figure*}
%    \centering
%    \subfloat[Caption a.]{\includegraphics[width=\linewidth,height=0.2\textheight,keepaspectratio]{plots/radiated_power_pattern}
%        \label{fig:beam-pattern}}
%    \qquad
%    \subfloat[Caption b.]{\includegraphics[width=\linewidth,height=0.2\textheight,keepaspectratio]{fig/experiments_att_geometry_type_01}
%        \label{fig:geometry}}
%    \caption{(a) The idealized azimuth radiation pattern of our $16 \times 16$ transmit \upa steered broadside. The elevation pattern is identical. The \gls{eirp} is $60$ dBm, and the $3$ dB beamwidth is roughly $7^\circ$. The pattern of the receive \upa steered broadside is identical. (b) The 3-D geometry relative to each phased array.}
%    \label{fig:subfigs-array}
%\end{figure*}

%\begin{align}
%\dirset = \braces{(\theta,\phi) : \theta \in [0,2\pi), \phi \in [0,\pi)}
%\end{align}

\begin{figure}
    \centering
    \includegraphics[width=\linewidth,height=0.25\textheight,keepaspectratio]{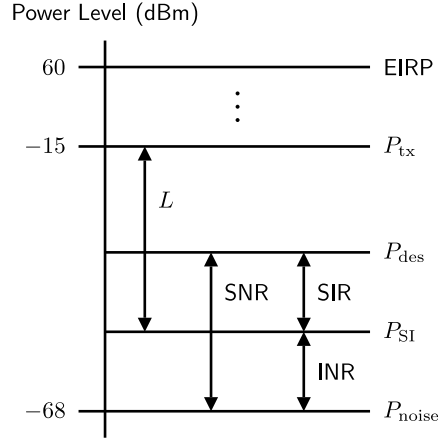}
    \caption{A summary of the key power levels and power ratios associated with our measurement system.}
    \label{fig:breakdown}
\end{figure}

The key power levels and power ratios associated with our measurement setup are summarized in \figref{fig:breakdown}.
When steering the transmit array toward some $(\thetatx,\phitx)$ and receive array toward $(\thetarx,\phirx)$, the power of self-interference coupled between the arrays is
% out of the receive array can be described as
\begin{align}
\powersi\parens{\thetatx,\phitx,\thetarx,\phirx} = \powertx \cdot L\parens{\thetatx,\phitx,\thetarx,\phirx}\inv \label{eq:spider-man}
% \bars{\vw\parens{\thetarx,\phirx}\trans \times \mH \times \vw\parens{\thetatx,\phitx}}^2
\end{align}
at the receive array output, where $\powertx = -15$ dBm is the power into the transmit array and 
\begin{align}
L\parens{\thetatx,\phitx,\thetarx,\phirx} = \frac{1}{\bars{\vw\parens{\thetarx,\phirx}\trans \mH \vf\parens{\thetatx,\phitx}}^2} \label{eq:hulk}
\end{align}
is the effective isolation between the transmit array input and receive array output established by transmit beamforming weights $\vf$ and receive weights $\vw$; $\mH \in \setmatrixcomplex{256}{256}$ is the (unknown) over-the-air self-interference channel matrix between the arrays and has scaling that accounts for the inherent path loss along with transmit and receive gains.
In other words, self-interference power $\powersi$ includes the spatial coupling between the transmit and receive beams with the over-the-air channel along with large-scale power gains introduced by the transmit array (e.g., \acrlongpl{pa}) and receive array (e.g., \acrlongpl{lna}).

\begin{figure}
    \centering
    \includegraphics[width=\linewidth,height=0.1\textheight,keepaspectratio]{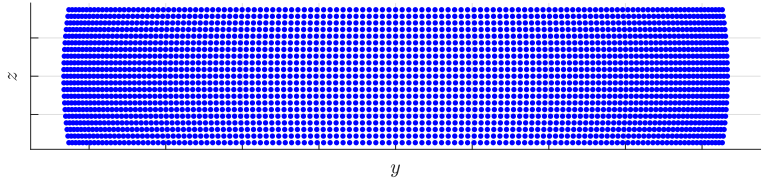}
    \caption{The set of 2,541 transmit directions $\txdirsetcb$ used during measurement, shown here as their respective projections onto the $y$-$z$ plane, where looking into the page represents steering outward from the array (the positive $x$ direction). The set of receive directions $\rxdirsetcb$ is identical, densely spanning $-60^\circ$ to $60^\circ$ in azimuth and $-10^\circ$ to $10^\circ$ in elevation, each in $1^\circ$ steps.}
    \label{fig:beams-full}
\end{figure}

Sets of $\Ntx$ transmit directions and $\Nrx$ receive directions
\begin{align}
\txdirsetcb &= \braces{\parens{\thetatx\idx{i},\phitx\idx{i}}}_{i=1}^{\Ntx}, \qquad
\rxdirsetcb = \braces{\parens{\thetarx\idx{j},\phirx\idx{j}}}_{j=1}^{\Nrx}
\end{align}
are specified prior to executing our measurement campaign.
% The isolation offered between each combination of transmit-receive beam pairs is measured to form an $\Nrx \times \Ntx$ matrix $\mL$.
We measure the self-interference power between each transmit and receive steering combination for a total of $\Ntx \times \Nrx$ measurements.
%The isolation afforded between the $m$-th transmit beam $\vw\parens{\thetatx\idx{m},\phitx\idx{m}}$ and $n$-th receive beam $\vw\parens{\thetarx\idx{n},\phirx\idx{n}}$ is
%\begin{align}
%\entry{\mL}{n,m} = \frac{1}{\bars{\vw\parens{\thetarx\idx{n},\phirx\idx{n}}\trans \times \mH \times \vw\parens{\thetatx\idx{m},\phitx\idx{m}}}^2}
%\end{align}
%where $\mH \in \setmatrixcomplex{256}{256}$ is the (unknown) over-the-air self-interference channel matrix between our arrays.
%We use $L\parens{\thetatx,\phitx,\thetarx,\phirx}$ to represent the isolation measured for some transmit direction $(\thetatx,\phitx) \in \txdirsetcb$ and receive direction $(\thetarx,\phirx) \in \rxdirsetcb$.
% \subsection{Transmit and Receive Beams}
Depicted in \figref{fig:beams-full}, the results in this work are based on measurements whose transmit and receive directions are each distributed uniformly in azimuth from $-60^\circ$ to $60^\circ$ with $1^\circ$ spacing and in elevation from $-10^\circ$ to $10^\circ$ with $1^\circ$ spacing.
This amounts to $\Ntx = \Nrx = 121 \times 21 = 2541$ directions for transmission and for reception, totaling $2541 \times 2541 \approx 6.5$ million self-interference power measurements.
% The measured transmit/receive beam directions are shown in \figref{fig:beams-full}.
%Let $\LL$ be the set of all nearly 6.5 million isolation measurements as
%\begin{align}
%\LL = \braces{L\parens{\thetatx,\phitx,\thetarx,\phirx} : \parens{\thetatx,\phitx} \in \txdirsetcb, \parens{\thetarx,\phirx} \in \rxdirsetcb}.
%\end{align}

% Our isolation measurements are captured at the output of the receive array and referenced to the power delivered to the input of the transmit array, which is $\Ptx = -15$ dBm.
Referencing the measured self-interference power $\powersi$ to the noise floor of the receive array, the \gls{inr} for given transmit and receive directions can be written as
\begin{align}
% \todB{\inr\parens{\thetatx,\phitx,\thetarx,\phirx}} = \todBm{\powersi\parens{\thetatx,\phitx,\thetarx,\phirx}} - \todBm{\powernoise}
\inr\parens{\thetatx,\phitx,\thetarx,\phirx} = \frac{\powersi\parens{\thetatx,\phitx,\thetarx,\phirx}}{\powernoise}
\end{align}
where $\powernoise = -68$ dBm is the noise power at the receive array output over $100$ MHz.
\gls{inr} is an important quantity for describing if a full-duplex system is self-interference-limited ($\inr \gg 0$ dB), noise-limited ($\inr \ll 0$ dB), or somewhere in between.
For full-duplex, we desire low $\inr$, roughly $\inr \leq 0$ dB in most cases, to ensure self-interference does not erode full-duplexing gain.
The set of all nearly 6.5 million measured \ginr values we write as
\begin{align}
\II = \braces{\inr\parens{\thetatx,\phitx,\thetarx,\phirx} : \parens{\thetatx,\phitx} \in \txdirsetcb, \parens{\thetarx,\phirx} \in \rxdirsetcb}
\end{align}
and refer to the \ginr measured when transmitting with the $i$-th transmit beam and receiving with the $j$-th receive beam as
\begin{align}
\inrij = \inr\parens{\thetatx\idx{i},\phitx\idx{i},\thetarx\idx{j},\phirx\idx{j}}.
\end{align}
A \textit{desired} signal having received power $\powerdes$ (at the output of the receive array) would see a \gls{sinr} of
\begin{align}
\sinr\parens{\thetatx,\phitx,\thetarx,\phirx} 
&= \frac{\powerdes}{\powernoise + \powersi\parens{\thetatx,\phitx,\thetarx,\phirx}} = \frac{\snr}{1 + \inr\parens{\thetatx,\phitx,\thetarx,\phirx}}
\end{align}
where \snr is the \gls{snr} of the desired signal.
Notice that $\sinr$ depends on the level of self-interference incurred when steering the transmitter toward $(\thetatx,\phitx)$ and the receiver toward $(\thetarx,\phirx)$; of course, $\snr$ would practically also be a function of $(\thetarx,\phirx)$.
This work is solely concerned with measuring $\inr$ (self-interference), from which desired signals having some $\snr$ can be evaluated in a full-duplex sense.
% This work is solely concerned with measuring self-interference power $\powersi\parens{\thetatx,\phitx,\thetarx,\phirx}$---equivalently $\inr\parens{\thetatx,\phitx,\thetarx,\phirx}$---from which desired signals having some $\snr$ can be evaluated in a full-duplex sense.
We would like to point out that all measurements collected in this campaign are for a fixed setup as described; valuable future work would investigate the impact of system parameters such as beam shape (e.g., beamwidth and side lobe levels), array sizes, and panel geometries.

\section{High-Level Summary and Spatial Insights} \label{sec:summary}

\begin{figure}
    \centering
    \includegraphics[width=\linewidth,height=0.26\textheight,keepaspectratio]{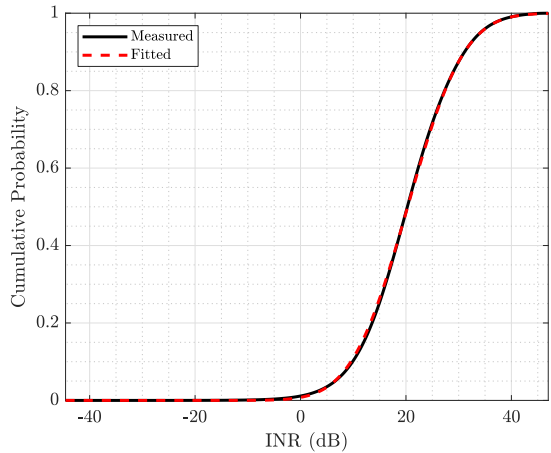}
    \caption{The \gls{cdf} of the nearly 6.5 million measured \gls{inr} values and its fitted log-normal distribution.}
    \label{fig:cdf}
\end{figure}

Perhaps the best summary of our measurements is the \gls{cdf} of the nearly 6.5 million measured \ginr values in \figref{fig:cdf}.
The maximum and minimum measured \ginr were nearly $46.99$ dB and $-44.57$ dB, respectively.
The measured \ginr typically falls between $0$ dB and $40$ dB, with median \ginr at $20.27$ dB.
Nearly $99$\% of beam pairs offer an \ginr greater than $0$ dB, where self-interference power exceeds noise power.
Around $90$\% of beam pairs yield an $\inr \geq 10$ dB, where self-interference is at least ten times as strong as noise.
Just over $2$\% of beam pairs offer an $\inr \leq 3$ dB.
% Various percentiles are summarized in \tabref{tab:statistics-full}.
% Naturally, this \gls{cdf} of $\inr$ would shift left/right if transmit power were to decrease/increase or noise power were to increase/decrease, for instance; those wishing to appropriately translate our measurements to systems with different power levels can refer to \figref{fig:breakdown}.
% Naturally, this \gls{cdf} of $\inr$ would shift left/right if transmit power were to decrease/increase or noise power were to increase/decrease, for instance, by shrinking/widening the gap between self-interference and noise; those wishing to appropriately translate our measurements to systems with different power levels can refer to \figref{fig:breakdown}.
Naturally, this \gls{cdf} of $\inr$ would shift left/right if transmit power were to decrease/increase or noise power were to increase/decrease, for instance; those wishing to appropriately translate our measurements to systems with different power levels can refer to \figref{fig:breakdown}.

\begin{remark}
    At first glance, the \gls{cdf} of the measured \inr values seems quite pessimistic from a full-duplex perspective, considering most beam pairs yield self-interference levels that are well above the noise floor.
    Multi-panel full-duplex \mmwave systems similar to ours will typically be overwhelmed with self-interference when choosing a random transmit and receive beam, motivating the need for additional means to reduce self-interference.\footnote{We explore one method for reducing self-interference via small shifts of the transmit and receive beams in \secref{sec:angular-spread}.}
    Nonetheless, it is a welcome sight to observe that there exist select beam pairs that do in fact offer \ginr levels sufficiently low for full-duplex---we explore the directional nature of these beam pairs shortly in \subsecref{subsec:max-med-min}. 
    Valuable future work would explore solutions to reduce self-interference and investigate how system design choices can potentially reduce \ginr levels while maintaining service to users.
\end{remark}

We found that the \gls{cdf} in \figref{fig:cdf} can be well approximated by a log-normal distribution, shown as a dashed line in \figref{fig:cdf}.
That is to say that the measured \ginr values in dB approximately follow a normal distribution as
\begin{align}
\todB{\II} \overset{\mathrm{fit}}{\sim} \distgauss{\mu}{\sigma^2} \label{eq:full-log-normal}
\end{align}
where $\mu = 20.32$ and $\sigma^2 = 70.69$ are the fitted mean and variance of the normal distribution.
Like the \gls{cdf} in \figref{fig:cdf}, changes to large-scale parameters that impact \ginr---such as the transmit power or noise power---can be accounted for in $\mu$ to shift the fitted normal distribution left or right.
Albeit limited, engineers can make first-order statistical approximations of self-interference via this log-normal distribution when drawing \iid \ginr values as $\todB{\inr} \sim \distgauss{\mu}{\sigma^2}$. % using \eqref{eq:full-log-normal}.
For instance, the probability that the \ginr of a {random} transmit-receive beam pair will fall below $\gamma$ (in linear units) can be well approximated as
\begin{align}
\prob{\inr \leq \gamma} 
% = \prob{\todB{\inr} \leq \todB{\gamma}}
= \frac{1}{2}\brackets{1 + \erf{\frac{\todB{\gamma}-\mu}{\sigma \cdot \sqrt{2}}}}.
\end{align}

% \input{tab/tab-statistics-full}

%\begin{align}
%P_i(\inr) = \frac{1}{\Nrx} \cdot \sum_{j=1}^{\Nrx} \ind{\todB{\inrijmin} \leq 0}
%\end{align}

%\begin{figure*}[!t]
%    \centering
%    \subfloat[Observed by each transmit beam.]{\includegraphics[width=0.475\linewidth,height=\textheight,keepaspectratio]{plots/full/median_max_min_per_tx_beam}\label{fig:max-med-min-a}}
%    \quad
%    \subfloat[Observed by each receive beam.]{	\includegraphics[width=0.475\linewidth,height=\textheight,keepaspectratio]{plots/full/median_max_min_per_rx_beam}\label{fig:max-med-min-b}}
%    \caption{For each transmit beam and receive beam, shown are the median, maximum, and minimum isolation across all receive and transmit beams, respectively.}
%    \label{fig:mean-median-max-min-full}
%\end{figure*}

\begin{figure*}[!t]
	\centering
	\subfloat[Measured \ginr.]{\includegraphics[width=0.475\linewidth,height=\textheight,keepaspectratio]{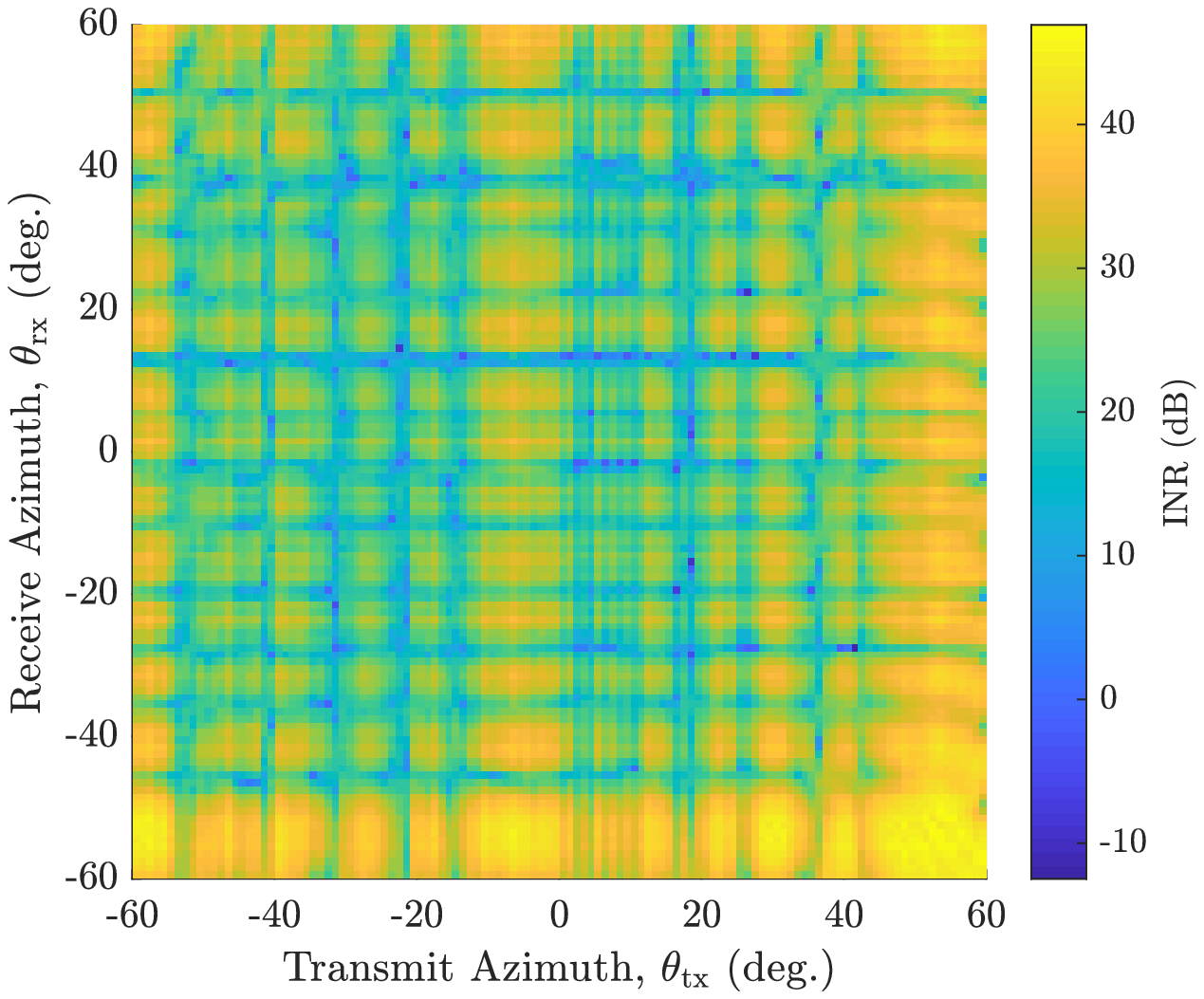}
		\label{fig:azimuth-plane-a}}
	\quad
	\subfloat[Simulated \ginr with spherical-wave channel.]{\includegraphics[width=0.475\linewidth,height=\textheight,keepaspectratio]{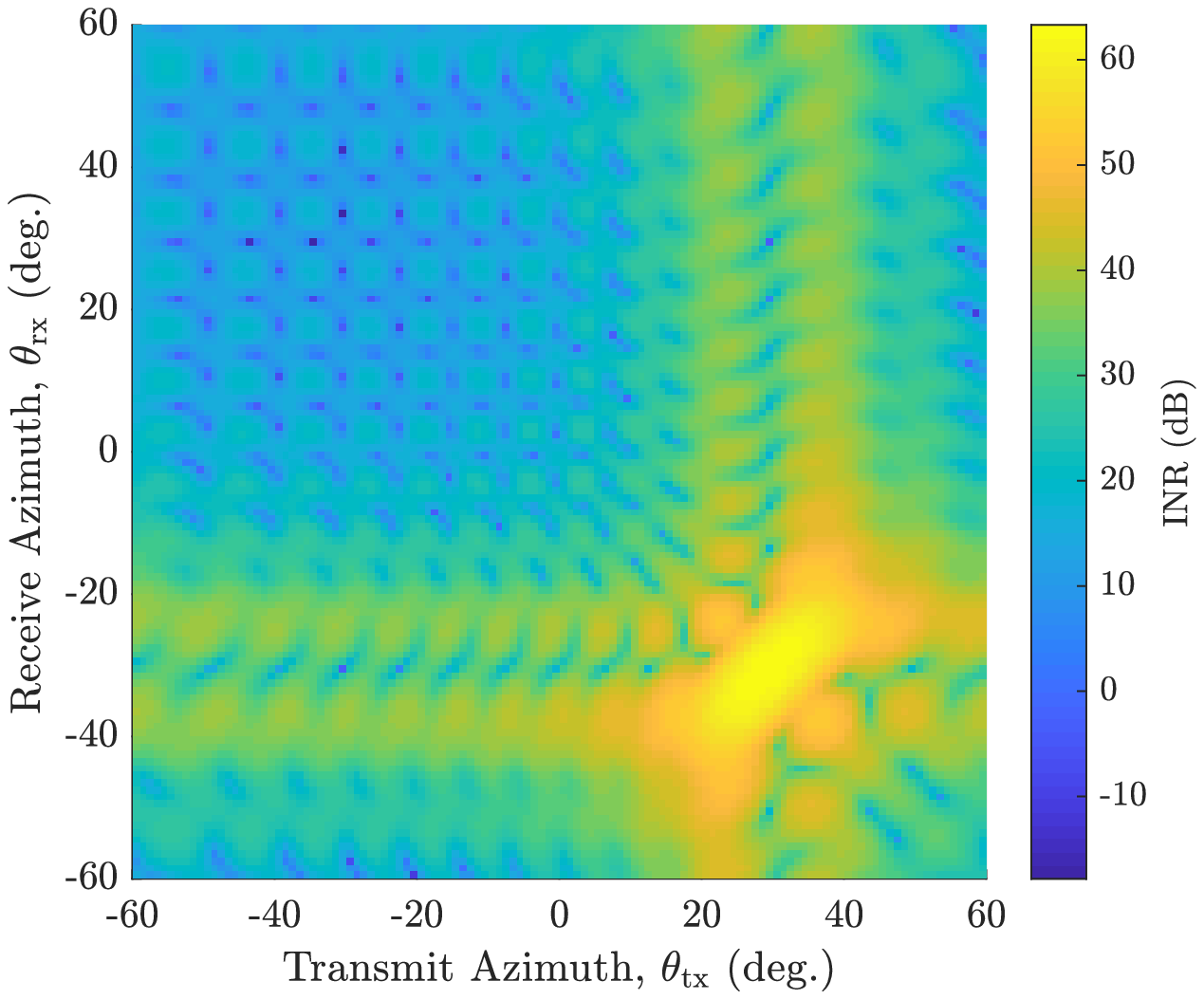}
		\label{fig:azimuth-plane-b}}
	\caption{(a) Measured \ginr as a function of azimuthal transmit and receive directions when $\phitx = \phirx = 0^\circ$. (b) The simulated counterpart of (a) when using a spherical-wave near-field self-interference channel model \cite{spherical_2005}, which clearly does not align well with what was measured. This motivates the need for a new, measurement-backed channel model for \mmwave self-interference.}
	\label{fig:azimuth-plane}
\end{figure*}

\subsection{Is an Idealized Near-Field Self-Interference Channel Model Realistic?}
A natural question to ask before proceeding is if the aforementioned spherical-wave \mimo channel model \cite{spherical_2005}---an idealized near-field propagation model---aligns with our measurements.
If it does, self-interference power values could be realized deterministically via the product of transmit and receive beamforming weights $\vf$ and $\vw$ with a \mimo channel matrix $\mH$ based on the spherical-wave model.
Unfortunately, however, we found that the spherical-wave \mimo channel model does not align with our measurements.
Consider \figref{fig:azimuth-plane}, where we plot the measured \ginr values across the azimuth plane and the simulated counterpart using the spherical-wave \mimo channel model.
Notice that the two are starkly different, indicating that this idealized near-field channel model---which has been used so frequently as a means to evaluate \mmwave full-duplex---does not translate to practical systems, which pose a number of nonidealities stemming from array enclosures, mounting infrastructure, and non-isotropic antenna elements, for instance.
% Notice that the two are starkly different, indicating that this idealized near-field channel model---which has been used so frequently as a means to evaluate \mmwave full-duplex---is not realistic.
This motivates the need for a practical, measurement-backed \mimo channel model for \mmwave self-interference, which we plan to address in future work.
% This motivates us to continue outlining our statistical model of \mmwave self-interference, and to do so, we begin by presenting our approach to estimating $\muij$.

\subsection{Maximum, Median, and Minimum \ginr for Particular Transmit Beams and Receive Beams}
\label{subsec:max-med-min}
\begin{figure*}[t]
    \centering
    \subfloat[Observed by each transmit beam.]{\includegraphics[width=0.475\linewidth,height=\threecolfigheightfrac\textheight,keepaspectratio]{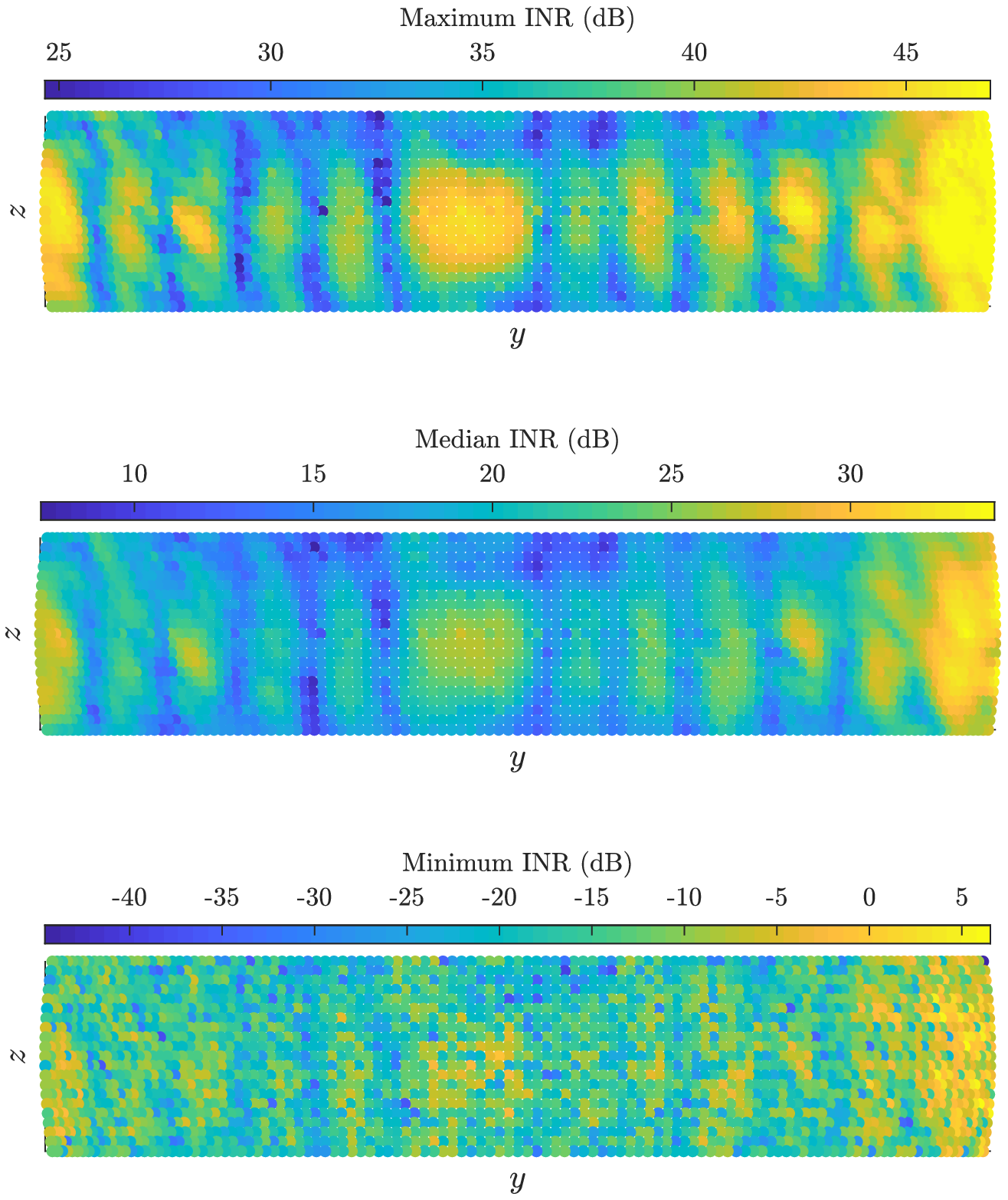}
        \label{fig:max-med-min-a}}
    \quad
    \subfloat[Observed by each receive beam.]{\includegraphics[width=0.475\linewidth,height=\threecolfigheightfrac\textheight,keepaspectratio]{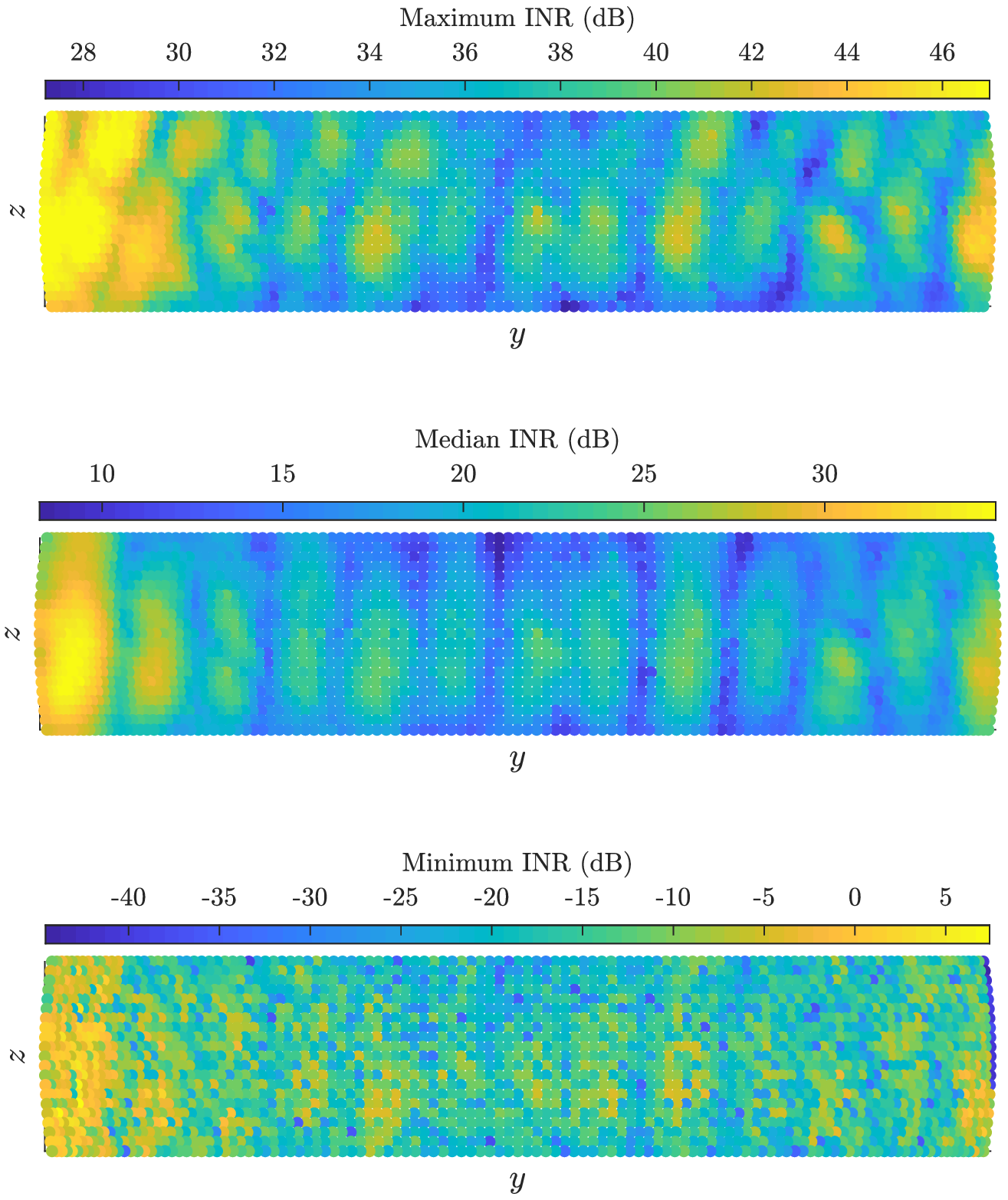}
        \label{fig:max-med-min-b}}
    \caption{For each transmit beam and receive beam, shown are the maximum, median, and minimum \ginr across all receive and transmit beams, respectively.}
    \label{fig:max-med-min}
\end{figure*}

The \gls{cdf} in \figref{fig:cdf} and its corresponding fitted distribution are certainly useful \textit{statistically} but do not provide any \textit{spatial} insight on self-interference.
As such, we now hone in on narrower perspectives to better visualize and interpret our measurements spatially.
First, let us begin by considering \figref{fig:max-med-min-a}, which shows the maximum, median, and minimum \ginr observed by each transmit beam across all receive beams; each dot corresponds to a transmit beam's projection onto the $y$-$z$ plane (i.e., its direction from the perspective of the transmit array).
In other words, the maximum \ginr observed by the $i$-th transmit beam (the $i$-th dot) is simply
% \begin{align}
% \inrimax = 
$\max_{j}  \inrij$,
% \end{align}
for example, with median and minimum expressed analogously.
\figref{fig:max-med-min-b} similarly shows these statistics observed when receiving a particular direction.
% Note that the color scale is unique to each individual plot in both figures.
Referencing \figref{fig:max-med-min-b}, we can see that the median \ginr per receive beam ranges from approximately $8$ dB to $35$ dB.
The maximum \ginr observed at each receive beam is at least around $28$ dB and at most over $46$ dB, while the minimum \ginr is at least around $-45$ dB and at most around $7$ dB.
% It does seem evident, however, that the steering toward the extreme right edge also tends to offer less isolation, albeit much less severely.
In a similar fashion, we examine these statistics for each transmit beam in \figref{fig:max-med-min-a}, which tell a similar story both visually and numerically as the receive side.

\begin{remark}
There are a few important things to take away from \figref{fig:max-med-min-a} and \figref{fig:max-med-min-b}.
As intuition may suggest based on \figref{fig:chamber}, the results of \figref{fig:max-med-min-a} and \figref{fig:max-med-min-b} indicate that:
% (i) transmitting to the right (toward the receiver) tends to offer less isolation and (ii) receiving to the left (toward the transmitter) tends to offer less isolation.
%As intuition may suggest based on \figref{fig:chamber}, the results of \figref{fig:max-med-min-a} and \figref{fig:max-med-min-b} indicate that:
\begin{itemize}
    \item transmitting to toward the receiver tends to couple more self-interference
    \item receiving toward the transmitter tends to couple more self-interference.
    % \item transmitting around broadside (an azimuth and elevation of zero) tends to couple more self-interference
\end{itemize}
% steering on the left edge of our measurement region towards the transmit array tends to couple more self-interference by virtually all three of these statistics.
% Transmitting rightward, toward the receiver, tends to offer less isolation in general.
Considering minimum \ginr is at most around $7$ dB in both, we see that (i) even when steering our transmitter toward the receiver, there exist some receive beam(s) that offer low \ginr (at most around $7$ dB) and (ii) even when steering our receiver toward the transmitter, there exist some transmit beam(s) that offer low \ginr (at most around $7$ dB).
%\begin{itemize}
%    \item even when steering our transmitter toward the receiver, there exists some receive beam(s) that offer low \ginr (at most around $7$ dB).
%    \item even when steering our receiver toward the transmitter, there exists some transmit beam(s) that offer low \ginr (at most around $7$ dB).
%\end{itemize}
This suggests that---while transmitting toward the receiver and receiving toward the transmitter \textit{generally} results in more self-interference---there exist receive beams and transmit beams that \textit{can} offer low \ginr.
% Upon inspection, these low-\ginr beam pairs seem to be randomly scattered throughout the beam space; we explore this further in \secref{sec:angular-spread}.
In \secref{sec:angular-spread}, we observe that low-\ginr beam pairs appear to be distributed throughout space. 
We have ongoing work that investigates if these low-\ginr beam pairs can in fact be used to serve users with high beamforming gain while simultaneously offering reduced self-interference, facilitating full-duplex operation.
% in fact useful for sustaining full-duplex communication by offering low self-interference and simultaneously delivering high beamforming gain.
In a similar fashion, observing maximum \ginr illustrates that there also consistently exists transmit-receive combinations that can lead to high self-interference.
From this, we can conclude that there are not transmit beams nor receive beams that \textit{universally} offer low or high \ginr---though there exist those that {tend} to.
Rather, the amount of self-interference coupled depends heavily on one's choice of transmit beam \textit{and} receive beam.
\end{remark}

\begin{remark}
Additional takeaways include the fact that we observe strong similarities between the transmit and receive profiles, which validates some degree of channel symmetry.
However, there do exist noteworthy differences, particularly the strong self-interference present when transmitting around broadside but not when receiving.
The high self-interference coupled when transmitting around broadside is not necessarily expected nor easily explained; it can perhaps be attributed to coupling behind the arrays due to mounting hardware and array enclosures. % to the presence of side lobes and their possible enhancement in this near-field setting and/or is due .
Also, we see significantly more variation across $y$ than $z$, suggesting that the azimuth of the steering direction plays a greater role than elevation, which one may expect since our transmitter and receiver are separated in azimuth but not in elevation.
While it may seem obvious that transmitting toward the receiver and receiving toward the transmitter would couple the most self-interference, it was not clear that this would be the case since the transmit and receive arrays exist in the near-field of one another.
The far-field distance of our arrays is approximately $2.4$ meters based on the rule-of-thumb $2D^2 / \lambda$ \cite{balanis}, while our arrays are separated by only $30$ cm.
The reactive/radiating near-field boundary, on the other hand, is around a mere $23$ cm based on the rule-of-thumb $0.62\sqrt{D^3/\lambda}$ \cite{balanis}, suggesting that our arrays live just within the radiating near-field of one another. 
Operating in this near-field regime, the highly directional beams created by our \upas are not necessarily ``highly directional'' from the perspective of one another \cite{roberts_wcm}.
\end{remark}

\subsection{Meeting \ginr Thresholds with Particular Transmit Beams and Receive Beams}

\begin{figure*}[t]
    \centering
    \subfloat[Observed by each transmit beam.]{\includegraphics[width=0.475\linewidth,height=\threecolfigheightfrac\textheight,keepaspectratio]{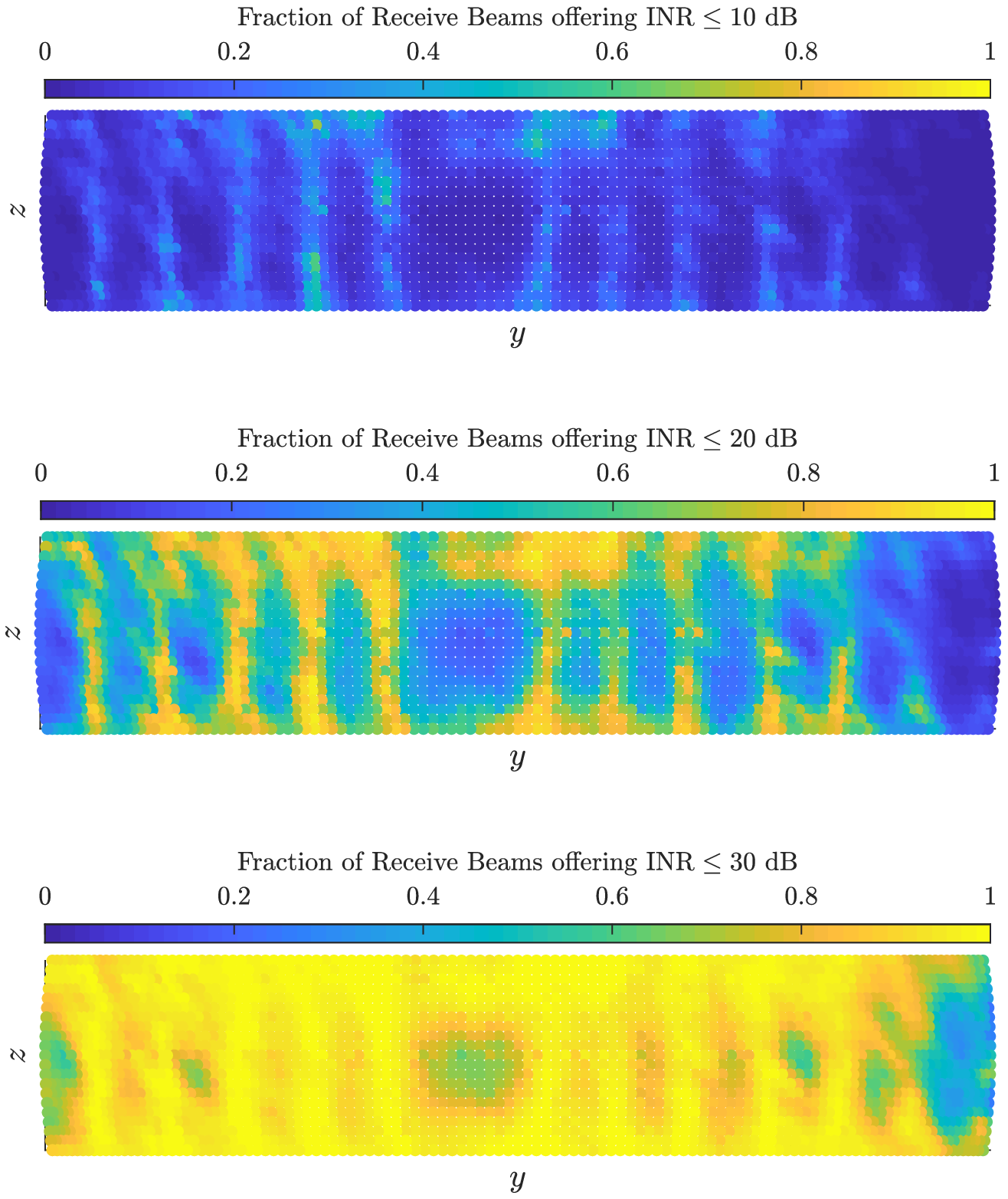}
        \label{fig:fractions-a}}
    \quad
    \subfloat[Observed by each receive beam.]{\includegraphics[width=0.475\linewidth,height=\threecolfigheightfrac\textheight,keepaspectratio]{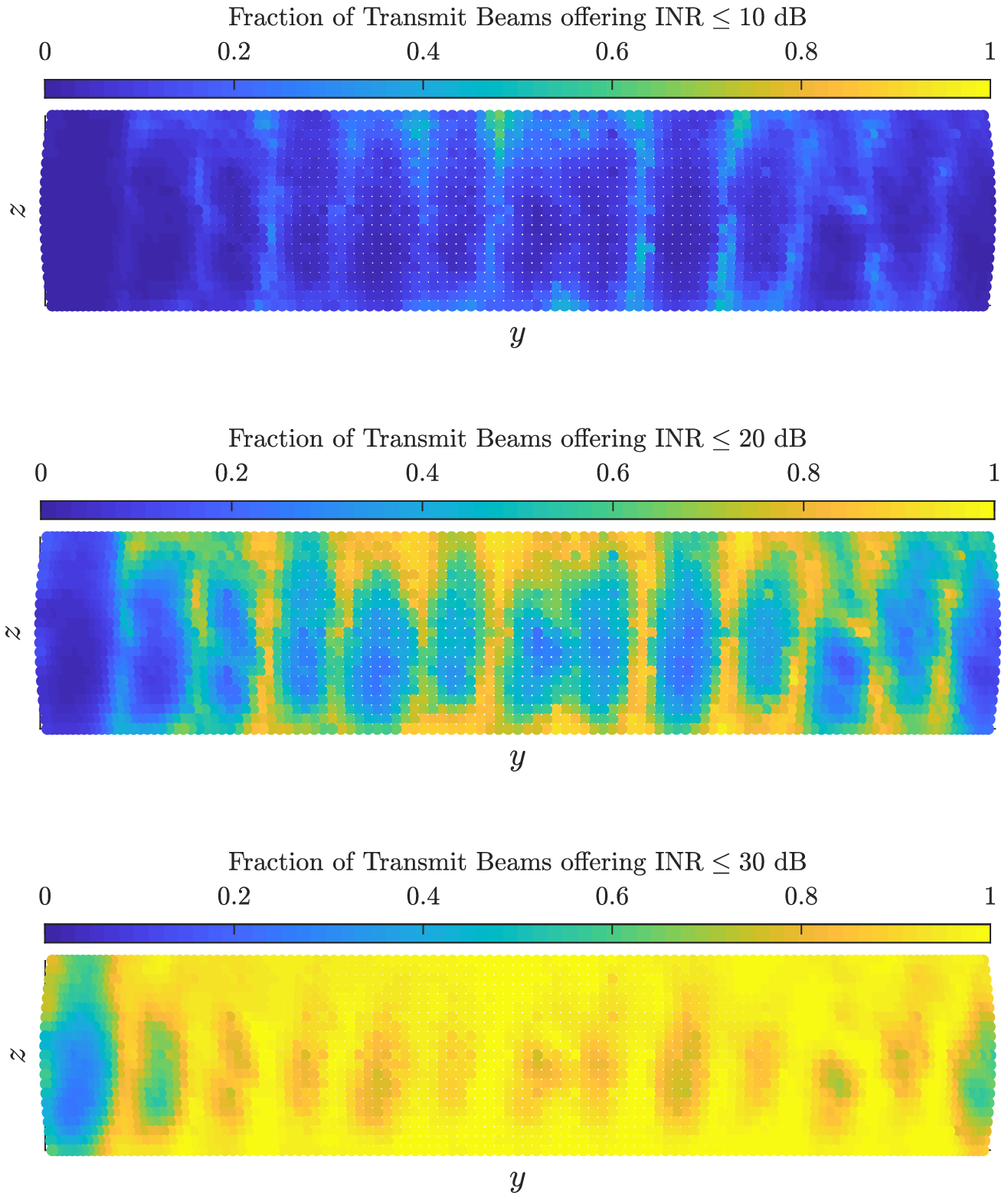}
        \label{fig:fractions-b}}
    \caption{For each transmit beam and receive beam, shown is the fraction of receive beams and transmit beams, respectively, that meet various \ginr thresholds.}
    \label{fig:fractions}
\end{figure*}

Having looked at the maximum, median, and minimum \ginr for particular transmit beams and receive beams, we now examine the fraction of beams that offer certain levels of \ginr.
In \figref{fig:fractions-a}, for each transmit beam, we look at the fraction of receive beams that offer at most $10$ dB, $20$ dB, and $30$ dB of \ginr.
Similarly, in \figref{fig:fractions-b}, for each receive beam, we look at the fraction of transmit beams that offer these same levels of \ginr.
From the top plot of each, we see that a modest \ginr threshold of $10$ dB (where self-interference is ten times stronger than noise) cannot be met very reliably by any transmit beam nor any receive beam.
At best, select beams can only meet this target \ginr around $50$\% of the time, with the vast majority falling quite short of this.
Naturally, as the \ginr threshold rises to $20$ dB, the fraction of beams that can meet this threshold increases.
Transmit and receive options emerge that offer an \ginr of at most $20$ dB over $50$\% of the time, with some approaching $100$\%.
Still, however, the transmit beams steering rightward toward the receiver and the receive beams steering leftward toward the transmitter struggle to offer an \ginr below $20$ dB.
Increasing the \ginr threshold further, nearly all transmit and receive beams can confidently offer an \ginr within $30$ dB, though those least likely to do so are the rightward transmit beams, leftward receive beams, and broadside transmit beams.

%These two figures begin to further explain where a large portion of high-\ginr beam pairs come from: most of their density (visible in the \gls{cdf} in \figref{fig:cdf}) lay in transmit beams steering toward the receive array and in receive beams steering toward the transmit array.
%These directions correspond to the transmit and receive directions that couple high amounts of self-interference \textit{almost regardless} of the receive and transmit direction, respectively.
%% In other words, these rightward transmit beams and leftward receive beams correspond to the columns and rows of light blue beam pairs, respectively, highlighted in \figref{fig:matrix-thresh-full}.
%For example, when transmitting to the upper right, only around $10$--$20$\% of receive beams offer more than 35 dB of isolation, as evidenced by the top plot in \figref{fig:fractions-a}.

% Less dramatically, a large portion of the moderately high to high isolation beam pairs can be seen as coming from  select transmit and receive directions.
% In fact, these directions seem to be fairly agnostic to elevation, existing as vertical strips of bright dots. % , which is likely attributed to the fact that our horizontally separated arrays differ in azimuth but not in elevation, leading to much more variation as a function of azimuth.

\begin{remark}
The most promising transmit beams and receive beams in meeting an \ginr threshold of $20$ dB, for example, can be seen as thin vertical strips of bright yellow.
These vertical strips, which were also visible as low-\ginr beams in \figref{fig:max-med-min}, are likely attributed to nulls in the transmit and receive beam patterns.
Notice, however, since the statistics of \figref{fig:fractions} were taken over all transmit/receive beams, it shows that the transmit nulls are robust to some degree, somewhat reliably offering lower \ginr regardless of the receive beam being used (and vice versa).
These transmit and receive beams offering lower \ginr across large fractions of receive beams and transmit beams, respectively, correspond to the approximate transmit and receive nulls at the channel input and output (i.e., approximate right and left null spaces of $\mH$), respectively.
Recall, from \figref{fig:max-med-min-a} and \figref{fig:max-med-min-b}, we did not see any transmit beams or receive beams that \textit{universally} provided high isolation.
If indeed these vertical strips are attributed to nulls in the beam patterns, this suggests that the self-interference channel between the transmit and receive arrays is quite {directional}, which somewhat further bucks the thought that near-field interaction dominates their coupling.
Our future work will explore this to better understand the coupling nature of the arrays.
\end{remark}

%These vertical strips isolation separated by strips of low isolation are speculated to be caused by the alignment and misalignment of side lobes (which are perhaps not well defined in this near-field region).
%These transmit and receive beams offering high isolation across large fractions of receive beams and transmit beams, respectively, correspond to the \textit{approximate} transmit and receive nulls at the channel input and output (i.e., approximate left and right null spaces of $\mH$), respectively, that offer high isolation.
%Recall, from \figref{fig:max-med-min-a} and \figref{fig:max-med-min-b}, we did not see any transmit beams or receive beams that \textit{universally} provided high isolation.

% We see that transmitting around broadside leads to low isolation across many receive beams and transmit beams, respectively.
% These results are not necessarily expected nor easily explained; it can perhaps be attributed to the presence of side lobes and their enhancement in this near-field setting.

\subsection{\ginr for Particular Transmit-Receive Beam Pairs}
\begin{figure*}[t]
    \centering
    \subfloat[Observed by each transmit beam.]{\includegraphics[width=0.475\linewidth,height=\twocolfigheightfrac\textheight,keepaspectratio]{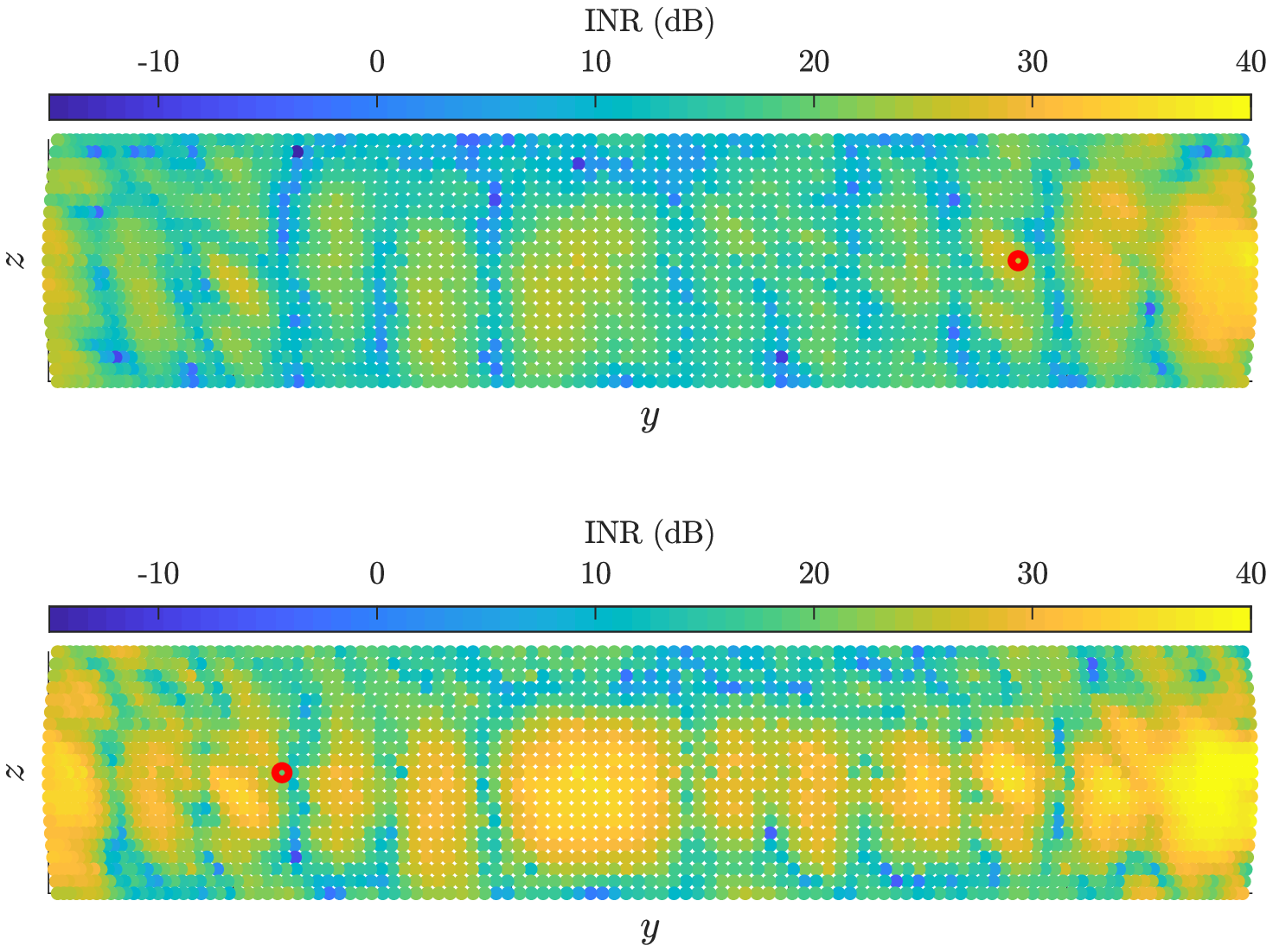}
        \label{fig:raw-inr-a}}
    \quad
    \subfloat[Observed by each receive beam.]{\includegraphics[width=0.475\linewidth,height=\twocolfigheightfrac\textheight,keepaspectratio]{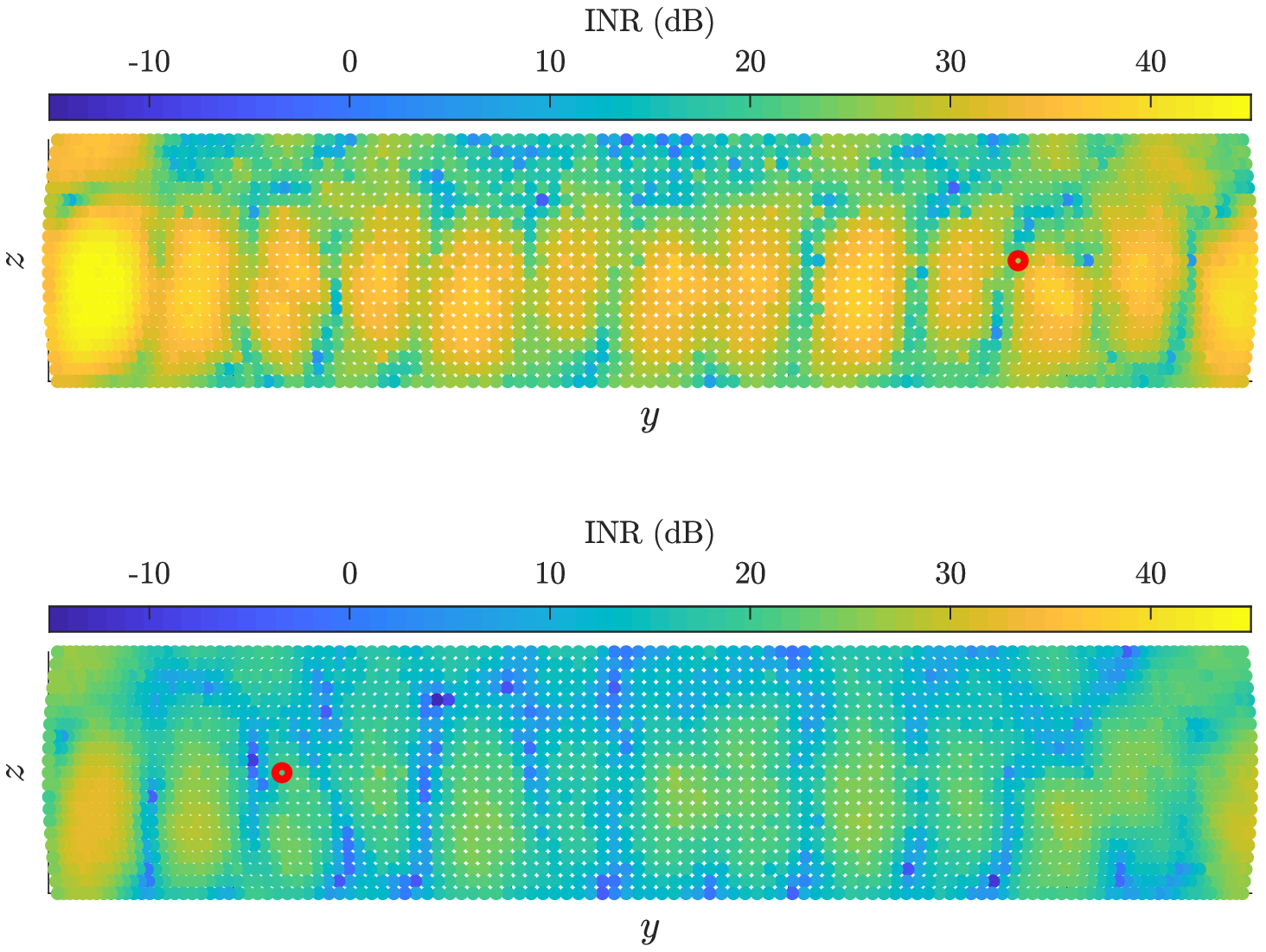}
        \label{fig:raw-inr-b}}
    \caption{For each transmit beam and receive beam, shown is the measured \ginr across all receive and transmit beams, respectively. The red circle indicates the (a) receive direction and (b) transmit direction.}
    \label{fig:raw-inr}
\end{figure*}

Honing in further, we now look at the isolation achieved at each transmit beam \textit{for a particular receive beam} and at each receive beam \textit{for a particular transmit beam}, as shown in \figref{fig:raw-inr-a} and \figref{fig:raw-inr-b}.
Let us first consider the \ginr observed across receive beams for particular transmit beams; imagine fixing the transmit beam and sweeping the receive beam to measure \ginr at each.
In \figref{fig:raw-inr-b}, we have selected two transmit directions: toward the receive array (top plot) and away from the receive array (bottom plot).
For each, have shown the \ginr measured between the transmit beam and each receive beam option.
Shown in the top plot of \figref{fig:raw-inr-b}, when transmitting rightward toward the receiver (whose direction is shown as a red circle), we see fairly high \ginr across the receive profile.
Large orange/yellow spots make up most of the receive profile, highlighting just how difficult it may be to find a receive beam that offers low \ginr for this particular transmit beam.
There exist some low-\ginr receive options narrowly in between large spots of orange or at high and low elevation.

% Note that we have used different color scales across the five figures, though the scale is fairly consistent except when transmitting broadside.
% The range of isolation values across receive beams is fairly consistent for each of the five transmit directions, varying only by about $5$ dB or so on the lower end and around $15$ dB on the upper end.
% As expected, when the transmit beam changes, the isolation profile across receive beams also changes.
% When transmitting in any of these three directions, we observe a wide range of isolation across receive beams.

Now, looking at the bottom plot of \figref{fig:raw-inr-b}, when steering the transmitter leftward away from the receiver, the \ginr profile across receive beams expectedly changes.
The \ginr profile sees a widespread decrease of about $10$ dB or more and options for extremely low \ginr are more available.
Still, receiving leftward toward the transmitter remains the least attractive option and reinforces that isolation may {tend} to be lower when transmitting rightward toward the receiver and when receiving leftward toward the transmitter---but this is not {universally} the case.
Looking at both plots in \figref{fig:raw-inr-b}, the receive beam that offers minimum \ginr varies with transmit beam, which further backs our claim that there are not receive beams that universally offer low \ginr.
Moreover, the low-\ginr receive directions are typically quite narrow in the sense that small changes in receive direction can lead to significant changes in isolation.
For instance, when transmitting rightward toward the receiver, the \ginr across receive beams varies by about $60$ dB, and we see that shifting a receive beam by only $1^\circ$ to $2^\circ$ in azimuth and/or elevation can lead to changes of $20$--$30$ dB or more in \ginr.
Notice that this sensitivity to steering direction is much more apparent with low-\ginr beams than high-\ginr ones.

Similarly, in \figref{fig:raw-inr-a}, we have selected two receive directions and, for each, have shown the \ginr measured between the receive beam and each transmit beam.
Analogous conclusions can be drawn as with \figref{fig:raw-inr-b}, though there are useful comments to make.
% First, we note that the range of isolation seen across the five receive beams varies by around $7$ dB on the lower end and about $16$ dB on the upper end; this is slightly more variability than in \figref{fig:raw-inr-b} but is fairly close.
% When receiving in the upper left (fourth plot), we can see that a near isolation-maximizing transmit direction is in approximately broadside; the similar observation made in \figref{fig:raw-inr-b} further illustrates the symmetry seen in our measurements and, thus, the self-interference channel.
Again, varying with each receive beam, there exists an \ginr-minimizing transmit beam.
Notice that even when the receive beam is steered away from the transmit array (to the right; top plot), transmitting toward the receive array (to the right) still inflicts substantial self-interference.
We can clearly see that simply steering the transmitter away from the receiver \textit{or} steering the receiver away from the transmitter does not offer widespread low \ginr.
% Even when both are done so simultaneously, we do not necessarily see high isolation, though perhaps we are more likely to.
Comparing \figref{fig:raw-inr-a} and \figref{fig:raw-inr-b}, we observe a certain degree of symmetry.
Transmitting toward the receiver (top \figref{fig:raw-inr-b}) is similar to receiving toward the transmitter (bottom \figref{fig:raw-inr-a}).
Transmitting away from the receiver (bottom \figref{fig:raw-inr-b}) is similar to receiving away from the transmitter (top \figref{fig:raw-inr-a}).
% Transmitting broadside (middle \figref{fig:raw-inr-b}) is similar to receiving broadside (middle \figref{fig:raw-inr-a}).
This further verifies a sense of spatial symmetry of our self-interference channel $\mH$.

% When transmitting to the left (away from the receiver; bottom plot), we see that receive beams offer moderately high or high isolation more often.
% Receiving to the extreme left (toward the transmitter), leads to low isolation, unavoidably.

\begin{remark}
\figref{fig:raw-inr-a} and \figref{fig:raw-inr-b} highlight that there exist large-scale (global) trends in the amount of self-interference coupled between the transmit and receive arrays, since general steering direction of the transmitter and receiver can play a significant role in the \ginr profile.
In addition, they also illustrate the local phenomena present in the \ginr profile: small shifts in steering direction can have drastic impacts on the degree of self-interference coupled.
\figref{fig:raw-inr-a} and \figref{fig:raw-inr-b} showed this sensitivity of the transmit beam and receive beam separately---in the next section, we investigate this sensitivity when the transmit beam \textit{and} receive beam both see small shifts in their steering direction.
\end{remark}

\section{Quantifying the Angular Spread of mmWave Self-Interference} \label{sec:angular-spread}

\begin{figure}
    \centering
    \includegraphics[width=\linewidth,height=0.18\textheight,keepaspectratio]{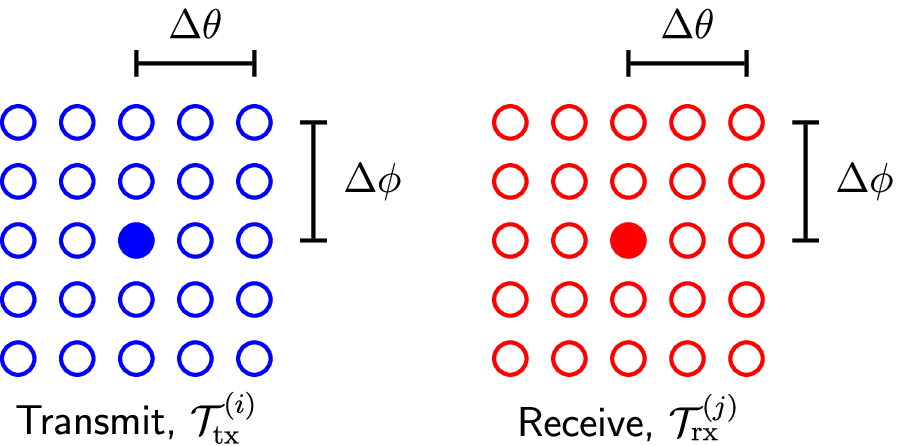}
    \caption{The transmit and receive steering neighborhoods $\txdirsetmeas\idx{i}$ and $\rxdirsetmeas\idx{j}$, where the filled circles indicate the nominal steering directions $\parens{\thetatx\idx{i},\phitx\idx{i}}$ and $\parens{\thetarx\idx{j},\phirx\idx{j}}$ and the unfilled circles indicate directions comprising the $\parens{\Delta\theta,\Delta\phi}$-neighborhood surrounding each. In this case, the $25$ directions in each yields $25^2 = 625$ transmit-receive steering combinations.}
    \label{fig:neighborhood}
\end{figure}

% \subsection{Preliminaries}
In this section, we inspect how self-interference varies with small changes in transmit and receive directions.
Let us begin by defining $\anglediff{\alpha,\beta}$ as the absolute difference between two angles $\alpha,\beta$ (in degrees), written as
\begin{align}
\anglediff{\alpha,\beta} =
\begin{cases}
\zeta, & \zeta \leq 180^\circ \\
360^\circ - \zeta, & \zeta > 180^\circ \\
\end{cases}
\end{align}
where $\zeta = \bars{\alpha - \beta} \ \mathrm{mod} \ 360^\circ$ and $\mathrm{mod}$ is the modulo operator.
Let $\txdirsetmeas\idx{i}$ and $\rxdirsetmeas\idx{j}$ be the $(\Delta\theta,\Delta\phi)$-neighborhoods around the $i$-th transmit direction and $j$-th receive direction, respectively, defined as
\begin{align}
\txdirsetmeas\idx{i}\nbr &= \braces{(\theta,\phi) \in \txdirsetcb : \anglediff{\theta,\thetatx\idx{i}} \leq \Delta\theta, \anglediff{\phi,\phitx\idx{i}} \leq \Delta\phi} \\
\rxdirsetmeas\idx{j}\nbr &= \braces{(\theta,\phi) \in \rxdirsetcb : \anglediff{\theta,\thetarx\idx{j}} \leq \Delta\theta, \anglediff{\phi,\phirx\idx{j}} \leq \Delta\phi}
\end{align}
and illustrated in \figref{fig:neighborhood}.
For some $(\Delta\theta,\Delta\phi)$ in degrees, the cardinality of these sets is
\begin{align} \label{eq:tx-rx-card}
\card{\txdirsetmeas\idx{i}\nbr}, \card{\rxdirsetmeas\idx{j}\nbr} \leq (2\cdot \Delta\theta + 1) \cdot (2 \cdot \Delta\phi + 1)
\end{align}
with equality when not at the edge of the measurement space (which is typical); the crude use of $\Delta\theta$ and $\Delta\phi$ here is thanks to our $1^\circ$ spacing of $\txdirsetcb$ and $\rxdirsetcb$.
% Let $L(\thetatx,\phitx,\thetarx,\phirx)$ be the measured isolation observed when transmitting toward $(\thetatx,\phitx)$ and receiving toward $(\thetarx,\phirx)$.

\begin{figure}
    \centering
    \includegraphics[width=\linewidth,height=0.3\textheight,keepaspectratio]{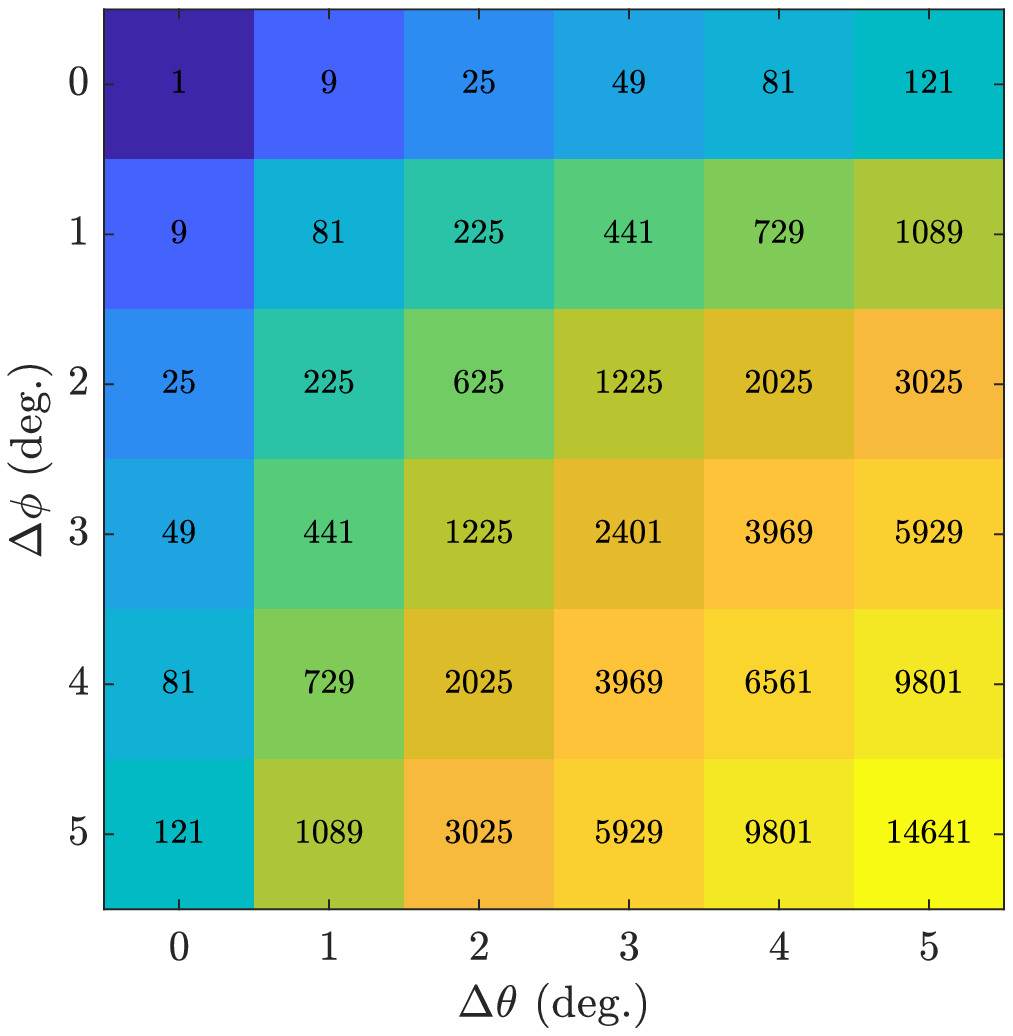}
    \caption{The number of transmit-receive beam pairs for a typical \nbr-neighborhood under our $\nbroneone$ resolution.} %This is the upper bound on the cardinality of $\Iij\nbr$.}
    \label{fig:neighborhood-size}
\end{figure}

Then, let $\IIij(\Delta\theta,\Delta\phi)$ be the set of measured \gls{inr} values across the $(\Delta\theta,\Delta\phi)$-neighborhood surrounding the $(i,j)$-th transmit-receive beam pair, expressed as
\begin{align}
\IIij(\Delta\theta,\Delta\phi) = \braces{\inr(\thetatx,\phitx,\thetarx,\phirx) : \parens{\thetatx,\phitx} \in \txdirsetmeas\idx{i}\nbr, \parens{\thetarx,\phirx} \in \rxdirsetmeas\idx{j}\nbr}
\end{align}
where $\txdirsetmeas\idx{i}$ and $\rxdirsetmeas\idx{j}$ depends on the beam pair $(i,j)$ and neighborhood size $(\Delta\theta,\Delta\phi)$.
As the $(\Delta\theta,\Delta\phi)$-neighborhoods are widened, the cardinality of $\IIij$ grows, which is simply the product of that of $\txdirsetmeas\idx{i}$ and $\rxdirsetmeas\idx{j}$.
\begin{align}
\card{\IIij(\Delta\theta,\Delta\phi)} = \card{\txdirsetmeas\idx{i}\nbr} \cdot \card{\rxdirsetmeas\idx{j}\nbr} \label{eq:nbr-card}
% \parens{(2\cdot \Delta\theta + 1) \cdot (2 \cdot \Delta\phi + 1)}^2
\end{align}
Based on \eqref{eq:tx-rx-card}, the upper bound of \eqref{eq:nbr-card} is tabulated for various $\nbr$ in \figref{fig:neighborhood-size}, which grows with order $\mathcal{O}\parens{\Delta\theta^2 \cdot \Delta\phi^2}$. % and depends heavily our $\nbroneone$ measurement resolution.

The minimum \ginr and maximum \ginr offered by beam pairs across the $(\Delta\theta,\Delta\phi)$-neighborhood surrounding the $(i,j)$-th beam pair can be expressed as simply
\begin{align}
\inrijmin(\Delta\theta,\Delta\phi) &= \min\braces{\IIij(\Delta\theta,\Delta\phi)} \label{eq:batman} \\
\inrijmax(\Delta\theta,\Delta\phi) &= \max\braces{\IIij(\Delta\theta,\Delta\phi)}.
\end{align}
Using these, the \gls{inr} range (in dB) we define as
\begin{align}
\inrijrng(\Delta\theta,\Delta\phi) = \todB{\inrijmax(\Delta\theta,\Delta\phi)} - \todB{\inrijmin(\Delta\theta,\Delta\phi)} \geq 0
\end{align}
which captures how much the \gls{inr} can vary over the $(\Delta\theta,\Delta\phi)$-neighborhood surrounding the $(i,j)$-th beam pair.
% Thus, when $\Delta\theta = 2 \cdot 60^\circ = 120^\circ$ and $\Delta\phi = 2 \cdot 10^\circ = 20^\circ$, the set $\LLij$ contains the entire set of nearly 6.5 million measurements for any $i,j$.
By examining $\inrijmin\nbr$, $\inrijmax\nbr$, and $\inrijrng\nbr$ for each transmit-receive steering combination $(i,j)$ and for variably sized neighborhoods, we can gain insight into the angular spread of self-interference.
We point out that, since our measurements were taken with $1^\circ$ resolution in azimuth and elevation, there exists the potential to see greater \ginr range, lower minimum \ginr, and/or higher maximum \ginr if sub-\nbroneone resolutions were used; as such, the results herein can be considered a potentially conservative measure on these statistics over small neighborhoods.

\subsection{INR Range over Various Neighborhoods}

\begin{figure*}
    \centering
    \subfloat[$\inrijrng(\Delta\theta,\Delta\phi)$.]{\includegraphics[width=0.475\linewidth,height=0.26\textheight,keepaspectratio]{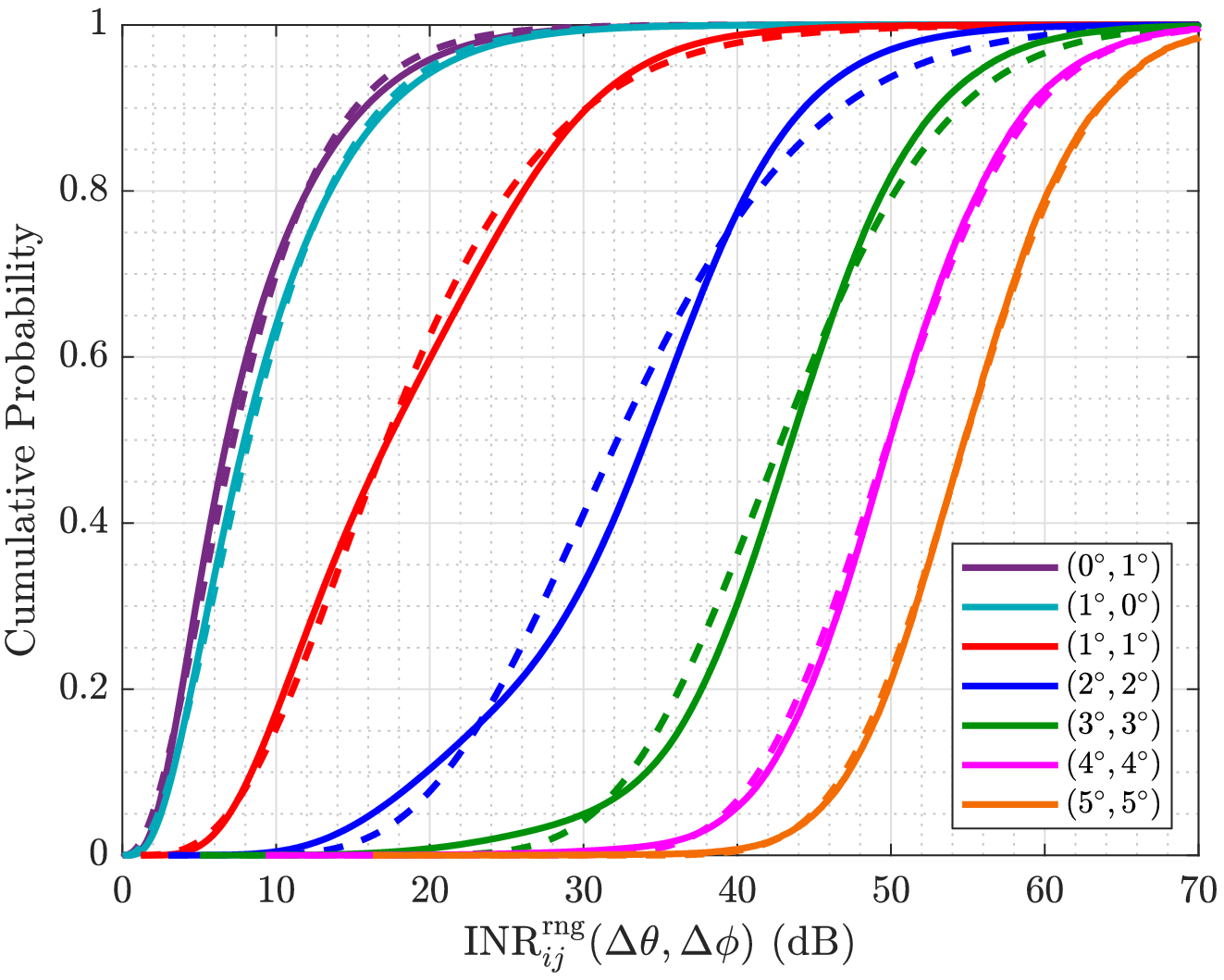}
        \label{fig:inr-rng-a}}
    \quad
    \subfloat[$\alpharng(\Delta\theta,\Delta\phi)$ and $\betarng(\Delta\theta,\Delta\phi)$.]{\includegraphics[width=0.475\linewidth,height=0.26\textheight,keepaspectratio]{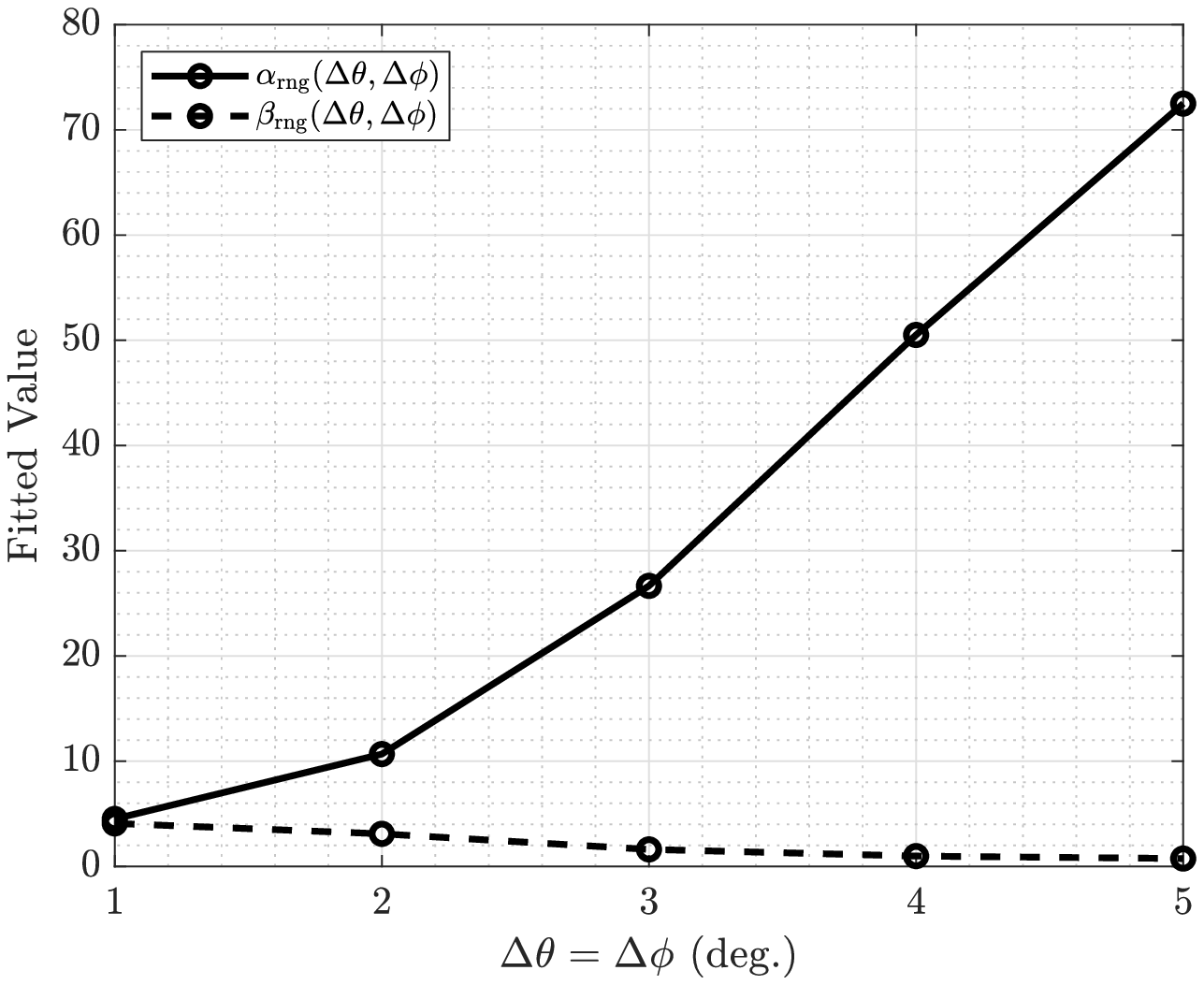}
        \label{fig:inr-rng-b}}
    \caption{(a) The \gls{cdf} of $\inrijrng$ across all nearly 6.5 million transmit-receive beam pairs for various neighborhood sizes $(\Delta\theta,\Delta\phi)$. Fitted distributions for each are shown as dashed lines. (b) The fitted parameters $\alpharng(\Delta\theta,\Delta\phi)$ and $\betarng(\Delta\theta,\Delta\phi)$ for various $\Delta\theta = \Delta\phi$.}
    \label{fig:inr-rng}
\end{figure*}

In \figref{fig:inr-rng-a}, we plot the \gls{cdf} of \gls{inr} range for variably sized $\parens{\Delta\theta,\Delta\phi}$-neighborhoods across all measured direction pairs (i.e., each \gls{cdf} contains nearly 6.5 million points).
% We define the isolation range across a $\parens{\Delta\theta,\Delta\phi}$-neighborhood around the $i$-th transmit direction and $j$-th receive direction as
% $\parens{\thetatx\idx{i},\phitx\idx{i}}$ and $\parens{\thetarx\idx{j},\phirx\idx{j}}$ as
% \begin{align}
% \max\braces{\todB{\LLij}} - \min\braces{\todB{\LLij}}
% \end{align}
% which captures how much the isolation (or self-interference power or $\inr$) can vary with slight shifts in the transmit and/or receive steering directions.
% $\parens{\Delta\theta,\Delta\phi}$-neighborhood.
% We have considered neighborhood sizes up to $2^\circ$ in both azimuth and elevation.
As shown in \figref{fig:inr-rng-a}, moving a beam pair by only $1^\circ$ in either azimuth \textit{or} elevation can lead to notable changes in \gls{inr}: around $25$\% of beam pairs observe over $10$ dB of \gls{inr} range in either case.
As the neighborhood size increases, we naturally observe a wider range of \gls{inr}.
$50$\% of beam pairs see more than $17$ dB of variability in \gls{inr} across a $(1^\circ,1^\circ)$-neighborhood.
In other words, if we consider a beam pair at random and look around its $(1^\circ,1^\circ)$-neighborhood, we would expect the \gls{inr} to vary by $17$ dB or more.
Across a $(2^\circ,2^\circ)$-neighborhood, $80$\% of beam pairs see around $25$ dB or more of variability in \gls{inr}.
Notice that there exists slightly more variability in \gls{inr} across azimuth than across elevation, evidenced by the \nbronezero- and \nbrzeroone-neighborhoods---perhaps due to the horizontal separation of our transmit and receive arrays.

%\begin{figure}
%    \centering
%    \includegraphics[width=\linewidth,height=0.35\textheight,keepaspectratio]{plots_v2/angular/main_apple_v20_838}
%    \caption{}
%    \label{fig:inr-rng}
%\end{figure}

%\begin{figure*}
%    \centering
%    \subfloat[\gls{inr} range, $\inrijrng$.]{\includegraphics[width=0.48\linewidth,height=0.27\textheight,keepaspectratio]{plots_v2/angular/main_apple_v01_01.eps}
%        \label{fig:relative-isolation-range-all}}
%    \quad
%    \subfloat[Minimum \gls{inr}, $\inrijmin$.]{\includegraphics[width=0.48\linewidth,height=0.27\textheight,keepaspectratio]{plots_v2/angular/main_apple_v01_02.eps}
%        \label{fig:minimum-inr-all}}
%    \caption{The \glspl{cdf} of (a) \gls{inr} range $\inrijrng$ and (b) minimum \gls{inr} $\inrijmin$ for various $\parens{\Delta\theta,\Delta\phi}$-neighborhoods at each of the 6.5 million measured transmit-receive beam pairs.}
%    % \label{fig:subfigs-}
%\end{figure*}

To provide engineers with statistics on $\inrijrng$ for variably sized $(\Delta\theta,\Delta\phi)$-neighborhoods, we have fit a distribution to $\braces{\inrijrng(\Delta\theta,\Delta\phi)}$.
Specifically, we found that a Gamma distribution can be fitted to each of the \glspl{cdf} in \figref{fig:inr-rng-a} as follows
\begin{align}
\braces{\todB{\inrijrng(\Delta\theta,\Delta\phi)} \ \forall \ i,j} \overset{\mathrm{fit}}{\sim} \distgamma{\alpharng(\Delta\theta,\Delta\phi)}{\betarng(\Delta\theta,\Delta\phi)}
\end{align}
where $\alpharng(\Delta\theta,\Delta\phi) > 0$ and $\betarng(\Delta\theta,\Delta\phi) > 0$ are the fitted shape and rate (inverse scale) of the Gamma distribution.
The fitted Gamma distributions for each neighborhood in \figref{fig:inr-rng-a} are shown as dashed lines.
In addition, we have plotted $\alpharng\nbr$ and $\betarng\nbr$ as functions of $\Delta\theta = \Delta\phi$ in \figref{fig:inr-rng-b}.
As $\Delta\theta = \Delta\phi$ increases, the shape parameter $\alpharng$ drastically increases and the rate parameter $\betarng$ decays toward zero, which is a reflection of the \glspl{cdf} in \figref{fig:inr-rng-a} shifting rightward.

In addition to those shown in \figref{fig:inr-rng}, we fitted unique Gamma distributions for $\Delta\theta, \Delta\phi \in \braces{0^\circ,1^\circ,\dots,5^\circ}$ and tabulated the fitted parameters $(\alpharng\nbr,\betarng\nbr)$ for each in \tabref{tab:neighborhood-fits-rng}.
Engineers wishing to realize the range in \ginr over a random \nbr-neighborhood or conduct statistical analyses related to such can refer to \tabref{tab:neighborhood-fits-rng} for adequate Gamma distribution parameters $(\alpharng\nbr,\betarng\nbr)$.
Then, the expected range in $\inr$ (in dB) over some $\nbr$-neighborhood, for instance, can be approximated as simply
\begin{align}
\ev{\inrijrng\nbr} = \frac{\alpharng\nbr}{\betarng\nbr}
\end{align}
based on the Gamma distribution, along with an assortment of other statistics readily computed.

\begin{remark}
    \figref{fig:inr-rng-a} highlights that the self-interference channel is not spatially smooth.
    Rather, small changes in steering direction can result in significant changes in the degree of self-interference coupled and, hence, significant changes in full-duplex performance.
    As such, \mmwave full-duplex systems cannot expect to reliably avoid self-interference by broadly steering transmit and receive beams.
    Instead, transmit and receive beams will need to be carefully (and \textit{jointly}) steered, as small errors in steering direction can lead to drastic changes in self-interference.
    % Instead, transmit and receive beams will likely need to be carefully (and \textit{jointly}) selected to confidently offer reduced self-interference---beams will need to be accurately and reliably steered, as small errors in steering direction can lead to drastic changes in self-interference.
\end{remark}

%\begin{figure}
%    \centering
%    \includegraphics[width=\linewidth,height=0.35\textheight,keepaspectratio]{plots_v2/angular/main_apple_v20_830}
%    \caption{Caption.}
%    \label{fig:}
%\end{figure}

\subsection{Minimum INR over Various Neighborhoods}

\begin{figure*}[t]
    \centering
    \subfloat[Observed by each transmit beam.]{\includegraphics[width=0.475\linewidth,height=\threecolfigheightfrac\textheight,keepaspectratio]{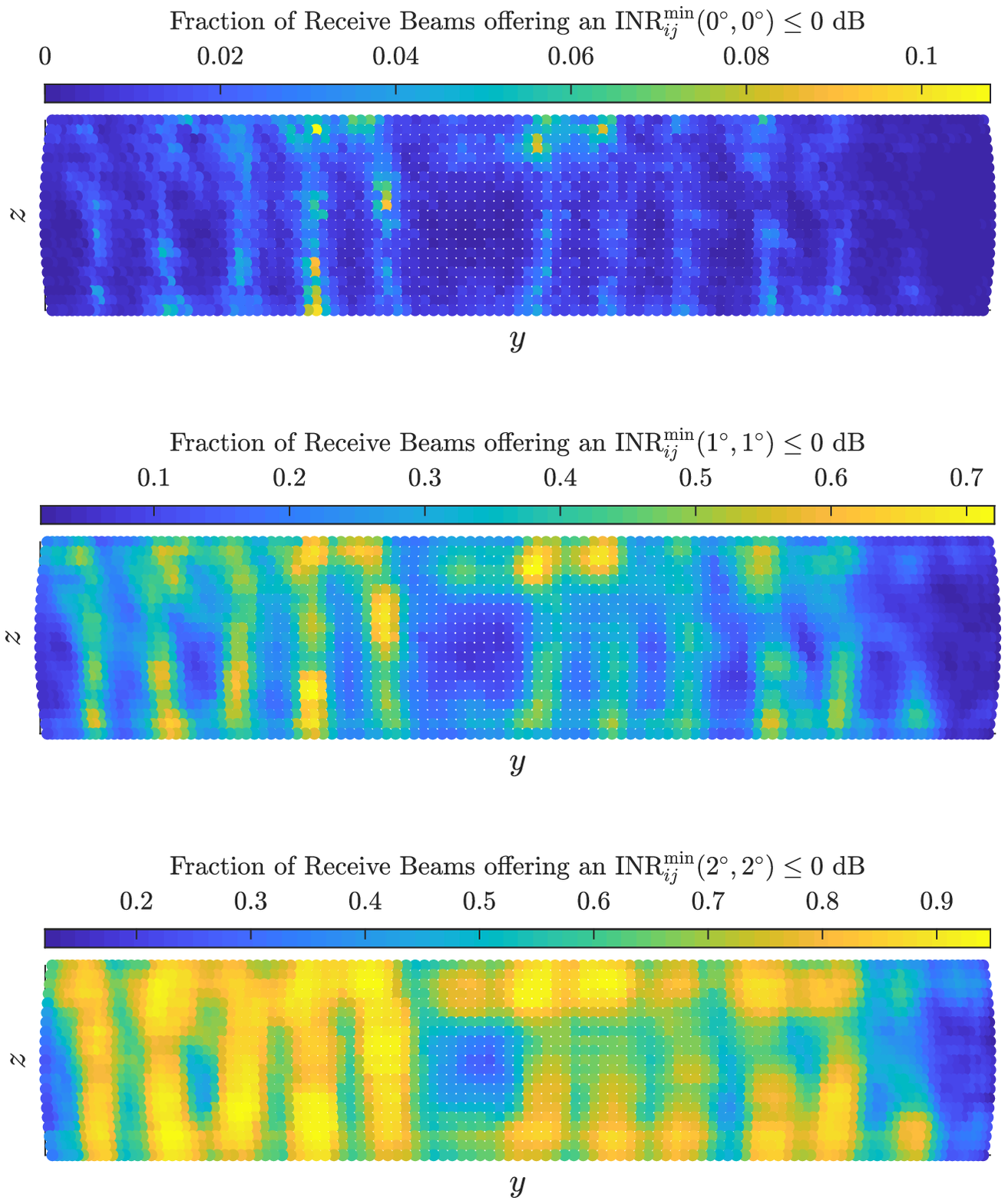}
        \label{fig:inr-min-nbr-a}}
    \quad
    \subfloat[Observed by each receive beam.]{\includegraphics[width=0.475\linewidth,height=\threecolfigheightfrac\textheight,keepaspectratio]{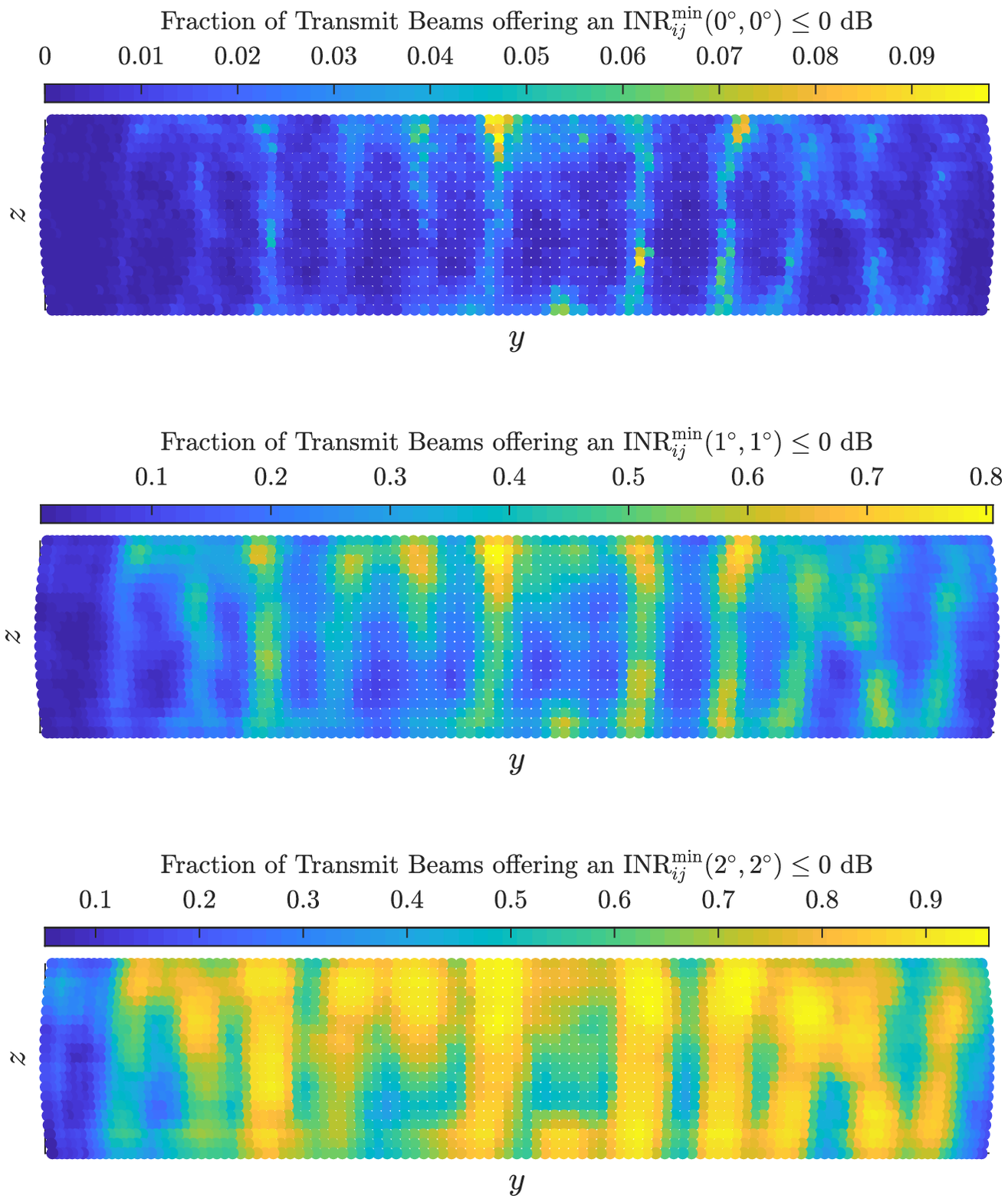}
        \label{fig:inr-min-nbr-b}}
    \caption{For each transmit beam and receive beam, shown are the fraction of receive beams and transmit beams, respectively, that can offer an \ginr of $0$ dB or less when allowed to deviate by various \nbr. Low \ginr becomes reliably within arm's reach by increasing $\nbr$.}
    \label{fig:inr-min-nbr}
\end{figure*}

Now, we examine the minimum \ginr that can be reached by each beam pair if allowed to deviate around some spatial neighborhood.
To illustrate this, we have included \figref{fig:inr-min-nbr}.
In \figref{fig:inr-min-nbr-a}, at each transmit beam, we show the fraction of receive beams that can offer a minimum \ginr of $0$ dB or less.
We do this for various neighborhood sizes \nbr, where a \nbrzerozero-neighborhood is simply no deviation.
With a \nbrzerozero-neighborhood, most transmit-receive beams cannot meet the \ginr threshold of $0$ dB.
As the neighborhood grows to \nbroneone, we see that, for some transmit beams, a large fraction of receive beams can reach an \ginr of $0$ dB (and vice versa).
With \nbrtwotwo of freedom, we see the clouds of yellow grow as more receive beams offer an \ginr of $0$ dB for even more transmit beams.
\figref{fig:inr-min-nbr} illustrates the significant changes observed in \ginr due to slight shifts of the transmit and receive beams and shows that \ginr levels suitable for full-duplex are in fact within arm's reach.

\begin{figure*}
    \centering
    \subfloat[$\inrijmin(\Delta\theta,\Delta\phi)$.]{\includegraphics[width=0.475\linewidth,height=\textheight,keepaspectratio]{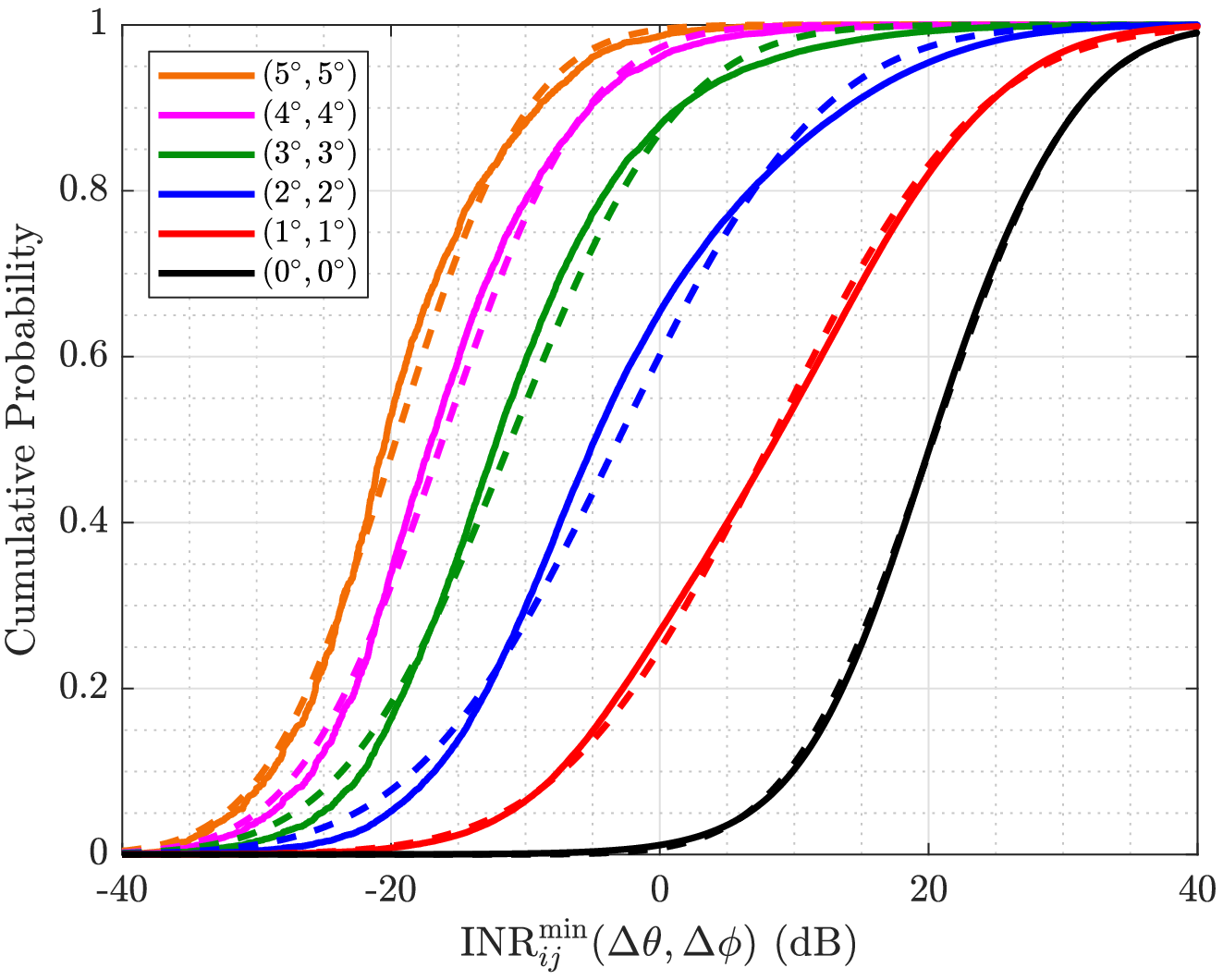}
        \label{fig:inr-min-a}}
    \quad
    \subfloat[$\mumin(\Delta\theta,\Delta\phi)$ and $\varmin(\Delta\theta,\Delta\phi)$.]{\includegraphics[width=0.475\linewidth,height=\textheight,keepaspectratio]{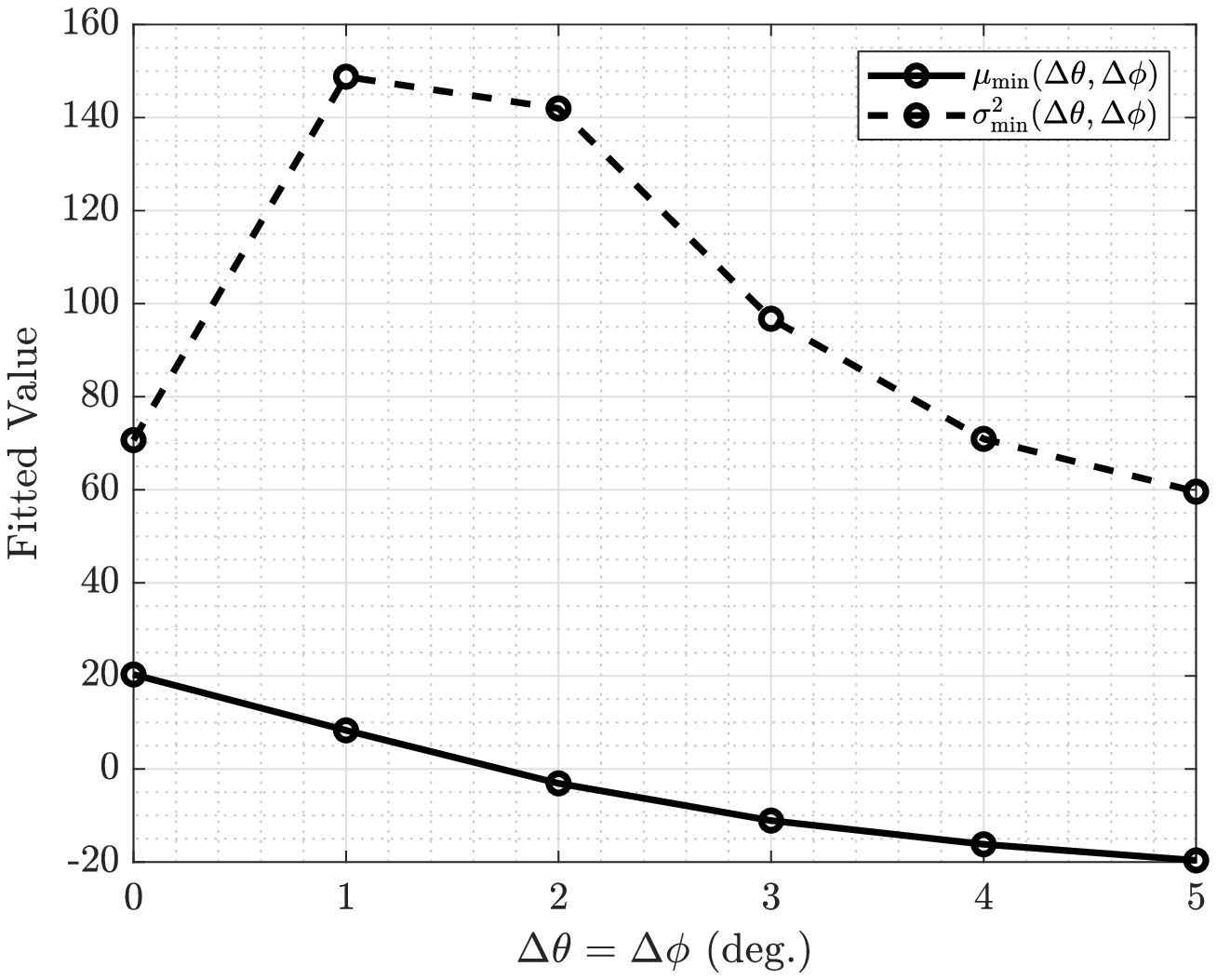}
        \label{fig:inr-min-b}}
    \caption{(a) The \gls{cdf} of $\inrijmin(\Delta\theta,\Delta\phi)$ for various $(\Delta\theta,\Delta\phi)$. For each, the fitted distribution is shown as a dashed line. (b) The fitted parameters $\mumin(\Delta\theta,\Delta\phi)$ and $\varmin(\Delta\theta,\Delta\phi)$ for various $\Delta\theta = \Delta\phi$.}
    \label{fig:inr-min}
\end{figure*}

Consider \figref{fig:inr-min-a}, where we plot the \gls{cdf} of $\braces{\inrijmin}$ for variably sized $\parens{\Delta\theta,\Delta\phi}$-neighborhoods across all measured beam pairs.
% Since \inr is minimized when isolation is maximized, this is equivalent to choosing the maximum isolation in each $\LLij$.
The dashed black line in \figref{fig:inr-min-a} is simply the \gls{cdf} of $\braces{\inrij}$ for each beam pair since $\Delta\theta = \Delta\phi = 0^\circ$ (shown previously in \figref{fig:cdf}).
% , which shows that the vast majority of the nearly 6.5 million beam pairs yield an $\inr \gg 0$ dB.
% On its own, this dashed line is quite demotivating from a full-duplex perspective since a vast majority of direction pairs result in $\inr \gg 0$ dB, suggesting that user location greatly dictates self-interference levels.
When considering small neighborhoods around each beam pair, however, much more promising results are observed.
When shifting beams by no more than $1^\circ$ in azimuth and elevation, the probability of reaching $\inr \leq 0$ dB (where self-interference is no stronger than noise) grows to over $25$\% from around $1$\%.
With $2^\circ$, it grows to over $65$\%; that is to say that over $65$\% of beam pairs are within $(2^\circ,2^\circ)$ of a beam pair that offers $\inr \leq 0$ dB.

\begin{remark}
    Clearly, \gls{inr} can be greatly improved with slight shifts in the steering directions of the transmit and/or receive beams.
    From this, we draw an important conclusion: steering directions that do not \textit{inherently} offer high isolation (i.e., low \inrij) are likely spatially near ones that do.
    These are highly encouraging results for the potential of \mmwave full-duplex since they suggest that self-interference can be greatly reduced while making very minor adjustments to the transmit and receive steering directions.
    This is reinforced further by the fact that our beams have a $3$ dB beamwidth around $7^\circ$, meaning slight deviations will hopefully not sacrifice too much beamforming gain when making these adjustments.
    To reach these low-\ginr beam pairs, however, it may require searching over many beam pairs within a small spatial neighborhood, as highlighted by \figref{fig:neighborhood-size}.
    For instance, with our $1^\circ$ resolution, a typical $\nbrtwotwo$-neighborhood contains $625$ transmit-receive beam pairs, highlighting that there may be practical hurdles in locating low-\ginr beam pairs within a given neighborhood.
    Exploring how side lobes, beamwidth, relative array geometry, and mounting infrastructure play a role in this small-scale variability would be valuable future work.
\end{remark}

To conduct statistical analyses of $\inrijmin$, we have fit a distribution to $\braces{\inrijmin(\Delta\theta,\Delta\phi)}$ for variably sized $(\Delta\theta,\Delta\phi)$-neighborhoods.
Specifically, we found that a normal distribution can be fitted to each of the \glspl{cdf} in \figref{fig:inr-min-a} as follows
\begin{align}
\braces{\todB{\inrijmin(\Delta\theta,\Delta\phi)} \ \forall \ i,j} \overset{\mathrm{fit}}{\sim} \distgauss{\mumin(\Delta\theta,\Delta\phi)}{\varmin(\Delta\theta,\Delta\phi)}
\end{align}
where $\mumin(\Delta\theta,\Delta\phi)$ and $\varmin(\Delta\theta,\Delta\phi)$ are the fitted mean and variance of the normal distribution.
The dashed lines in \figref{fig:inr-min-a} depict each neighborhood's fitted distribution, and \figref{fig:inr-min-b} show the fitted parameters for various $\Delta\theta = \Delta\phi$.
Shown as the solid line in \figref{fig:inr-min-b}, the mean $\inrijmin\nbr$ steadily decreases by about $10$ dB per unit increase in $\Delta\theta = \Delta\phi$ before beginning to saturate. 
The dashed line shows the variance of the fit, which initially increases and then decreases, which suggests that $\nbroneone$ and $\nbrtwotwo$ of deviation can offer a reduction in \ginr but by highly variable amounts---evident also by their distributions in \figref{fig:inr-min-a}.
The variance decreases as the majority of beam pairs can reach similar levels of \ginr with \nbrthreethree or greater of deviation; the distributions in \figref{fig:inr-min-a} become more upright.
% not universally so---a substantial number beam pairs cannot reach low

For neighborhood sizes where $\Delta\theta, \Delta\phi \in \braces{0^\circ,1^\circ,\dots,5^\circ}$, we tabulated the fitted parameters $(\mumin\nbr,\varmin\nbr)$ in \tabref{tab:neighborhood-fits-min}.
To conduct statistical analyses related to minimum \gls{inr}, engineers can refer to \tabref{tab:neighborhood-fits-min} for adequate distribution parameters $\mumin\nbr$ and $\varmin\nbr$.
For instance, to realize the minimum \gls{inr} over a random neighborhood of size $\nbr$, engineers can simply draw 
\begin{align}
\todB{\inrmin\nbr} \sim \distgauss{\mumin\nbr}{\varmin\nbr}. \label{eq:inr-min-drawn-from-normal}
\end{align}
Plenty of statistics and statistical functions are readily available for the normal distribution which can be used to conduct system performance analyses. %  for convenience, we provide two below.
For instance, when $\todB{\inrmin\nbr} \sim \distgauss{\mumin\nbr}{\varmin\nbr}$, the probability that the \nbr-neighborhood surrounding a random beam pair exhibits a minimum \ginr of at most $\zeta$ is
\begin{align}
% \prob{\inrmin; \Delta\theta,\Delta\phi} = \frac{1}{\stdmin\nbr \sqrt{2\pi}} \cdot \exp\parens{-\frac{1}{2} \parens{\frac{\todB{\inrmin}-\mumin\nbr}{\stdmin\nbr}}^2}
\prob{\inrmin \leq \zeta; \Delta\theta,\Delta\phi} = \frac{1}{2}\brackets{1 + \erf{\frac{\todB{\zeta}-\mumin\nbr}{\stdmin\nbr \cdot \sqrt{2}}}}. \label{eq:inr-min-cdf-normal}
\end{align}

%\begin{theorem} \label{thm:inr-min}
%When $\todB{\inrmin\nbr} \sim \distgauss{\mumin\nbr}{\varmin\nbr}$, the probability that the \nbr-neighborhood surrounding a random beam pair exhibits a minimum \ginr of $\inrmin$ is
%\begin{align}
%% \prob{\inrmin; \Delta\theta,\Delta\phi} = \frac{1}{\stdmin\nbr \sqrt{2\pi}} \cdot \exp\parens{-\frac{1}{2} \parens{\frac{\todB{\inrmin}-\mumin\nbr}{\stdmin\nbr}}^2}
%\prob{\inrmin; \Delta\theta,\Delta\phi} = \frac{\exp\parens{-\frac{1}{2} \parens{\frac{\todB{\inrmin}-\mumin\nbr}{\stdmin\nbr}}^2}}{\stdmin\nbr \cdot \sqrt{2\pi}}
%\end{align}
%\end{theorem}
%
%\begin{corollary}
%The probability that the \nbr-neighborhood surrounding a random beam pair exhibits a minimum \ginr of at most $\zeta$ is
%\begin{align}
%% \prob{\inrmin; \Delta\theta,\Delta\phi} = \frac{1}{\stdmin\nbr \sqrt{2\pi}} \cdot \exp\parens{-\frac{1}{2} \parens{\frac{\todB{\inrmin}-\mumin\nbr}{\stdmin\nbr}}^2}
%\prob{\inrmin \leq \zeta; \Delta\theta,\Delta\phi} = \frac{1}{2}\brackets{1 + \erf{\frac{\todB{\zeta}-\mumin\nbr}{\stdmin\nbr \cdot \sqrt{2}}}}.
%\end{align}
%\end{corollary}

% \input{tab/tab-neighborhood-fits-min}

\begin{figure}
    \centering
    \includegraphics[width=\linewidth,height=0.25\textheight,keepaspectratio]{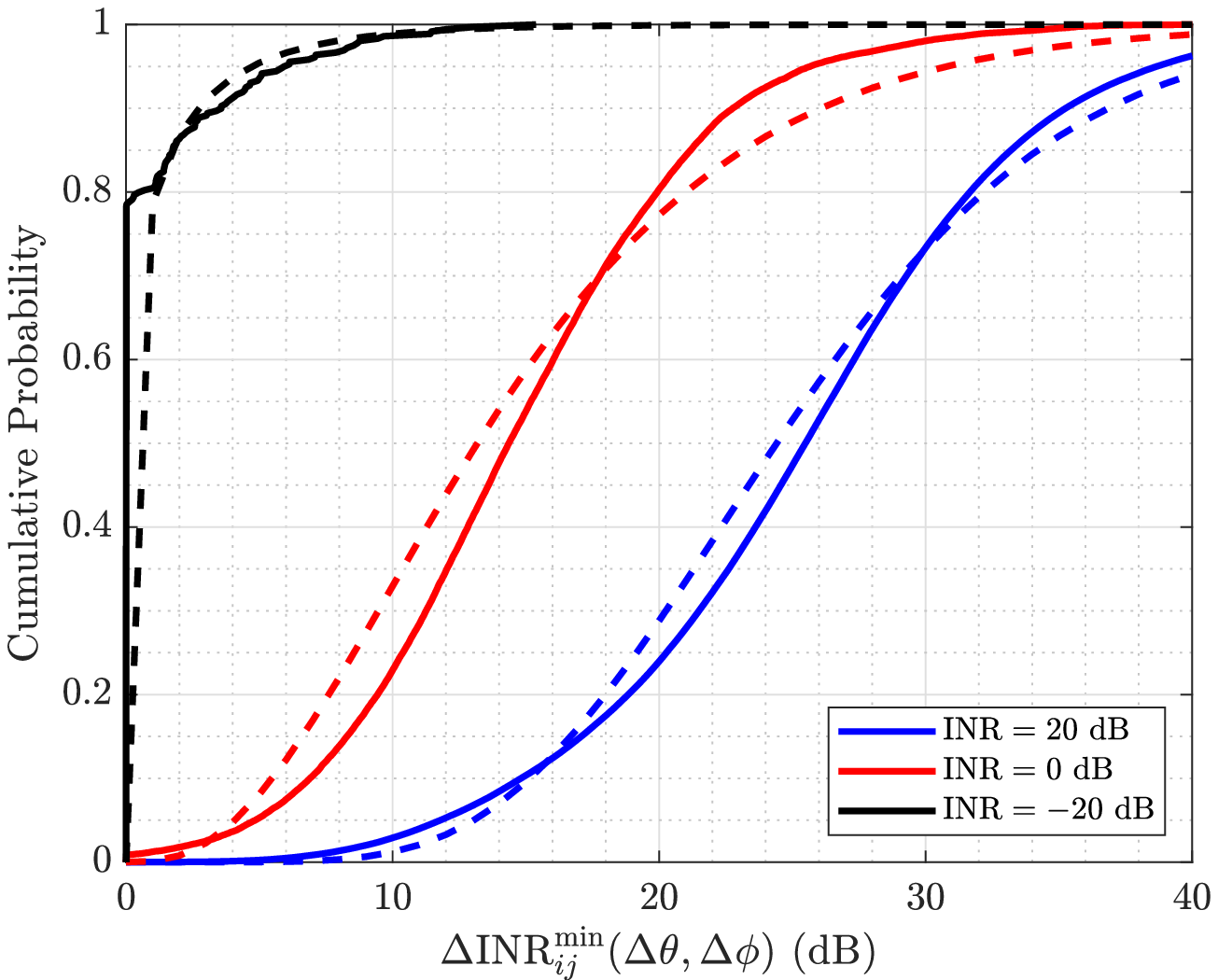}
    \caption{The \gcdf of $\braces{\Delta\inrijmin\parens{\Delta\theta,\Delta\phi,\inrij} : \inrij \approx \inr}$ for various $\inr$, where $\nbr = \nbrtwotwo$. Their fitted counterparts are shown as dashed lines.}
    \label{fig:delta-inr-min}
\end{figure}

% \textit{As a function of \ginr.} 
While fitting the \glspl{cdf} in \figref{fig:inr-min-a} directly to normal distributions is useful, it does not capture how the distribution of $\inrijmin$ varies as a function of $\inrij$.
In other words, it does not provide insight on if high-\gls{inr} beam pairs are as likely to be within arm's reach of low \ginr as low-\gls{inr} beam pairs are, for instance.
To provide more detailed statistics on $\inrijmin$ based on $\inrij$, we begin by computing
\begin{align}
\Delta\inrijmin\parens{\Delta\theta,\Delta\phi,\inrij} = \todB{\inrij} - \todB{\inrijmin(\Delta\theta,\Delta\phi)} \geq 0
\end{align}
which is the difference (in dB) between the inherent \ginr offered by the $(i,j)$-th beam pair and the minimum \ginr of its surrounding $\nbr$-neighborhood.
We subsequently form the set % $\DDmin\parens{\Delta\theta,\Delta\phi,\inr}$ as
% Map from $\inr$ to an $\inrijmin$ distribution over 
\begin{align}
% \IImin\parens{\Delta\theta,\Delta\phi,\inr} = 
\DDmin\parens{\Delta\theta,\Delta\phi,\inr} = 
\braces{\Delta\inrijmin\parens{\Delta\theta,\Delta\phi,\inrij} : \inrij \approx \inr} 
% \sim \distgauss{\mu(\inr,\Delta\theta,\Delta\phi)}{\sigma^2(\inr)}
% \braces{\todB{\inrijmin(\Delta\theta,\Delta\phi)} : \inrij \approx \inr } \sim \distgauss{\mu(\inr,\Delta\theta,\Delta\phi)}{\sigma^2(\inr)}
\end{align}
which is the set of all $\Delta\inrijmin(\Delta\theta,\Delta\phi,\inrij)$ values for beam pairs offering an $\inrij$ of approximately $\inr$.
The approximation here is merely used to ensure $\DDmin$ has a sufficient number of points in it to successfully fit it, as will become clear.
We found that the distribution of $\DDmin$ could be well approximated by a Gamma distribution as
\begin{align}
% \todB{\IImin\parens{\Delta\theta,\Delta\phi,\inr}} \sim \distgauss{\mumin(\Delta\theta,\Delta\phi,\inr)}{\varmin(\Delta\theta,\Delta\phi,\inr)}
{\DDmin\parens{\Delta\theta,\Delta\phi,\inr}} \overset{\mathrm{fit}}{\sim} \distgamma{\alphamin(\Delta\theta,\Delta\phi,\inr)}{\betamin(\Delta\theta,\Delta\phi,\inr)}
\end{align}
where $\alphamin(\Delta\theta,\Delta\phi,\inr)$ and $\betamin(\Delta\theta,\Delta\phi,\inr)$ are the fitted shape and rate parameters, parameterized by $\inr$ in addition to $\nbr$.
% We conducted this fitting for $\inr$ in the range $[-20,40]$ dB to ensure enough measured data points were present in $\DDmin$ for satisfactory fitting.
Alongside the true \gpcdf of various $\DDmin$, we plotted their fitted distributions using dashed lines in \figref{fig:delta-inr-min} for various $\inr$ and a \nbrtwotwo-neighborhood.
This illustrates that extremely low-\ginr beam pairs tend to see less reduction in \ginr over their neighborhoods compared to that of high-\ginr beam pairs. 
This is somewhat expected but also indicates that low-\ginr beam pairs are not congregated together but rather spread out throughout our transmit-receive space. 

In \tabref{tab:neighborhood-fits-min-inr}, we tabulated $(\alphamin,\betamin)$ for $\inr \in \braces{-20,-10,\dots,40}$ dB to provide engineers with $\alphamin(\Delta\theta,\Delta\phi,\inr)$ and $\betamin(\Delta\theta,\Delta\phi,\inr)$ for particular $\inr$ values.
To approximate $\alphamin(\Delta\theta,\Delta\phi,\inr)$ and $\betamin(\Delta\theta,\Delta\phi,\inr)$ for any $\inr \in [-20,40]$ dB, weighted interpolation can be used, for instance.
It is our hope that this means to realize $\Delta\inrijmin$ for particular $\inrij$ is useful for statistical analyses, simulation, and system evaluation.
For instance, one may draw $\inr$ from the global distribution (i.e., the \gcdf in \figref{fig:cdf}) as
% \begin{align}
$\todB{\inr} \sim \distgauss{\mu}{\sigma^2}$
% \end{align} 
and then use it when referencing \tabref{tab:neighborhood-fits-min-inr} to fetch $\alphamin(\Delta\theta,\Delta\phi,\inr)$ and $\betamin(\Delta\theta,\Delta\phi,\inr)$ based on some neighborhood size $\nbr$.
% A weighted average may be used to interpolate between $\inr$ values listed in \tabref{tab:neighborhood-fits-min-inr}.
From there, a realization of $\inrmin(\Delta\theta,\Delta\phi,\inr)$ can be drawn as
\begin{align}
\todB{\inrmin(\Delta\theta,\Delta\phi,\inr)} \sim \todB{\inr} -  \underbrace{\distgamma{\alphamin(\Delta\theta,\Delta\phi,\inr)}{\betamin(\Delta\theta,\Delta\phi,\inr)}}_{\Delta\inrmin\nbrinr}
\end{align}
which is the minimum \ginr over the $\nbr$-neighborhood surrounding a beam pair offering a nominal \ginr of $\inr$.
Statistical analysis can be conducted using the Gamma distribution, for instance, \gcdf as follows.
% The \gls{pdf} of the Gamma distribution fitted to $\Delta\inrmin\nbr$ follows the distribution
% When $\Delta\inrmin\nbrinr \sim \distgamma{\alphamin\nbrinr}{\betamin\nbrinr}$, the probability that the \nbr-neighborhood surrounding a random beam pair having an \ginr of $\inr$ exhibits a minimum \ginr of $\zeta$ or less is
When $\Delta\inrmin\nbrinr$ follows a Gamma distribution with parameters $\alphamin\nbrinr$ and $\betamin\nbrinr$, the probability that the \nbr-neighborhood surrounding a random beam pair having an \ginr of $\inr$ exhibits a minimum \ginr of $\zeta$ or less is
\begin{align}
\prob{\inrmin \leq \zeta; \Delta\theta,\Delta\phi,\inr} = \frac{\gamma\parens{\alphamin\nbrinr,\betamin\nbrinr \cdot \todB{\zeta}}}{\Gamma(\alphamin\nbrinr)} \label{eq:inr-min-cdf-gamma}
\end{align}
where $\Gamma(\cdot)$ is the Gamma function and $\gamma(\cdot,\cdot)$ is the lower incomplete Gamma function.

\subsection{Maximum INR over Various Neighborhoods}
\begin{figure*}
    \centering
    \subfloat[$\inrijmax(\Delta\theta,\Delta\phi)$.]{\includegraphics[width=0.475\linewidth,height=\textheight,keepaspectratio]{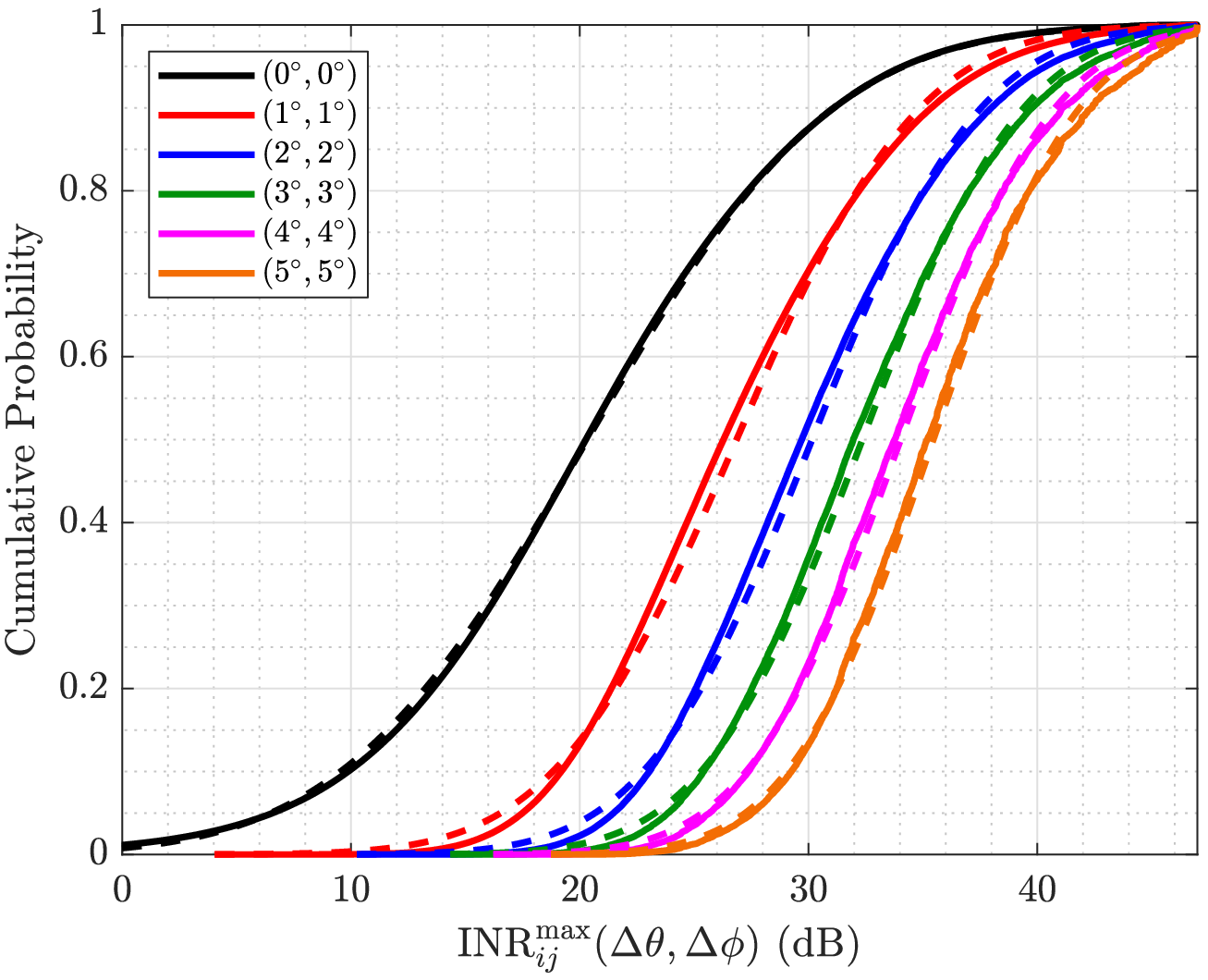}
        \label{fig:inr-max-a}}
    \quad
    \subfloat[$\mumax(\Delta\theta,\Delta\phi)$ and $\varmax(\Delta\theta,\Delta\phi)$.]{\includegraphics[width=0.475\linewidth,height=\textheight,keepaspectratio]{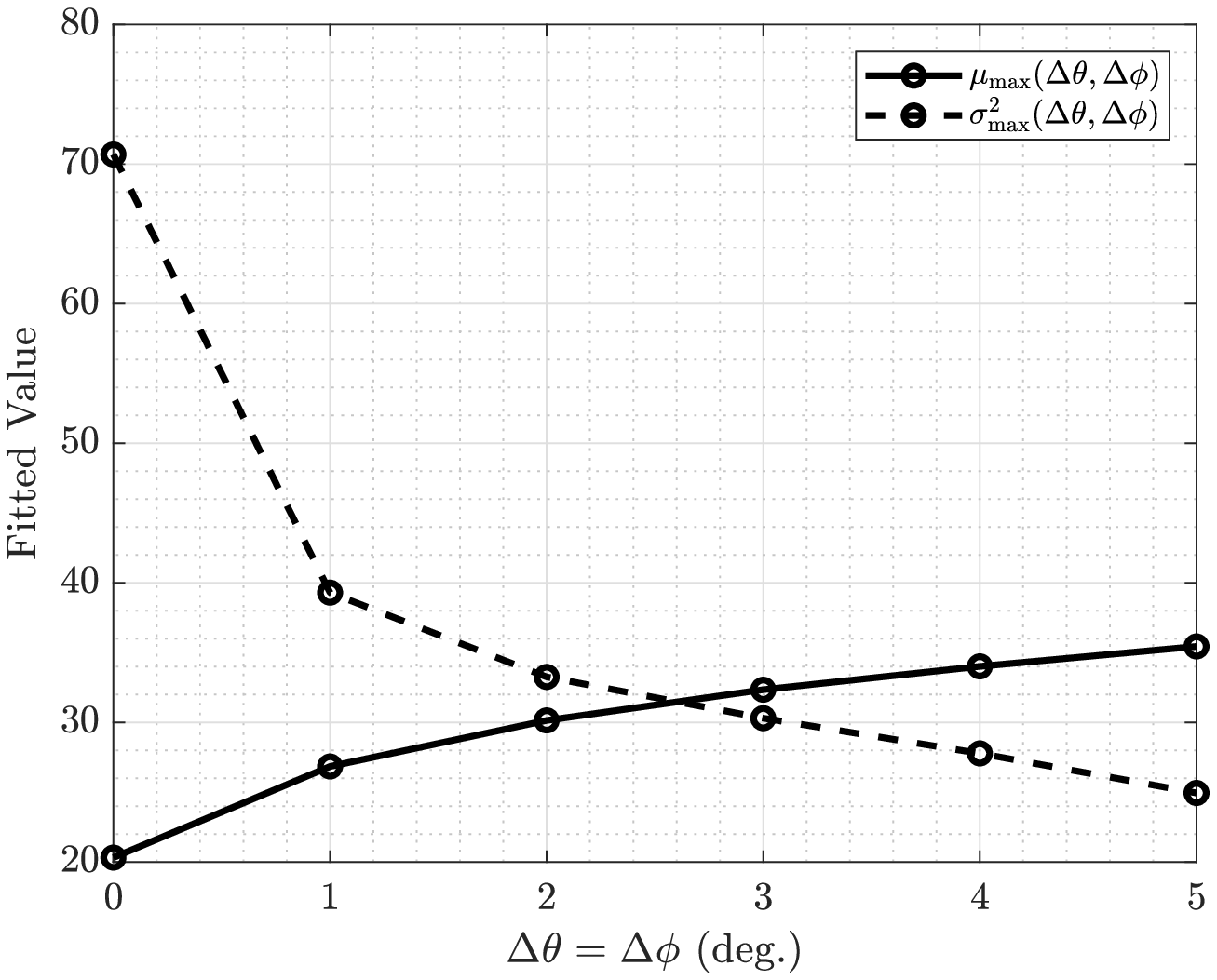}
        \label{fig:inr-max-b}}
    \caption{(a) The \gls{cdf} of $\inrijmax(\Delta\theta,\Delta\phi)$ for various $(\Delta\theta,\Delta\phi)$. (b) The fitted parameters $\mumax(\Delta\theta,\Delta\phi)$ and $\varmax(\Delta\theta,\Delta\phi)$ for various $\Delta\theta = \Delta\phi$.}
    \label{fig:inr-max}
\end{figure*}

As was done for {minimum} \ginr over various neighborhoods, we have conducted an analysis and modeling for {maximum} \ginr over various neighborhoods.
In \figref{fig:inr-max-a}, we plot the \gcdf of $\inrijmax\nbr$ of all nearly 6.5 million beam pairs for variably sized neighborhoods.
Shown in black is the $\nbrzerozero$-neighborhood, which is simply the \gcdf of $\inrij$.
Deviating by at most $\nbroneone$, the median $\inrijmax$ sees about a $6$ dB increase from about $20$ dB to $26$ dB.
Notice that the lower tail stops around $4$ dB, meaning all nearly 6.5 million beam pairs are within $\nbroneone$ of a beam pair offering an \ginr of $4$ dB or more.
Deviating by at most $\nbrtwotwo$, the median $\inrijmax$ increases to $30$ dB and the lower tail stops above $10$ dB.
This trend continues with diminishing gains as the neighborhood is widened.

\begin{remark}
    Before, our analysis of $\inrijmin$ highlighted that \ginr levels more attractive for full-duplex operation can be reached by small shifts in transmit and receive steering direction.
    \figref{fig:inr-max-a} highlights that small shifts in steering direction can likewise degrade (increase) \ginr.
    In fact, even beam pairs with very low \ginr inherently are highly sensitive, considering a $\nbroneone$ shift (at most) can lead to \ginr levels well above $0$ dB, where full-duplex systems would be overwhelmingly self-interference-limited.
    As such, small shifts in the transmit and/or receive beams have the potential to drive self-interference to levels unfit for full-duplex.  %, unless supplemental self-interference cancellation is used.
    Therefore, if attempting to steer along high-isolation beam pairs to mitigate self-interference, there needs to be fairly high accuracy in doing so, potentially motivating high-resolution phase shifters, for instance.
\end{remark}

Like before, we fit a distribution to the \gpcdf in \figref{fig:inr-max-a} to offer engineers a statistical tool for $\inrijmax$.
We fit a normal distribution as follows
\begin{align}
\braces{\todB{\inrijmax(\Delta\theta,\Delta\phi)} \ \forall \ i,j} \overset{\mathrm{fit}}{\sim} \distgauss{\mumax(\Delta\theta,\Delta\phi)}{\varmax(\Delta\theta,\Delta\phi)}
\end{align}
where $\mumax(\Delta\theta,\Delta\phi)$ and $\varmax(\Delta\theta,\Delta\phi)$ are the fitted mean and variance of the normal distribution.
The dashed lines in \figref{fig:inr-max-a} depict each neighborhood's fitted distribution.
\figref{fig:inr-max-b} shows the resulting fitted $\mumax\nbr$ and $\varmax\nbr$ for various $\Delta\theta = \Delta\phi$.
Naturally, the mean $\mumax\nbr$ increases with neighborhood size but does so with diminishing gains.
The variance $\varmax\nbr$ sees a sharp decrease with $\Delta\theta = \Delta\phi = 1^\circ$ from $0^\circ$.
This highlights that so many of the nearly 6.5 million beam pairs are within a mere $\nbroneone$ of notably higher \ginr.
The variance continues to trend down as the neighborhood widens since high-\ginr beam pairs can be more reliably reached.
In addition to the select neighborhoods in \figref{fig:inr-max-a}, we have tabulated the fitted parameters $\parens{\mumax\nbr,\varmax\nbr}$ for a variety of $\nbr$ in \tabref{tab:neighborhood-fits-max}, as was done for $\inrijrng$ and $\inrijmin$.
Engineers can use these distributions to conduct a variety of statistical analyses related to $\inrijmax$ (e.g., worst-case analyses analogous to \eqref{eq:inr-min-cdf-normal}).
%  by analogously translating those for $\inrijmin$ to $\inrijmax$ (i.e., \eqref{eq:inr-min-cdf-normal}).
% Engineers can use these distributions to conduct worst-case analyses, for instance, by 
%For instance, the probability that a random $\nbr$-neighborhood will exhibit a maximum \ginr of at least $\gamma$ can be approximated as
%\begin{align}
%\prob{{\inrijmax} \geq {\gamma}} = 1 - \frac{1}{2} \brackets{1 + \erf{\frac{\todB{\gamma}-\mumax\nbr}{\stdmax \cdot \sqrt{2}}}}.
%\end{align}
% or for a variety of other statistical analyses related to $\inrijmax$.

% \input{tab/tab-neighborhood-fits-max}

% \input{tab/tab-neighborhood-fits-max-inr}

\begin{figure}
    \centering
    \includegraphics[width=\linewidth,height=0.25\textheight,keepaspectratio]{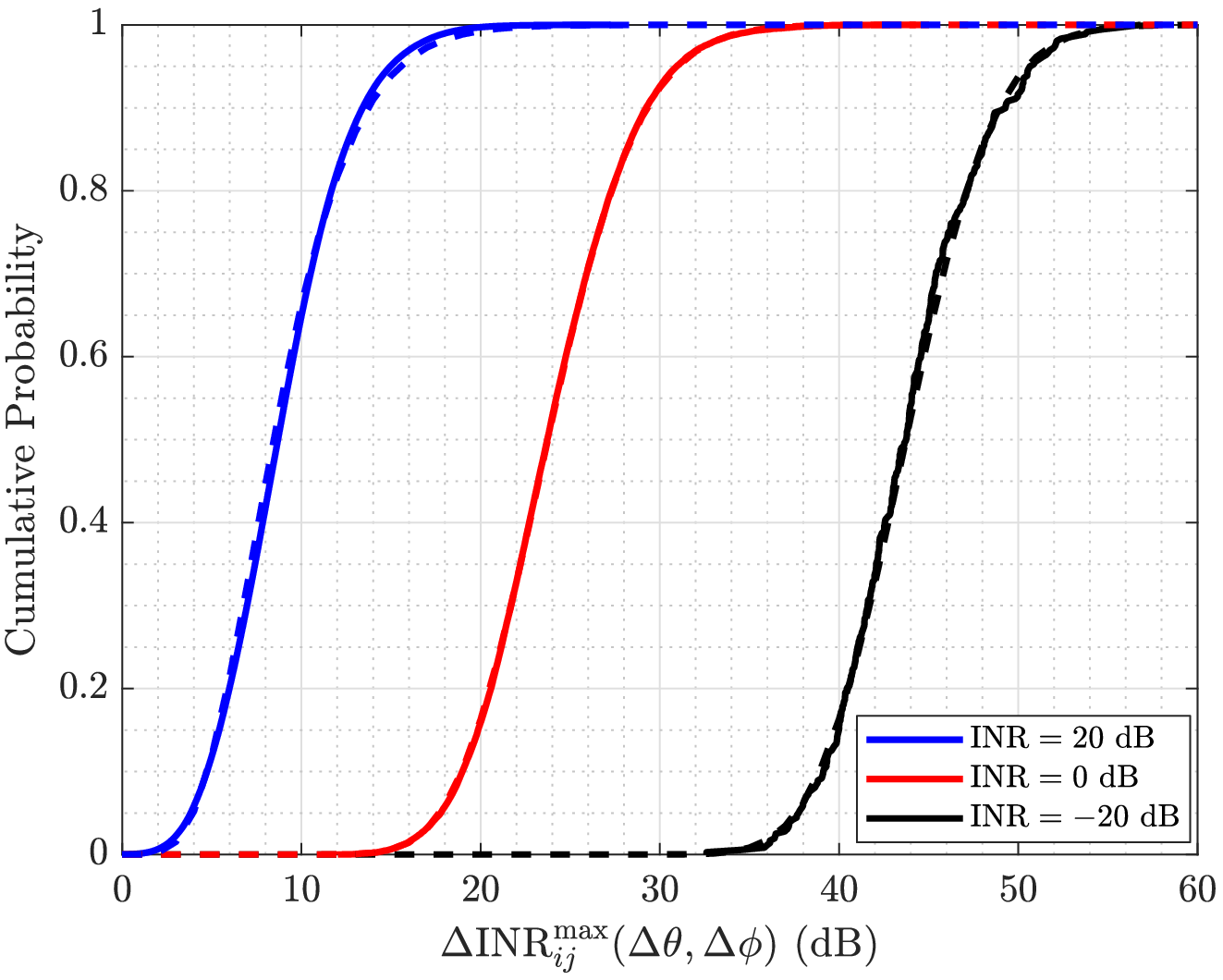}
    \caption{The \gcdf of $\braces{\Delta\inrijmax\nbrinrij : \inrij \approx \inr}$ for various $\inr$, where $\nbr = \nbrtwotwo$. Their fitted counterparts are shown as dashed lines. Beam pairs with inherently low \ginr are typically at a greater risk of large increases in \ginr caused by small shifts in the transmit and receive beams.}
    \label{fig:delta-inr-max}
\end{figure}

As was done with minimum \ginr, we conduct a statistical fit of $\inrijmax$ that depends on $\inrij$.
We define $\Delta\inrijmax\nbr$ as 
\begin{align}
\Delta\inrijmax\nbrinrij = \todB{\inrijmax(\Delta\theta,\Delta\phi)} - \todB{\inrij} \geq 0
\end{align}
which is the difference (in dB) between the maximum \ginr within the $\nbr$-neighborhood surrounding the $(i,j)$-th transmit-receive beam pair and the \ginr offered by that beam pair.
We form the set
% Map from $\inr$ to an $\inrijmin$ distribution over 
\begin{align}
\DDmax\parens{\Delta\theta,\Delta\phi,\inr} = 
\braces{\Delta\inrijmax\nbrinrij : \inrij \approx \inr} 
\end{align}
by collecting all $\Delta\inrijmax\nbrinrij$ for beam pairs offering an $\inrij$ of approximately $\inr$.
% The approximation here is again used to ensure $\DDmax$ has a sufficient number of points in it to successfully fit it.
We again use a Gamma distribution to approximate the distribution of $\DDmax$ as
\begin{align}
% \todB{\IImin\parens{\Delta\theta,\Delta\phi,\inr}} \sim \distgauss{\mumin(\Delta\theta,\Delta\phi,\inr)}{\varmin(\Delta\theta,\Delta\phi,\inr)}
{\DDmax\parens{\Delta\theta,\Delta\phi,\inr}} \overset{\mathrm{fit}}{\sim} \distgamma{\alphamax(\Delta\theta,\Delta\phi,\inr)}{\betamax(\Delta\theta,\Delta\phi,\inr)}
\end{align}
where $\alphamax(\Delta\theta,\Delta\phi,\inr)$ and $\betamax(\Delta\theta,\Delta\phi,\inr)$ are the fitted shape and rate parameters, parameterized by $\inr$ in addition to $\nbr$.
% We conducted this fitting for $\inr$ in the range $[-20,40]$ dB to ensure enough measured data points were present in $\DDmin$ for satisfactory fitting.
We plotted fitted distributions using dashed lines in \figref{fig:delta-inr-max} for various $\inr$ and a \nbrtwotwo-neighborhood.

In addition, we tabulated $(\alphamax,\betamax)$ for $\inr \in \braces{-20,-10,\dots,40}$ dB to provide engineers with statistical tools.
Again, weighted averaging may be used to interpolate between $\inr$ values listed in \tabref{tab:neighborhood-fits-max-inr}.
% means to approximate $\mumax(\Delta\theta,\Delta\phi,\inr)$ and $\varmin(\Delta\theta,\Delta\phi,\inr)$.
Like for minimum \ginr, $\Delta\inrijmax$ for particular $\inrij$ is can be realized using these fitted distributions as follows.
% For instance, 
One may draw $\inr$ from the global distribution (i.e., the \gcdf in \figref{fig:cdf}) as
% \begin{align}
$\todB{\inr} \sim \distgauss{\mu}{\sigma^2}$
% \end{align} 
and then use it when referencing \tabref{tab:neighborhood-fits-max-inr} to fetch $\alphamax(\Delta\theta,\Delta\phi,\inr)$ and $\betamax(\Delta\theta,\Delta\phi,\inr)$ based on some neighborhood size $\nbr$.
A realization of the maximum \ginr over the $\nbr$-neighborhood surrounding a beam pair offering a nominal \ginr of $\inr$ can be drawn as
\begin{align}
\todB{\inrmax(\Delta\theta,\Delta\phi,\inr)} \sim \todB{\inr} + \underbrace{\distgamma{\alphamax(\Delta\theta,\Delta\phi,\inr)}{\betamax(\Delta\theta,\Delta\phi,\inr)}}_{\Delta\inrmax\nbrinr}
\end{align}
which can facilitate statistical analyses; \eqref{eq:inr-min-cdf-gamma} can be straightforwardly translated from $\inrmin$ to $\inrmax$, for instance.
Note that \ginr range as a function of $\inr$ can be realized using $\inrmin\nbrinr$ and $\inrmax\nbrinr$.

\comment{
\begin{figure*}
    \centering
    \subfloat[Top $10$\%.]{\includegraphics[width=0.475\linewidth,height=\textheight,keepaspectratio]{plots_v2/angular/main_apple_v01_13}
        \label{fig:max-min-inr-top-bottom-a}}
    \quad
    \subfloat[Bottom $20$\%.]{\includegraphics[width=0.475\linewidth,height=\textheight,keepaspectratio]{plots_v2/angular/main_apple_v01_12}
        \label{fig:max-min-inr-top-bottom-b}}
    \caption{The \gls{cdf} of (a) maximum and (b) minimum $\inr$ for various $\parens{\Delta\theta,\Delta\phi}$-neighborhoods across (a) the $10$\% of beam pairs with the highest isolation and (b) the $20$\% of beam pairs with the lowest isolation.}
    \label{fig:max-min-inr-top-bottom}
\end{figure*}

\subsection{Across High- and Low-Isolation Beam Pairs}
Instead of considering all beam pairs, we now hone in on (i) the $10$\% of beam pairs offering the highest isolation (lowest $\inr$) and (ii) the $20$\% offering the lowest isolation (highest $\inr$).
From a full-duplex perspective, we naturally desire beam pairs offering high isolation.
It is useful to understand how deviating from high-isolation beam pairs or from low-isolation ones can lead to changes in isolation.
Consider \figref{fig:max-min-inr-top-bottom-a}, where we have plotted the \gls{cdf} of maximum \inr over various neighborhoods surrounding the top $10$\% of beam pairs.
The maximum \inr (in dB) offered by beam pairs across $\LLij$ is simply
\begin{align}
\max\braces{\todBm{\powertx} - \todB{\LLij} - \todBm{\powernoise}}.
\end{align}
The dashed line shows the $\inr$ offered inherently by these high-isolation beam pairs, topping out around $\inr  = 10$ dB and having a median $\inr$ of around $7$ dB.
The notable gap between the dashed line and the $(0^\circ,1^\circ)$- and $(1^\circ,0^\circ)$-neighborhoods shows that a mere $1^\circ$ shift in transmit and/or receive steering direction has the potential to increase $\inr$ levels drastically (around a $7$--$8$ dB shift in distribution).
% Shifting by $1^\circ$ in both azimuth and elevation leads to about $90$\% of beam pairs having an $\inr \geq 7$ dB.
Shifting even by this small amount has the potential to reach \inr levels that would typically make a full-duplex system overwhelmingly self-interference-limited, where $\inr \gg 0$ dB.
Shifting by $(1^\circ,1^\circ)$ yields a median maximum $\inr$ of around $19$ dB (a $12$ dB increase) and by $(2^\circ,2^\circ)$ yields around $24$ dB (a $17$ dB increase).

\begin{remark}
\figref{fig:max-min-inr-top-bottom-a} highlights that high-isolation beam pairs are typically highly sensitive to changes in steering direction. 
As such, small shifts in the transmit and/or receive beams have the potential to drive self-interference to levels unfit for full-duplex, unless supplemental self-interference cancellation is used.
Therefore, if attempting to steer along high-isolation beam pairs to mitigate self-interference, there needs to be fairly high accuracy in doing so, motivating the need for sufficiently high-resolution phase shifters, for instance.
\end{remark}

Now, in \figref{fig:max-min-inr-top-bottom-b}, we consider the $20$\% of beam pairs offering the lowest isolation (highest $\inr$), where we show the minimum \inr across various neighborhoods, described as in \eqref{eq:batman}.
Without deviating from these low-isolation beam pairs, the $\inr$ offered is over $27$ dB, meaning full-duplex operation would not be practical without additional self-interference cancellation measures.
Slight deviations of only $1^\circ$ in azimuth and/or elevation improve $\inr$ but not substantially;
shifting by $(1^\circ,1^\circ)$ at most nets a shift of about $-7$ dB in distribution (with exception of the tails).
With a $(2^\circ,2^\circ)$-neighborhood, more high-isolation beam pairs are within reach, as around $25$\% of beam pairs can fetch an $\inr \leq 3$ dB and around $45$\% can fetch an $\inr \leq 10$ dB.
% For instance, $0$\% o
% While a leftward shift in distribution is desirable, perhaps more substantial is the 

% While there remains a large portion of low-isolation beam pairs that are not within arm's reach of low \inr, 

\begin{remark}
The results in \figref{fig:max-min-inr-top-bottom-b} suggest that a significant number of beam pairs with inherently low isolation can potentially greatly improve full-duplex performance with minor shifts in the transmit and receive steering directions---shifts within the $3$ dB beamwidth of $7^\circ$.
This motivates a variety of future work investigating the importance and potential of beam/user selection in \mmwave full-duplex systems.
However, it is clear that a substantial fraction of low-isolation beam pairs are beyond $(2^\circ,2^\circ)$ from ones with notably higher isolation.
Systems requiring a certain $\inr$ threshold to operate successfully in full-duplex can see significant benefits with seemingly minor leftward shifts in distribution since the probability of meeting that threshold can drastically increase (e.g., meeting an $\inr \leq 3$ dB).
% There exists some asymmetry when comparing low-isolation and high-isolation beam pairs, seeing as larger shifts in distribution are present 
\end{remark}

}

\pagebreak

\section{Conclusion} \label{sec:conclusion}

We have collected nearly 6.5 million measurements of multi-panel self-interference at 28 GHz to better understand its spatial and statistical characteristics---providing the most comprehensive examination of such to date.
% Our measurements illustrate that the degree of self-interference coupled between colocated transmitting and receiving phased arrays is extremely sensitive to their steering directions, which can motivate beam steering solutions to mitigate self-interference. 
Our measurements illustrate that the degree of self-interference coupled between colocated transmitting and receiving phased arrays {tends} to be higher when the transmit and receive beams are steered toward one another but small shifts in steering direction (on the order of one degree) can lead to significant changes in such.
We have analyzed and statistically modeled this sensitivity, providing engineers with useful insights and statistical tools that can drive system design and evaluation, including those that may use analog and/or digital self-interference cancellation.
% Small shifts in steering direction can lead to drastic changes in the degree of self-interference coupled between transmit and receive beams.
% The amount of self-interference coupled \textit{tends} to be higher when the transmit and receive arrays are steered toward one another but small shifts in steering direction can lead to significant changes in such.
% This suggests that beamforming-based \mmwave full-duplex solutions will need to be strategic with their steering to avoid self-interference.
% We presented a stochastic model of \mmwave self-interference, allowing researchers to draw countless realizations of \mmwave self-interference that is spatially and statistically aligned with our measurements. % to evaluate proposed full-duplex solutions.
% Our model can produce realizations of self-interference that are comparable in a small- and large-scale sense to our measurements, making it a useful tool to design and evaluate practical \mmwave full-duplex systems.
% Finally, a toy example illustrated the impact self-interference has on potential full-duplex performance, highlighting that beam steering appears to be a promising full-duplex solution by leveraging the widespread variability in self-interference observed in our measurements---taking advantage of the fact that small shifts in steering direction can reduce self-interference without prohibitive loss in beamforming gain.
This measurement campaign sheds light on the efficacy of multi-panel \mmwave full-duplex systems, such full-duplex \iab proposed in 3GPP, and motivates strategic beam steering as a potential route to mitigate self-interference without prohibitively compromising beamforming gain.
% Our measurements and analyses can be useful references when developing \mmwave full-duplex solutions, including and beyond those relying solely on beamforming to mitigate self-interference.
Valuable future work would investigate the impacts of beam shape, array size, environmental reflections, and relative array geometry on self-interference.
Future directions capitalizing on this campaign include beam selection for \mmwave full-duplex, proposing a practically sound \mimo channel model for \mmwave self-interference, and prototyping full-duplex \mmwave systems.

\pagebreak

\appendix

\section{Tabulated Fitting Results}

\begin{table*}[!ht]
    % \small
    \centering
    \caption{The fitted parameters $\parens{\alpharng(\Delta\theta,\Delta\phi),\betarng(\Delta\theta,\Delta\phi)}$ for various $(\Delta\theta,\Delta\phi)$.}
    \label{tab:neighborhood-fits-rng}
    \begin{tabular}{|c|cccccc|}
        \hline
        \diagbox{$\Delta\phi$}{$\Delta\theta$} & $0^\circ$ & $1^\circ$ & $2^\circ$ & $3^\circ$ & $4^\circ$ & $5^\circ$ \\
        \hline
% $0^\circ$ & --- & $(2.7,3.4)$ & $(4.4,3.6)$ & $(6.7,3.1)$ & $(9.5,2.6)$ & $(12.5,2.2)$\\$1^\circ$ & $(2.6,3.2)$ & $(4.5,4.1)$ & $(6.9,3.8)$ & $(10.9,2.9)$ & $(16.0,2.2)$ & $(21.7,1.8)$\\$2^\circ$ & $(4.0,3.5)$ & $(6.6,3.8)$ & $(10.7,3.1)$ & $(17.6,2.2)$ & $(26.0,1.6)$ & $(34.7,1.3)$\\$3^\circ$ & $(5.8,3.2)$ & $(9.6,3.2)$ & $(16.3,2.3)$ & $(26.7,1.6)$ & $(38.1,1.2)$ & $(48.8,1.0)$\\$4^\circ$ & $(8.0,2.8)$ & $(13.9,2.5)$ & $(23.8,1.8)$ & $(37.3,1.3)$ & $(50.5,1.0)$ & $(61.8,0.9)$\\$5^\circ$ & $(10.4,2.4)$ & $(18.9,2.0)$ & $(31.7,1.4)$ & $(47.2,1.1)$ & $(61.2,0.9)$ & $(72.5,0.8)$\\
$0^\circ$ & --- & $(2.74,3.40)$ & $(4.42,3.64)$ & $(6.73,3.15)$ & $(9.50,2.62)$ & $(12.50,2.22)$\\$1^\circ$ & $(2.59,3.19)$ & $(4.52,4.10)$ & $(6.90,3.82)$ & $(10.90,2.94)$ & $(16.04,2.25)$ & $(21.69,1.80)$\\$2^\circ$ & $(4.04,3.48)$ & $(6.57,3.85)$ & $(10.69,3.11)$ & $(17.57,2.21)$ & $(25.96,1.64)$ & $(34.67,1.30)$\\$3^\circ$ & $(5.80,3.22)$ & $(9.63,3.16)$ & $(16.31,2.35)$ & $(26.67,1.63)$ & $(38.06,1.23)$ & $(48.77,1.01)$\\$4^\circ$ & $(7.98,2.81)$ & $(13.91,2.49)$ & $(23.80,1.77)$ & $(37.32,1.26)$ & $(50.52,0.99)$ & $(61.80,0.85)$\\$5^\circ$ & $(10.39,2.44)$ & $(18.85,2.00)$ & $(31.67,1.42)$ & $(47.24,1.05)$ & $(61.17,0.86)$ & $(72.50,0.76)$\\
        \hline
    \end{tabular}
\end{table*}

\begin{table*}[!ht]
    % \small
    \centering
    \caption{The fitted parameters $\parens{\mumin(\Delta\theta,\Delta\phi),\varmin(\Delta\theta,\Delta\phi)}$ for various $(\Delta\theta,\Delta\phi)$.}
    \label{tab:neighborhood-fits-min}
    \begin{tabular}{|c|cccccc|}
        \hline
         \diagbox{$\Delta\phi$}{$\Delta\theta$} & $0^\circ$ & $1^\circ$ & $2^\circ$ & $3^\circ$ & $4^\circ$ & $5^\circ$ \\
        \hline
        $0^\circ$ & $(20.32,70.69)$ & $(15.04,102.75)$ & $(10.34,114.62)$ & $(6.47,112.94)$ & $(3.53,107.37)$ & $(1.31,101.93)$\\$1^\circ$ & $(15.58,98.86)$ & $(8.32,148.79)$ & $(2.30,152.39)$ & $(-2.34,135.13)$ & $(-5.62,118.00)$ & $(-7.98,105.59)$\\$2^\circ$ & $(11.58,109.04)$ & $(3.15,153.04)$ & $(-3.07,141.96)$ & $(-7.56,117.66)$ & $(-10.61,99.81)$ & $(-12.80,88.34)$\\$3^\circ$ & $(8.23,106.83)$ & $(-0.88,137.50)$ & $(-6.93,119.30)$ & $(-11.07,96.78)$ & $(-13.87,82.49)$ & $(-15.89,73.80)$\\$4^\circ$ & $(5.50,99.76)$ & $(-3.98,118.15)$ & $(-9.74,99.32)$ & $(-13.57,81.48)$ & $(-16.16,70.95)$ & $(-18.05,64.68)$\\$5^\circ$ & $(3.39,93.25)$ & $(-6.27,103.55)$ & $(-11.76,86.70)$ & $(-15.37,72.63)$ & $(-17.82,64.54)$ & $(-19.63,59.62)$\\
        \hline
    \end{tabular}
\end{table*}

\begin{table*}[!ht]
    % \small
    \centering
    \caption{The fitted parameters $\parens{\alphamin\nbrinr,\betamin\nbrinr}$ for various \nbrinr.}
    \label{tab:neighborhood-fits-min-inr}
    \begin{tabular}{|c|ccccccc|}
        \hline
        & \multicolumn{7}{c|}{$\inr$ (dB)} \\
$(\Delta\theta,\Delta\phi)$ & $-20$ & $-10$ & $0$ & $10$ & $20$ & $30$ & $40$ \\ \hline $(1^\circ,1^\circ)$ & $(0.19,1.34)$ & $(0.21,11.37)$ & $(0.89,10.27)$ & $(3.66,4.05)$ & $(3.26,3.91)$ & $(3.22,2.80)$ & $(4.09,1.47)$ \\$(2^\circ,2^\circ)$ & $(0.18,4.91)$ & $(0.41,13.93)$ & $(3.06,4.83)$ & $(10.28,2.21)$ & $(8.67,2.93)$ & $(5.26,4.18)$ & $(4.15,4.03)$ \\$(3^\circ,3^\circ)$ & $(0.21,9.31)$ & $(0.87,10.20)$ & $(6.49,2.86)$ & $(17.38,1.58)$ & $(19.94,1.66)$ & $(12.18,2.74)$ & $(7.96,3.79)$ \\$(4^\circ,4^\circ)$ & $(0.26,12.40)$ & $(1.86,6.27)$ & $(10.40,2.07)$ & $(24.37,1.26)$ & $(31.12,1.21)$ & $(22.78,1.77)$ & $(20.63,1.94)$ \\$(5^\circ,5^\circ)$ & $(0.39,11.99)$ & $(2.98,4.66)$ & $(15.20,1.57)$ & $(31.79,1.05)$ & $(40.47,1.01)$ & $(30.88,1.45)$ & $(36.15,1.27)$ \\
        \hline
    \end{tabular}
\end{table*}

\begin{table*}[!ht]
    % \small
    \centering
    \caption{The fitted parameters $\parens{\mumax(\Delta\theta,\Delta\phi),\varmax(\Delta\theta,\Delta\phi)}$ for various $(\Delta\theta,\Delta\phi)$.}
    \label{tab:neighborhood-fits-max}
    \begin{tabular}{|c|cccccc|}
        \hline
         \diagbox{$\Delta\phi$}{$\Delta\theta$} & $0^\circ$ & $1^\circ$ & $2^\circ$ & $3^\circ$ & $4^\circ$ & $5^\circ$ \\
        \hline
$0^\circ$ & $(20.32,70.69)$ & $(24.36,44.95)$ & $(26.42,39.02)$ & $(27.63,36.81)$ & $(28.41,35.64)$ & $(29.06,34.66)$\\$1^\circ$ & $(23.84,49.63)$ & $(26.85,39.31)$ & $(28.64,35.37)$ & $(29.71,33.95)$ & $(30.41,33.26)$ & $(31.00,32.60)$\\$2^\circ$ & $(25.63,45.34)$ & $(28.42,37.00)$ & $(30.15,33.27)$ & $(31.18,31.99)$ & $(31.85,31.51)$ & $(32.41,31.00)$\\$3^\circ$ & $(26.89,43.38)$ & $(29.61,35.48)$ & $(31.33,31.57)$ & $(32.35,30.31)$ & $(33.01,29.93)$ & $(33.56,29.52)$\\$4^\circ$ & $(27.89,41.95)$ & $(30.61,33.81)$ & $(32.35,29.51)$ & $(33.38,28.13)$ & $(34.02,27.79)$ & $(34.57,27.40)$\\$5^\circ$ & $(28.72,40.61)$ & $(31.46,32.02)$ & $(33.23,27.27)$ & $(34.26,25.73)$ & $(34.90,25.36)$ & $(35.45,24.95)$\\
        \hline
    \end{tabular}
\end{table*}

\begin{table*}[!ht]
    % \small
    \centering
    \caption{The fitted parameters $\parens{\alphamax\nbrinr,\betamax\nbrinr}$ for various \nbrinr.}
    \label{tab:neighborhood-fits-max-inr}
    \begin{tabular}{|c|ccccccc|}
        \hline
        & \multicolumn{7}{c|}{$\inr$ (dB)} \\
$(\Delta\theta,\Delta\phi)$ & $-20$ & $-10$ & $0$ & $10$ & $20$ & $30$ & $40$ \\ \hline $(1^\circ,1^\circ)$ & $(97.36,0.39)$ & $(53.69,0.53)$ & $(22.22,0.83)$ & $(8.27,1.23)$ & $(3.94,1.44)$ & $(3.65,1.09)$ & $(3.73,0.72)$ \\$(2^\circ,2^\circ)$ & $(125.30,0.35)$ & $(74.38,0.46)$ & $(35.07,0.68)$ & $(14.26,1.04)$ & $(6.18,1.44)$ & $(4.83,1.28)$ & $(4.76,0.84)$ \\$(3^\circ,3^\circ)$ & $(132.69,0.36)$ & $(83.02,0.45)$ & $(42.63,0.64)$ & $(18.35,0.97)$ & $(7.76,1.44)$ & $(5.40,1.41)$ & $(5.37,0.88)$ \\$(4^\circ,4^\circ)$ & $(143.23,0.35)$ & $(92.33,0.43)$ & $(49.46,0.60)$ & $(22.61,0.89)$ & $(9.53,1.36)$ & $(5.96,1.46)$ & $(6.10,0.84)$ \\$(5^\circ,5^\circ)$ & $(160.40,0.32)$ & $(107.86,0.39)$ & $(58.95,0.53)$ & $(28.38,0.78)$ & $(12.12,1.19)$ & $(6.53,1.47)$ & $(6.94,0.79)$ \\
        \hline
    \end{tabular}
\end{table*}

\pagebreak

% \section*{References} \label{sec:bibliography}
% \printbibliography[heading=none]
\bibliographystyle{bibtex/IEEEtran}
\bibliography{bibtex/IEEEabrv,refs}

% \pagebreak

% \input{sec-biography.tex}

\end{document}